\documentclass[aps,preprint,showpacs,floatfix,groupedaddress]
{revtex4-1}
\usepackage{natbib} 
\usepackage{ulem}
\usepackage{graphicx}
\usepackage{amsmath}
\usepackage{textcomp}
\usepackage{mathtools}
\usepackage{bm}
\usepackage{hyperref}
\usepackage{amssymb}
\usepackage{multirow}

\usepackage{subfigure}

\newif\ifhyper
\hypertrue
\ifhyper
\hypersetup{
   citecolor = {blue},
   colorlinks = {true}, 
   linkcolor = {blue},
   urlcolor = {blue} 
}
\fi

\begin{document}

\title{Repetitive acoustic streaming patterns in sinusoidal shaped microchannels}

\author{Elnaz Attar Jannesar}
\author{Hossein Hamzehpour}%
\email{hamzehpour@kntu.ac.ir}
\affiliation{Department of Physics, K.N. Toosi University of Technology, Tehran 15875-4416, Iran}

\date{\today}

\begin{abstract}
Geometry of the fluid container plays a key role in the shape of acoustic streaming patterns. Inadvertent vortices can be troublesome in some cases, but if treated properly, the problem turns into a very useful parameter in acoustic tweezing or micromixing applications. In this paper, the effects of sinusoidal boundaries of a microchannel on acoustic streaming patterns are studied. Results show that while top and bottom sinusoidal walls are vertically actuated at the resonance frequency of basic hypothetical rectangular microchannel, some repetitive acoustic streaming patterns are recognized in classifiable cases. Such patterns can never be produced in rectangular geometry with flat boundaries. Relations between geometrical parameters and emerging acoustic streaming patterns lead us to propose formulas in order to predict more cases. Such results and formulations were not trivial at a glance. 



\end{abstract}

\maketitle
\section{Introduction}
\label{intro}
Ultrasound acoustic standing waves are used to generate two nonlinear acoustophoretic forces for manipulation of fluids and particles inside microfluidic systems \cite{Wiklund2012c}.
The acoustic radiation force tends to focus particles on the nodal or anti-nodal plane of the acoustic standing waves \cite{King1934,Yosioka1955,GorKov1962,Doinikov1997} while the Stokes drag force of the acoustic streaming velocity field, tends to defocus and spread out the suspended particles \cite{Rayleigh1884a,Schlichting1955a,Nyborg1953,Nyborg1958}. Critical particle diameter is determined as a crossover from the radiation force dominated region to the acoustic streaming-induced drag force dominated region \cite{Spengler2003,Barnkob2012}.
Particles with diameters larger than the critical diameter are enforced with the radiation force, as it caused by the scattering of acoustic waves from the surface of particles. On the other hand, tiny particles smaller than the critical size are affected by the acoustic streaming steady fluid flows, caused by viscous stresses in acoustic boundary layers.

The classical theory of Rayleigh streaming was established for shallow infinite parallel-plate channels \cite{Rayleigh1884a}. Schlichting streaming is modeled mathematically for single planar infinite rigid walls \cite{Schlichting1955a}. Many further studies have followed the same geometries \cite{Westervelt1953a,Hamilton2003,Nyborg1958,Rednikov2011a}. Muller et. al.\cite{Muller2013} have proposed a theoretical analysis of acoustic streaming with taking the effect of the vertical sidewalls into account. They published a complete description of microparticle acoustophoresis in combined with wall effects. 

For geometries more irregular than rectangular microchannels, the analytical studies are impossible and numerical simulations must be employed. To date, there has been a large quantity of literature published on this topic including analytical, numerical and experimental studies. Any changes in the boundary conditions, such as the geometry of the fluid container, dramatically affect the shape of the acoustic streaming patterns. Inadvertent vortices can be troublesome in some cases, but if treated properly, the problem turns into a very useful parameter in patterning or mixing applications \cite{Evander2012a,Wiklund2012c}. A microchannel with the sharp-edged sidewalls has been used as a micromixer \cite{Nama2014}. Same geometry proposes precise rotational manipulation of cells and other micrometer-sized biological samples \cite{Feng2018}. Also, additional study shows that mixing performance varies at different frequencies and tip angles \cite{Huang2013a}. Oscillations of tilted sharp-edge structures have suggested a programmable acoustofluidic pump \cite{Huang2014}. Function of oscillating microbubbles have considered in some other literatures. Acoustically driven sidewall-trapped microbubbles act as a fast microfluidic mixer \cite{Ahmed2009}. Microbubbles inside a horseshoe structure produce acoustic streaming vortices which are able to trap bacterial aggregations \cite{Yazdi2012}. In the other hand, focusing the sub-micrometer particles and bacteria both horizontally and vertically in the cross section of a microchannel is feasible using two-dimensional acoustic streaming phenomena. The single roll streaming flow is observed experimentally in a nearly-square channel, and acoustophoretic focusing of E. Coli bacteria and 0.6 µm particles is achieved \cite{Antfolk2014}. An ultrasonic device for micro-patterning and precision manipulation of micrometer-scale particles has been introduced using eight piezoelectric transducers shaped into an octagonal cavity \cite{bernassau2013interactive}. 
The effects of profiled surfaces on the boundary-driven streaming fields in 2D rectangular chambers have numerically investigated by Lei et. al \cite{Lei2018}. Their models predict that profiles with amplitudes comparable to the viscous boundary layer have the potential to dramatically enhance (and change the pattern of) acoustic streaming patterns. 
 
In this work, we numerically investigate the effects of different geometrical parameters on two-dimensional acoustic streaming patterns inside microchannels with acoustically oscillating sinusoidal walls in vertical direction. Some special acoustic streaming patterns emerge in the form of repetitive shapes in special cases that can never be produced in rectangular geometry with flat boundaries using one-dimensional oscillations. We propose a relation between such patterns and geometrical parameters that lead us to predict much more cases. Such results for sinusoidal geometry were not trivial at a glance.

The paper is organized as follows. In Sec.II we derive the governing equations that are solved numerically. This is followed in Sec.III by description to the numerical model and considering boundary conditions. In Sec.IV effects of three geometrical parameters are discussed that are the ratio of the side walls, symmetry or asymmetry of sinusoidal walls and geometrical wavelength of them. A formulation is suggested to make other cases predictable. Finally, an application is introduced numerically to trap sub-micron particles inside a sinusoidal microchannel in single tweezing points. Such trapping was never achieved in two-dimensional cases with only one directionally oscillation of boundaries. All conclusions stated in this paper can be leading points to optimize the performance of acoustofluidic devices.

\section{Theory}
\label{Th}
In the absence of external body forces and heat sources, there are three important governing equations in microfluidic systems as \cite{Muller2014}
\begin{subequations}
\begin{align}
&\partial_t{\rho}=\mbox{\bm{$\nabla$}}\cdot[-\rho \bm{v}],\label{eq1.a}\\
&\partial_t{(\rho \bm{v})}=\mbox{\bm{$\nabla$}}\cdot[\bm\sigma-\rho \bm{vv}],\label{eq1.b}\\
&\partial_t{(\rho \varepsilon+\tfrac12\rho 
v^2)}=\mbox{\bm{$\nabla$}}\cdot[k^{th}\mbox{\bm{$\nabla$}}T+\bm 
v\cdot\bm\sigma-\rho(\varepsilon+\tfrac12 v^2)\bm{v}]
\label{eq1.c}.
\end{align}
\label{eq1}
\end{subequations}
The continuity equation \ref{eq1.a} expresses conservation of mass where the mass current density is $\rho$, the Navier-Stokes equation \ref{eq1.b} expresses conservation of momentum where momentum current density is $\rho\bm{v}$, and the equation \ref{eq1.c} expresses conservation of energy where energy current density is $\rho(\varepsilon+\tfrac12 v^2)$ 
\cite{Landau1967,Muller2014}. $\epsilon$ is internal energy per unit mass, $\bm{v}$ is velocity of the fluid in the medium and $k^{th}$ is the thermal conductivity. Also, $\bm\sigma$ is  the stress tensor of the fluid (Cauchy stress tensor) as
\begin{equation}
\bm\sigma=\bm\tau -p\bm I=\eta[\mbox{\bm{$\nabla$}}\bm 
v+(\mbox{\bm{$\nabla$}}\bm v)^\textrm{T}]+[(\eta^B-\tfrac 
23\eta)\mbox{\bm{$\nabla$}}\cdot \bm{v}-p\,]\bm I,
\label{eq2}
\end{equation}
where the viscous stress tensor, $\bm\tau$, is expressed in terms of dynamic shear viscosity $\eta$ and bulk viscosity $\eta^B$. Additionally, $p$ is the pressure field and $\bm I$ is the unit tensor. 

Thermal effects will be neglected, because the thermal boundary layer thickness (thermal diffusion length), $\delta_{t}$, in fluids is much smaller than viscous boundary layer thickness (viscous penetration depth), $\delta_{\nu}$ that are defined as \cite{Muller2012}
\begin{subequations}
\begin{align}
&\delta_{t}=\sqrt{\frac{2D_{th}}{\omega}},\\
&\delta_{\nu}=\sqrt{\frac{2\nu}{\omega}},
\end{align}
\label{eq3}
\end{subequations}
where $D_{th}$ is the thermal diffusion constant, $\omega$ is angular frequency of the acoustic field and $\nu=\frac{\eta}{\rho}$ is the dynamic viscosity.

Considering the external acoustic field as a perturbation of the steady state of a fluid, All the fields can be expanded as $g=g_0+g_1+g_2$ taking first and second order (subscript 1 and 2, respectively) into account. We expand the non-linear fluid equations to second order.

For a medium with the speed of sound $c_0$, the magnitude of the perturbation can be characterized by the dimensionless acoustic Mach number \cite{Landau1967} as
\begin{equation}
Ma=\frac{v_1}{c_0}=\frac{\vert\rho_1\vert}{\rho_0}\ll1,
\label{eq4}
\end{equation}
where $\rho_0$ is the unperturbed density of the fluid and $\rho_1$ is the first order perturbation term of density. Ignoring thermal effects, the first-order perturbation approximation of governing equations in frequency domain are \cite{Muller2015} 
\begin{subequations}
\begin{align}
& \mbox{\bm{$\nabla$}}\cdot \bm{v}_1-i\omega\kappa_0 p_1=0,\label{eq5.a}\\
& \mbox{\bm{$\nabla$}}\cdot\bm\sigma_1+i\omega\rho_0\bm{v}_1=0,\label{eq5.b}
\end{align}
\label{eq5}
\end{subequations}
The acoustic pressure is calculated as $p_1=c_0^2 \rho_1$ and $\kappa_0=\frac{1}{\rho}\left(\frac{\partial \rho}{\partial p} \right)_s=\frac{1}{\rho_0 c_0^2}$ is isentropic compressibility where $c_0$ is the speed of sound in water at rest . Using the first order approximation of the perturbation theory, the stress tensor is calculated as
\begin{equation}
\bm\sigma_1=\bm\tau_1-p_1\bm I=\eta_0[\mbox{\bm{$\nabla$}}\bm 
v_1+(\mbox{\bm{$\nabla$}}\bm v_1)^\textrm{T}]+[(\eta_0^B-\tfrac 
23\eta_0)\mbox{\bm{$\nabla$}}\cdot \bm{v_1}-p_1]\bm{I},
\label{eq6}
\end{equation}
where zero indices refer to the properties at the equilibrium. 
The time averaged second-order perturbation approximation of governing equations in frequency domain are \cite{Muller2015}
\begin{subequations}
\begin{align}
& \mbox{\bm{$\nabla$}}\cdot{\left\langle \bm{v}_2 \right\rangle }+\kappa_0 
\left\langle \bm{v}_1\cdot \mbox{\bm{$\nabla$}}p_1\right\rangle =0,\label{eq7.a}\\
& \mbox{\bm{$\nabla$}}\cdot[\bm\sigma_2-\rho_0\left\langle 
\bm{v}_1\bm{v}_1\right\rangle]=0.\label{eq7.b}
\end{align}
\label{eq7}
\end{subequations}

The stress tensor of the fluid using the second-order perturbation theory is defined as,
\begin{equation}
\begin{aligned}
\bm\sigma_2=\bm\tau_2 &-p_2\bm I =\eta_0[\mbox{\bm{$\nabla$}}\bm 
v_2+(\mbox{\bm{$\nabla$}}\bm v_2)^\textrm{T}]\\&+[(\eta_0^B-\tfrac 
23\eta_0)(\mbox{\bm{$\nabla$}}\cdot \bm{v_2})\bm I]+\left\langle \eta_1[\mbox{\bm{$\nabla$}}\bm v_1+(\mbox{\bm{$\nabla$}}\bm 
v_1)^\textrm{T}] \right\rangle\\&+\left\langle[(\eta_1^B-\tfrac 
23\eta_1)(\mbox{\bm{$\nabla$}}\cdot \bm{v_1})\bm I]\right\rangle 
-p_2\bm I, 
\end{aligned}
\label{eq8}
\end{equation}
where $\eta_1^B$ and $\eta_1$ are the bulk viscosity and dynamic shear viscosity of the fluid, respectively. As mentioned above, in adiabatic thermodynamic approximation, the thermal term is ignored.
Resulting expression for the acoustic radiation force on a spherical particle with radius of $a$ is obtained by 
\cite{Settnes2012,Muller2012}

\begin{equation}
\mbox{$\bm{F}_{rad}$}=-\pi 
a^3\left[\frac{2\kappa_0}{3}Re[f_1^*p_1^{*in}\mbox{\bm{$\nabla$}}p_1^{in}]-\rho_0 
Re[f_2^*\bm{v}_1^{*in}\cdot\mbox{\bm{$\nabla$}}\bm{v}_1^{in}] \right],
\label{eq9}
\end{equation}
where $\bm v_1^{in}$ and $p_1^{in}$ are the first-order pressure and velocity fields of the incident acoustic wave evaluated at a particle position. Asterisks denote complex conjugations. The prefactors $f_1$ and $f_2$ are the so-called mono- and dipole scattering coefficients, respectively that in viscous fluid are calculated as \cite{Settnes2012,Muller2012}
\begin{subequations}
\begin{align}
&f_1(\tilde{\kappa})=1-\tilde{\kappa},&\textrm{with} \ \tilde{\kappa}=\frac{\kappa_p}{\kappa_0},\\
&f_2(\tilde{\rho},\tilde{\delta_{\nu }})=\frac{2\left[1-\Gamma(\tilde{\delta_{\nu 
}}) \right] (\tilde{\rho}-1)}{2\tilde{\rho}+1-3\Gamma(\tilde{\delta_{\nu }})},&\textrm{with} \  \tilde{\rho}=\frac{\rho_p}{\rho_0},\\
&\Gamma(\tilde{\delta_{\nu }})=-\frac{3}{2}\left[ 
1+\textrm{i}\left(1+\tilde{\delta_{\nu }} \right) \right] \tilde{\delta_{\nu 
}},&\textrm{with}\  \tilde{\delta_{\nu }}=\frac{\delta_{\nu }}{a},
\end{align}
\label{eq10}
\end{subequations}
where $\kappa_p$ and $\rho_p$ are the compressibility and density of the 
particles, respectively. 

The time-averaged streaming-induced drag force on a spherical particle of radius $a$ moving with velocity $\bm u$, far from the channel walls in a fluid with time averaged streaming velocity $\left\langle\bm{v}_2\right\rangle$ is given by \cite{Settnes2012}
\begin{equation}
\mbox{$\bm{F}_{drag}$}=6\pi\eta a (\left\langle \bm{v}_2\right\rangle -\bm u).
\label{eq11}
\end{equation}

The nonlinear acoustophoretic forces compete with each other. The crossover from the dominance of each force is defined through a critical particle radius. For a fixed spherical particle inside a rectangular microchannel, the crossover diameter is 
\cite{Muller2012}
\begin{equation}
2a_c=\sqrt{12\frac{\Psi}{\Phi}}\delta_{\nu},
\label{eq12}
\end{equation}
where $\Psi$ is a factor related to the channel geometry. The acoustophoretic contrast factor is 
calculated as $\Phi(\tilde{\kappa},\tilde{\rho},\tilde{\delta_{\nu }}) 
=\frac{1}{3}f_1(\tilde{\kappa})+\frac{1}{2}f_2^{r}(\tilde{\rho},\tilde{\delta_{\nu}})$ that contains material parameters. The monopole scattering coefficient, $f_1$, is real valued and depends only on the compressibility ratio between the particle and the fluid, $\tilde\kappa$ but the viscosity-dependent dipole scattering coefficient ,$f_2$, is in general a complex-valued number, and its real value is abbreviated as $f_2^r(\tilde{\rho},\tilde{\delta_{\nu }})=Re\left[f_2(\tilde{\rho},\tilde{\delta_{\nu }}) \right]$.

\section{Numerical model and boundary conditions}\label{NU}
In the following, a numerical model is presented for microchannels with actuating sinusoidal top and bottom walls. Finite element method is one of the most widely used numerical methods in computational simulations. In this study, the same method is utilized considering the weak form of the partial differential equations. 
\begin{figure*}[h]
	\begin{center} 
		\begin{tabular}{cc}
			\subfigure[]{\includegraphics[width=55mm]{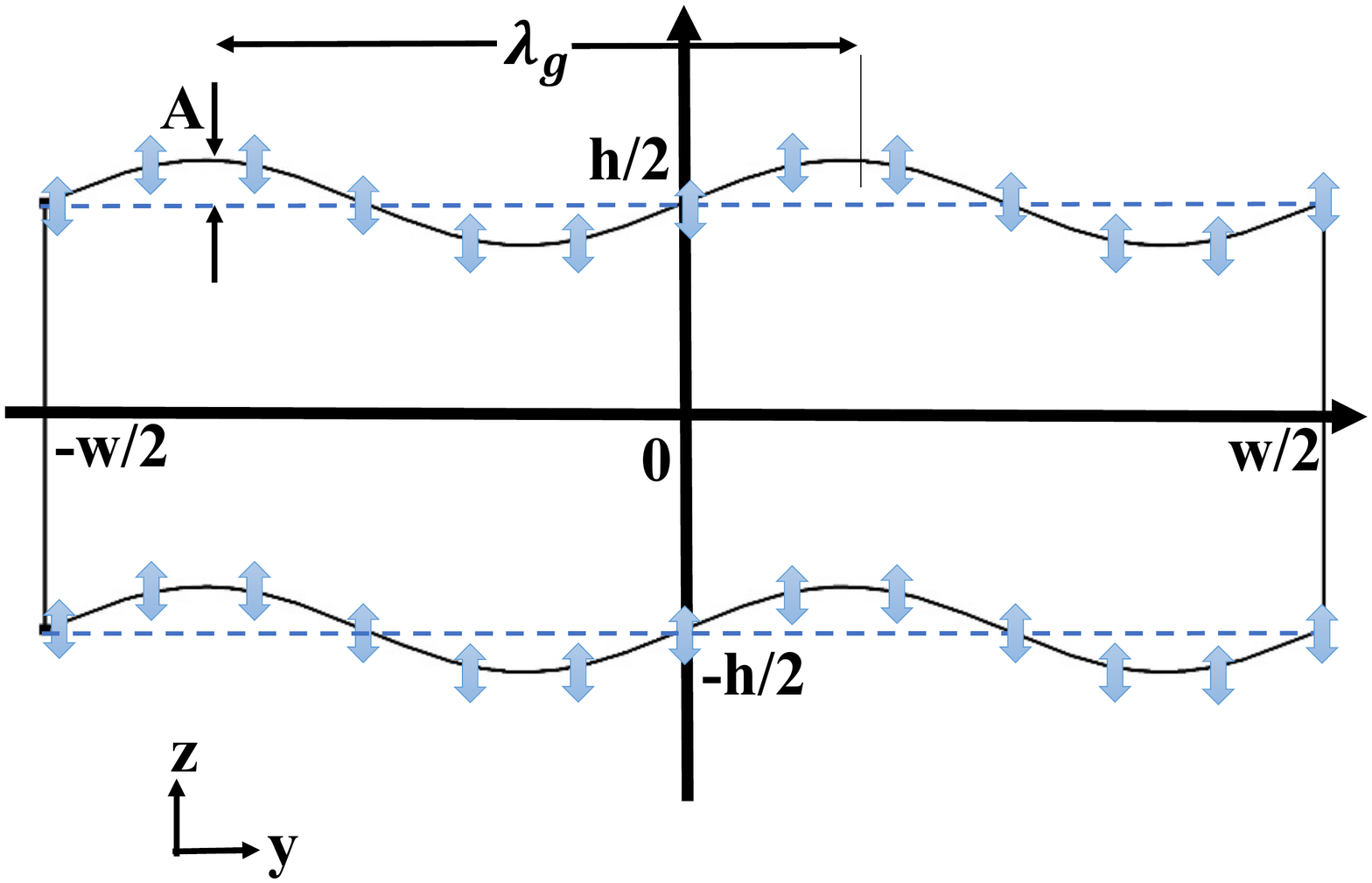}\label{fig1.a}}&
			\subfigure[]{\includegraphics[width=55mm]{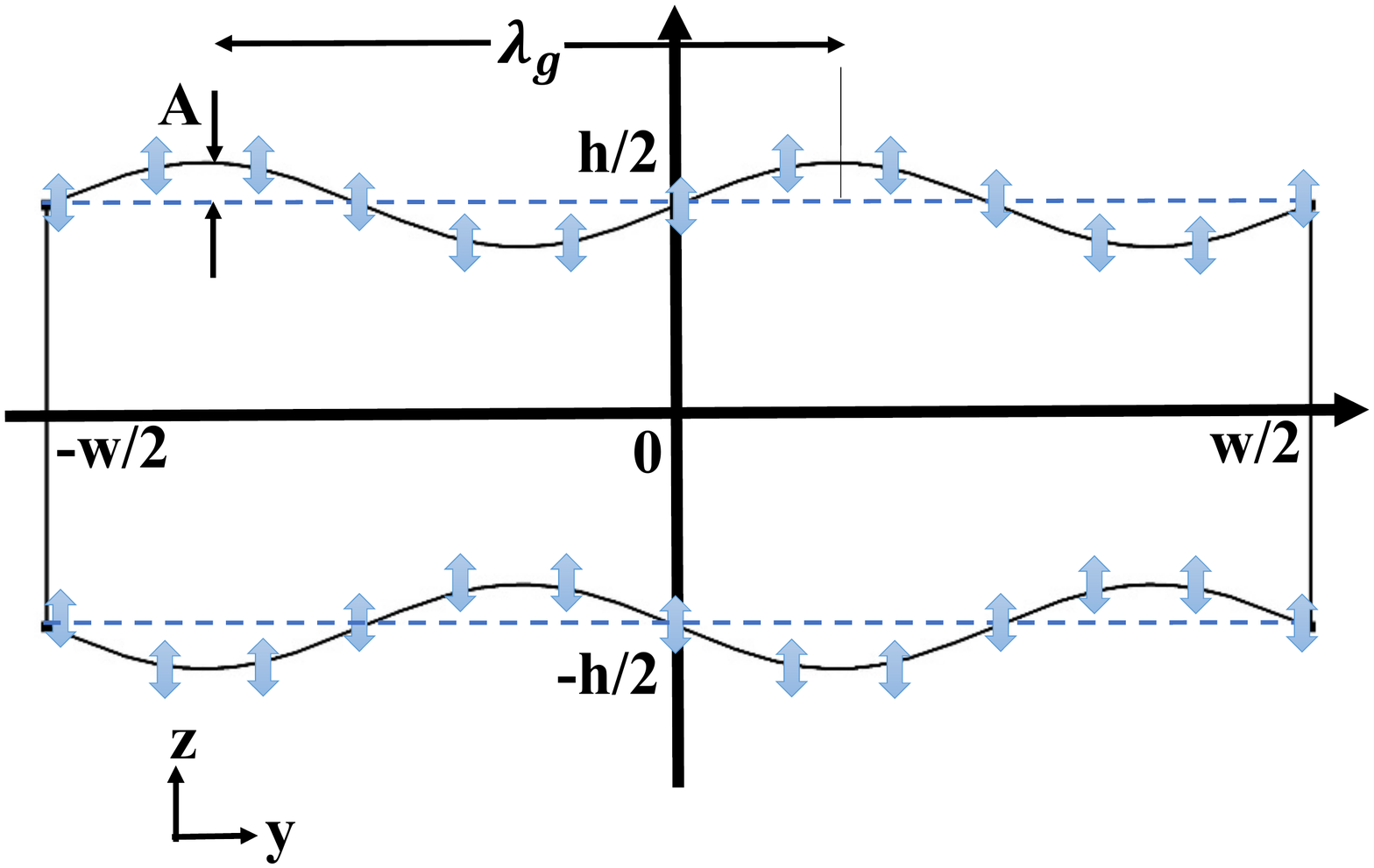}\label{fig1.b}}
		\end{tabular}
	\end{center}
\vspace{-0.5cm}
\caption{Two-dimensional schematics of microchannels in $yz$ plane is shown. Top and bottom walls are (a) symmetrically and (b) asymmetrically sinusoidal. Blue arrows show the oscillation direction of the actuated boundary walls at the resonance frequency of $f_v$.	
Hypothetical basic rectangular microchannels are shown by dashed lines with height of $h$ and width of $w$. $\lambda_g$ is defined as the geometrical wavelength of sinusoidal shaped boundaries.}
	\label {fig1}
\end{figure*}  

Two examples of sinusoidal microchannels are sketched in Fig\ref{fig1}. Top and bottom walls have symmetrical or asymmetrical sinusoidal forms given by the functions
\begin{equation}
z =\pm (\frac{h}{2}) \pm A \sin \left(\frac{2m\pi y}{w}+\phi_g\right)\;, \;m=\frac{1}{4}, \frac{1}{2}, \frac{3}{4}, 1, ...
\label{eq13}
\end{equation}
where $A$ is the amplitude of the sinusoidal boundaries, $w$ is the width of the microchannels, $h$ is the height of corresponding hypothetical rectangular microchannels, which is considered 100$\mu$m in this study, and $\phi_g$ is a geometrical phase parameter, typically equals to zero. Noteworthy, first $\pm$ sign in the Eq. \ref{eq11} refers to top or bottom boundaries, and the minus sign in second $\pm$ in equation belongs to the bottom walls of asymmetrical microchannels. 

The governing equations ($\ref{eq5}, \ref{eq7}$) are solved using finite element COMSOL Multiphysics v$5.2a$ 
software. We have used the mathematics-weak-form-PDE module of the software for both first and second order equations. The conducted steps are as follows: First, the flow equations are written as source-free flux formulation, $\bm\nabla\cdot\bm J+ F=0$; Then they are converted to the weak-form; Finally, the weak-form equations are solved by mathematics-weak-form-PDE module. In all cases, the zero-flux boundary condition, $\bm J \cdot \bm n =0$, is 
supposed; where $\bm n$ is the normal vector to the boundary surface. Also, all boundaries are considered as hard walls.

 Top and bottom walls are actuated by an external acoustic field at the frequency of $f_v$ which is the vertical resonance frequency of a hypothetical rectangular microchannel with a height of $h$. The boundary conditions of the first-order velocity field are
\begin{subequations}
	\begin{align}
	&\textrm{top--bottom}: &v_{y1}=0\;, \quad &v_{z1}=v_{bc}\sin(\omega 
t)\;,\\
	&\textrm{left--right}: &v_{y1}=0\;, \quad&v_{z1}=0\;,
	\end{align}
	\label{eq14}
\end{subequations}
where $v_{bc}=\omega d$, $\omega=2 \pi f$, and $d=0.1\;$nm. The displacement of the oscillating walls in the $z$ direction, $d$, is small enough to use the perturbation theory.

A zero-mass-flux boundary condition is considered for the second-order velocity field as
\begin{subequations}
	\begin{align}
	&\textrm{top-bottom}: &v_{y2}=0\;, \quad &v_{z2}=-\frac{\left\langle 
\rho_1 v_{z1} \right\rangle}{\rho_0}\;,\\
	&\textrm{left-right}: &v_{y2}=0\;, \quad&v_{z2}=0\;.
	\end{align}
	\label{eq15}
\end{subequations}

\begin{figure*}[h]
	\begin{center} 
		\begin{tabular}{c}
			\subfigure{\includegraphics[width=100mm]{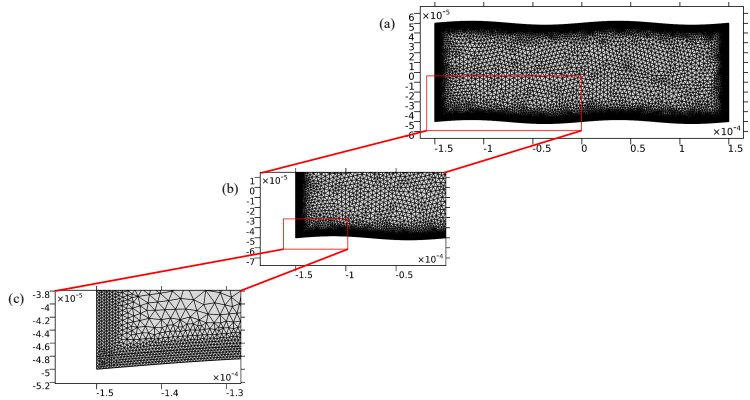}}
		\end{tabular}
	\end{center}
\vspace{-1cm}
	\caption{(a) Sketch of the spatial mesh of the sinusoidal computational domain in the $yz$ plain. (b) and (c) are two zoom-in scales on the mesh in the lower left corner.}
	\label{fig2}
\end{figure*}

The maximum mesh size in boundaries until 10$\delta_\nu$ and bulk are considered $0.5\mu m$ and $5\mu m$, respectively. The mesh element growth rate is 1.3 (see Fig. \ref{fig2}).

The fluid inside the microchannel is considered to be quiescent water. Also, the physical parameters for water at temperature of $T=25^{\circ}\mathrm{C}$ and pressure $p_0=0.1013$ MPa are shown in Table \ref{TB1}. 

\begin{table}
\caption{Physical parameters of  water at $T=25^{\circ}\mathrm{C}$ and 
$p_0=0.1013 {\;\rm  MPa}$} 
\label{TB1}
\begin{tabular}{lccc}
	\hline\hline
	Parameter&Symbol&Value&Unit\\
	\hline
	Mass density & $\rho_0$ & 9.970$\times10^{2}$ & kg/m$^3$\\
	Speed of sound & $c_0$ & 1.497$\times10^{3}$& m/s\\
	Shear viscosity & $\eta_0$ & 8.900$\times10^{-4}$ & Pa$\cdot$ s\\
	Bulk viscosity & $\eta_0^B$ & 2.485$\times10^{-3}$  & Pa$\cdot$ s\\
\hline
\end{tabular}
\end{table}

\section{Results and Discussion}\label{RES}
Acoustic streaming vortices can be troublesome in some cases, but if controlled and treated properly, the problem turns into a beneficial parameter in acoustic tweezing or micromixing applications. Concerning the fact that the acoustic streaming patterns are extremely sensitive to the geometry of the fluid container, extensive numerical calculations were carried out to study such effects on acoustic streaming patterns in two dimensions. In what follows we present the results and discuss their implications.

\subsection{Effective parameters on acoustic streaming patterns of sinusoidal microchannels}\label{RES-A}
The parameters which can affect streaming patterns in a microchannel with sinusoidal boundaries are the applied frequency, $f$, microchannel's width to height ratio, $n=\frac{w}{h}$, amplitude of the sinusoidal walls, $A$, symmetry or asymmetry of the sinusoidal walls, and  geometrical wavelength, $\lambda_g$. In this study, $f_v$ and $A$ are considered constant equal to $c/2h$ and $h/50$ respectively, where $c$ is the speed of sound in water. Effects of other parameters are investigated in details as follows.

\subsubsection{Effects of the microchannel's width to height ratio, $n$} \label{Res-A-1}
\begin{center}	
\begin{figure}[h!] 
\begin{tabular}{l|ll}	
\;& $p_1$ & $\langle v_2 \rangle$ \\
\hline
$n=1$&	
\subfigure[]{\includegraphics[width=18 mm]{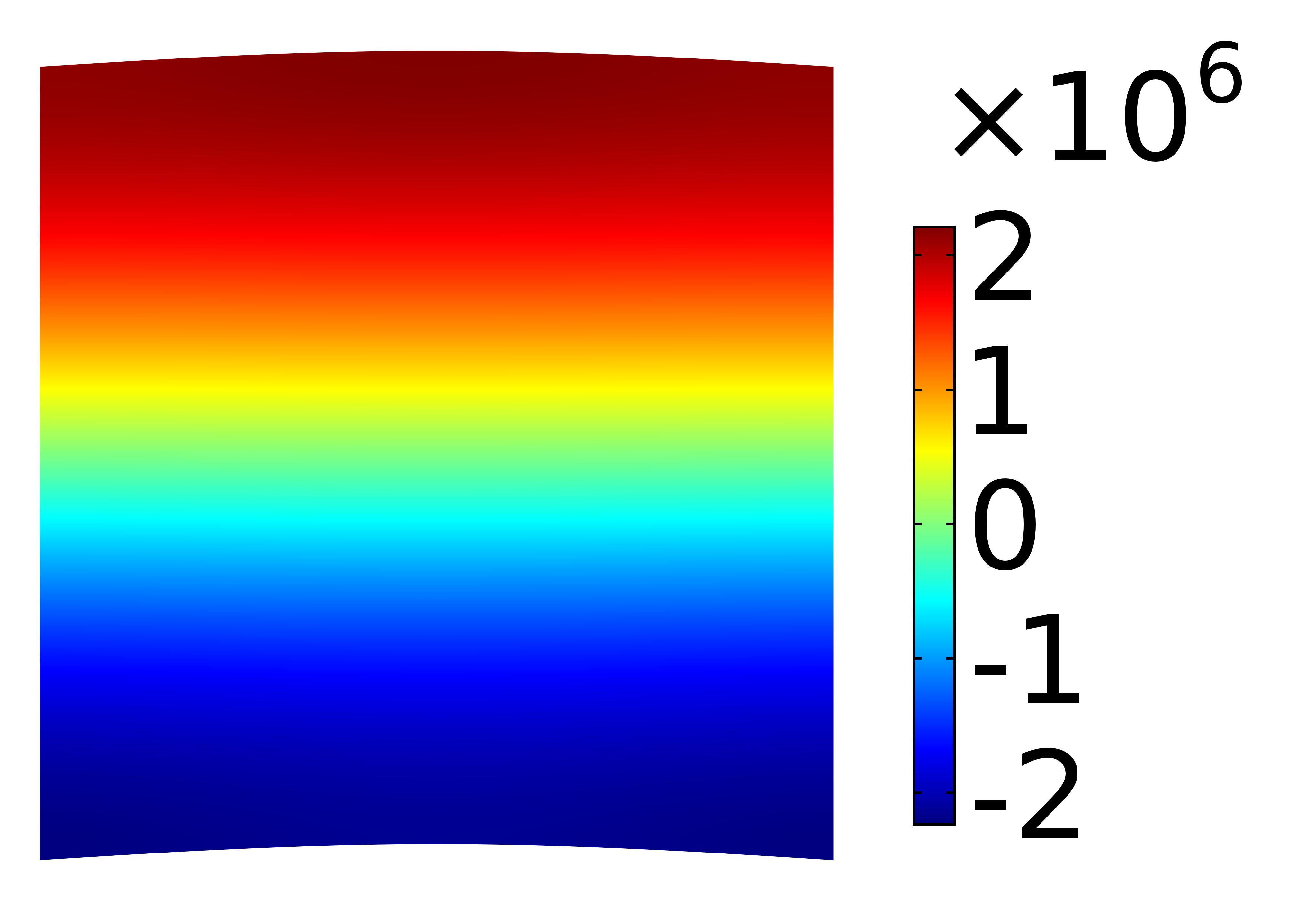}}&	
\subfigure[]{\includegraphics[width=18 mm]{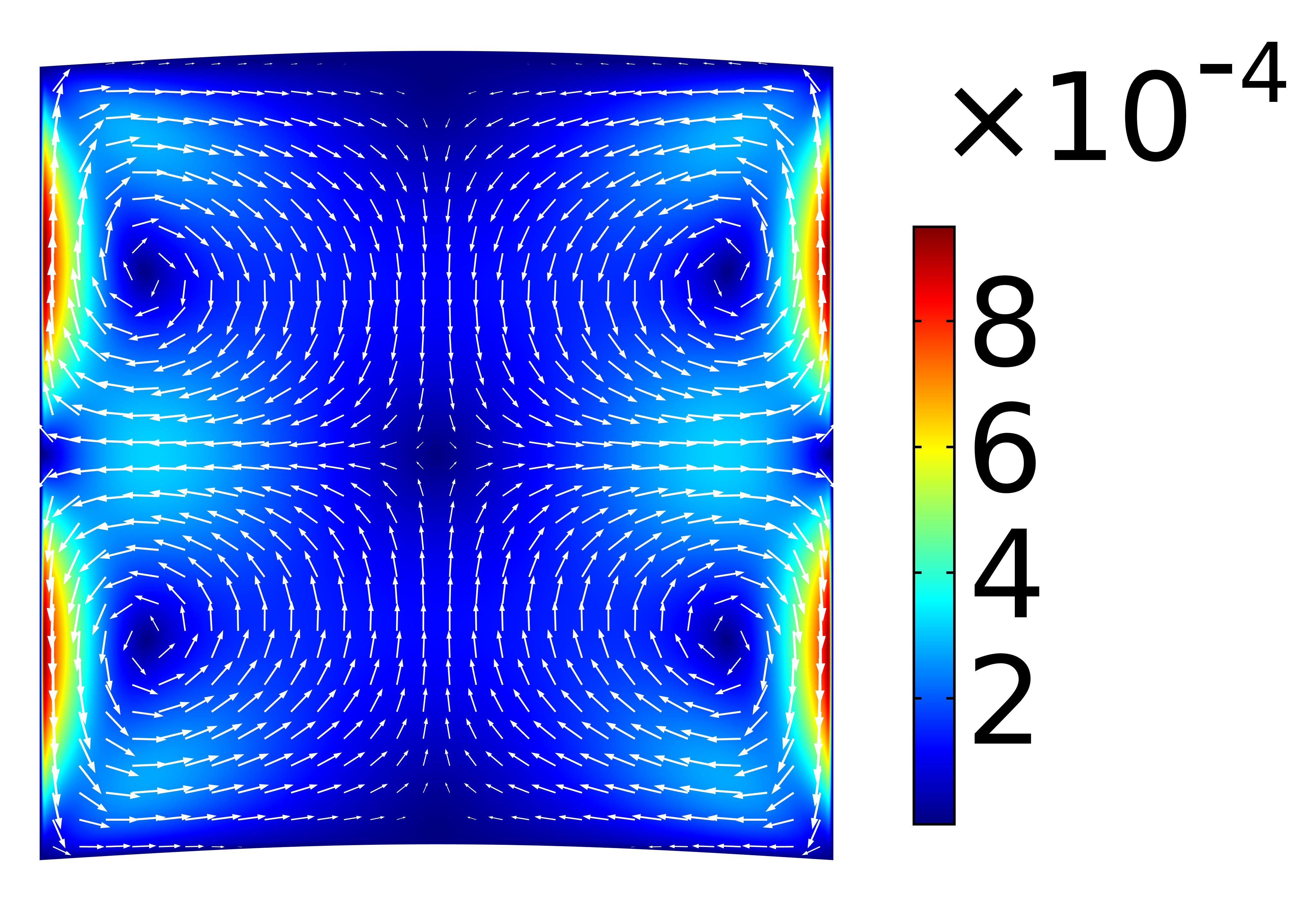}}\\
$n=2$&	
\subfigure[]{\includegraphics[width=35 mm]{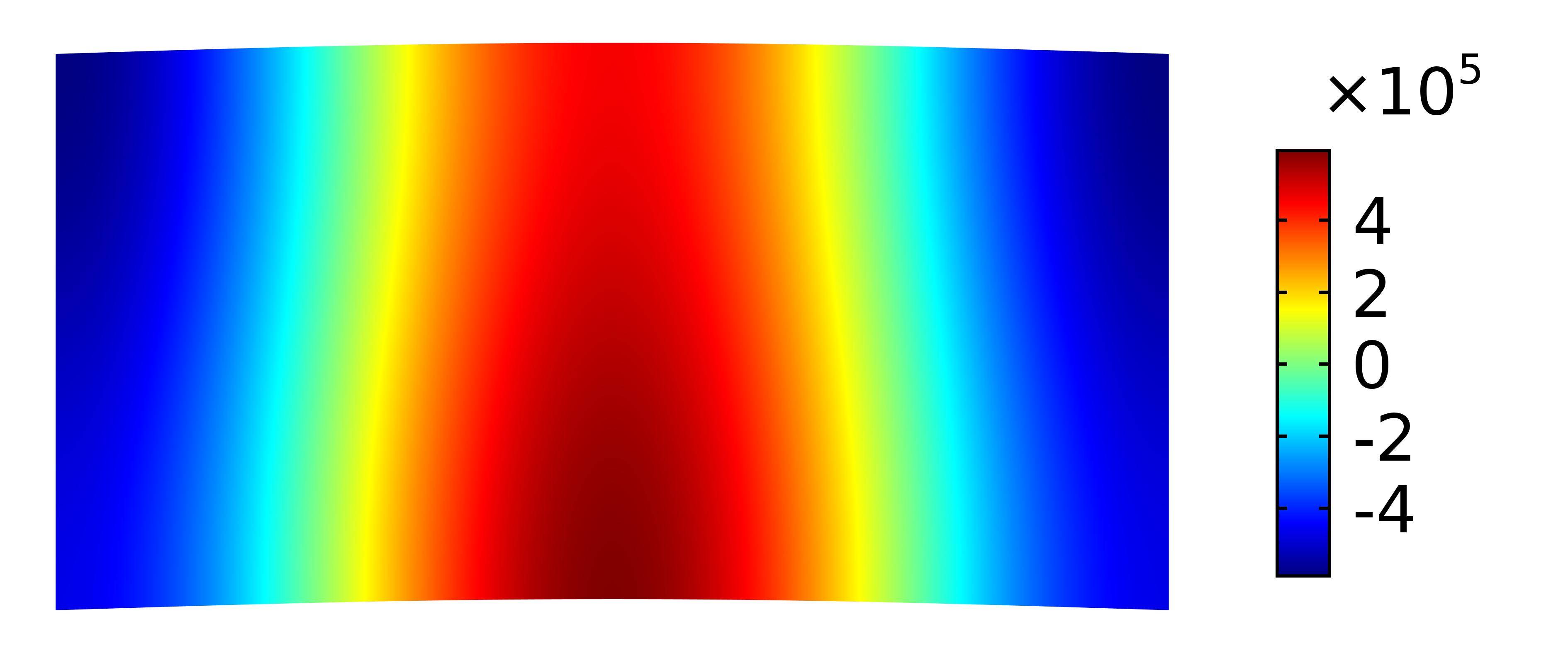}}&	
\subfigure[]{\includegraphics[width=35 mm]{sym-wide200-v2-mod1}}\\
$n=3$&	
\subfigure[]{\includegraphics[width=55 mm]{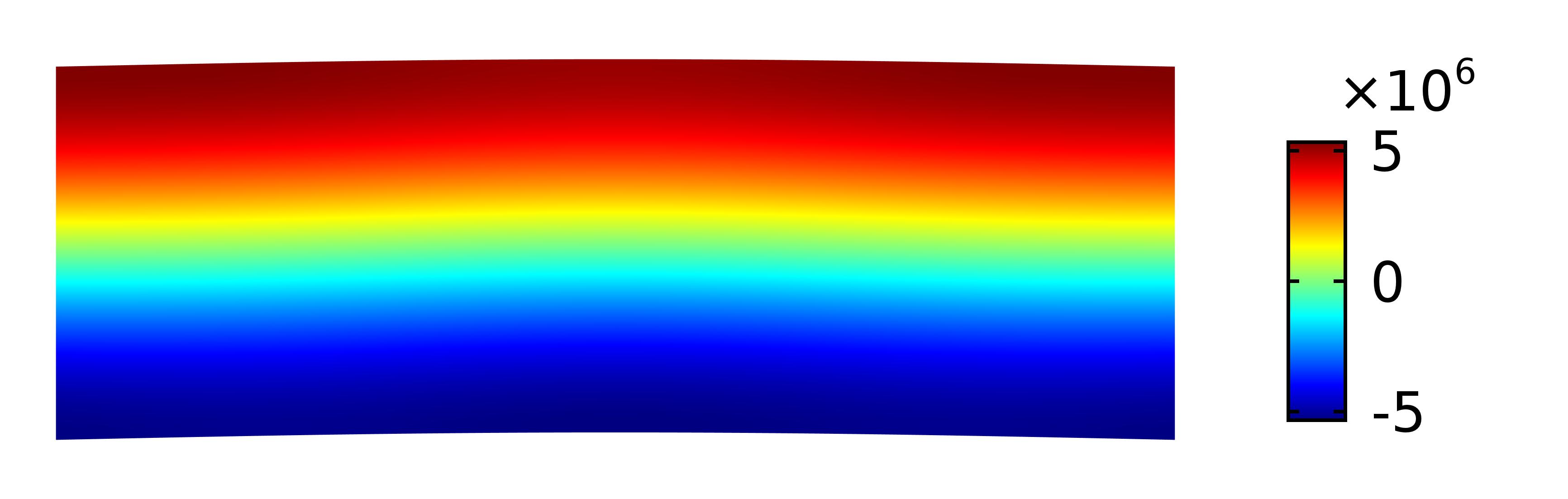}}&	
\subfigure[]{\includegraphics[width=55 mm]{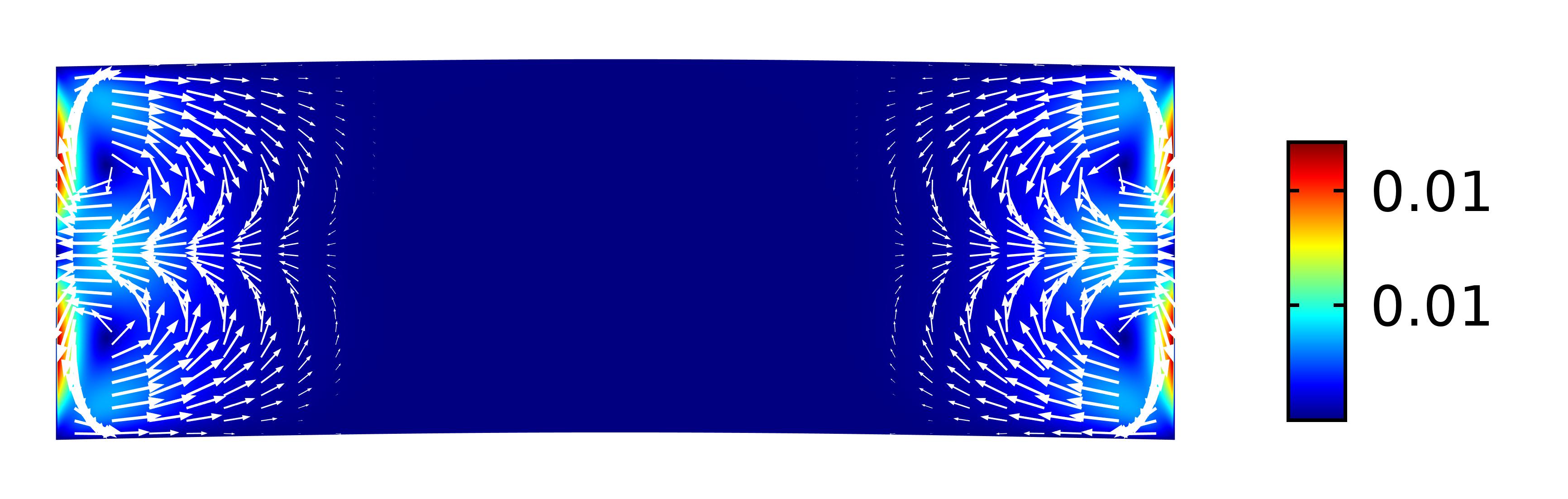}}\\
$n=4$&	
\subfigure[]{\includegraphics[width=70 mm]{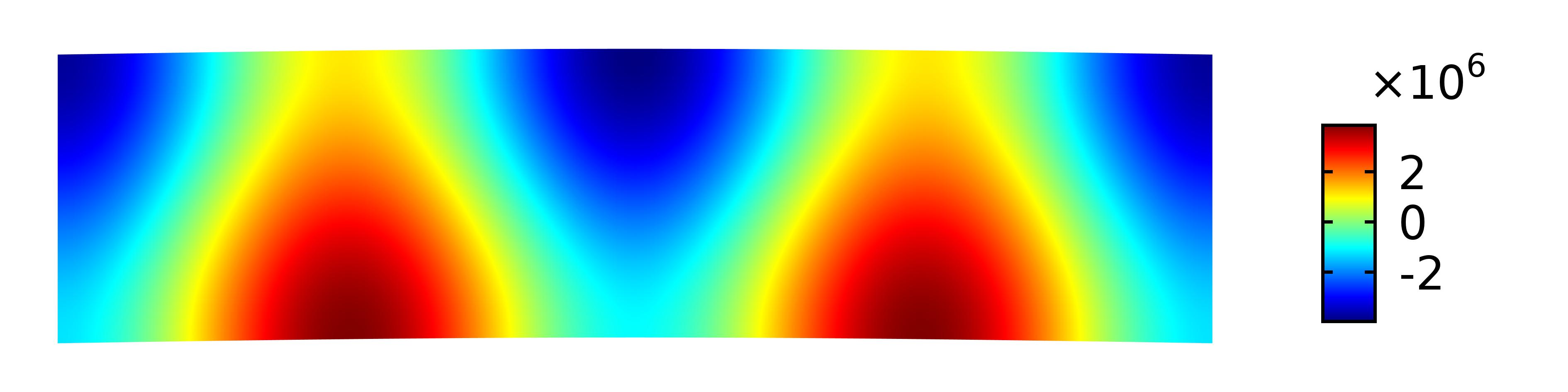}}&	
\subfigure[]{\includegraphics[width=70 mm]{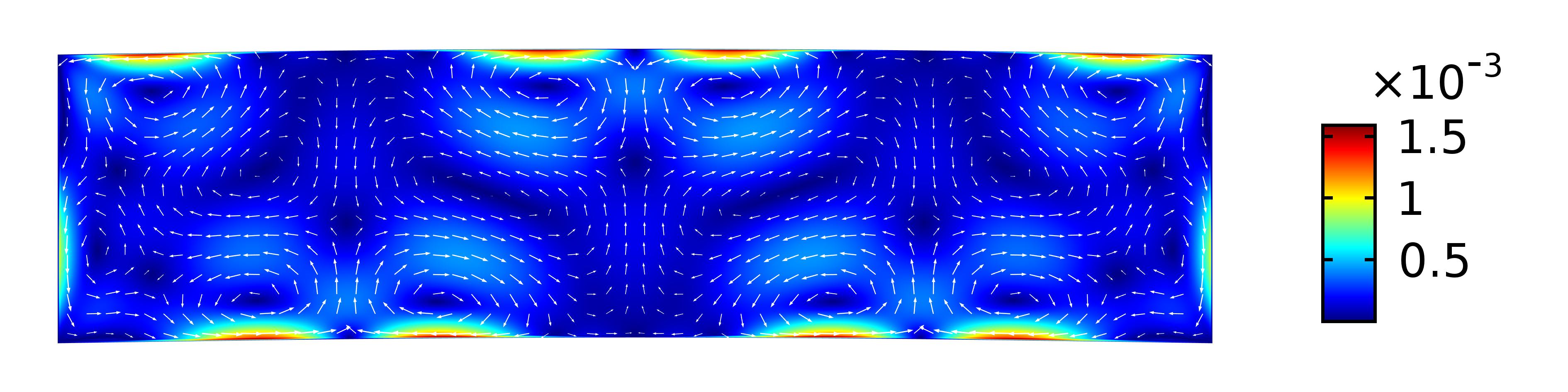}}\\
$n=5$&	
\subfigure[]{\includegraphics[width=80 mm]{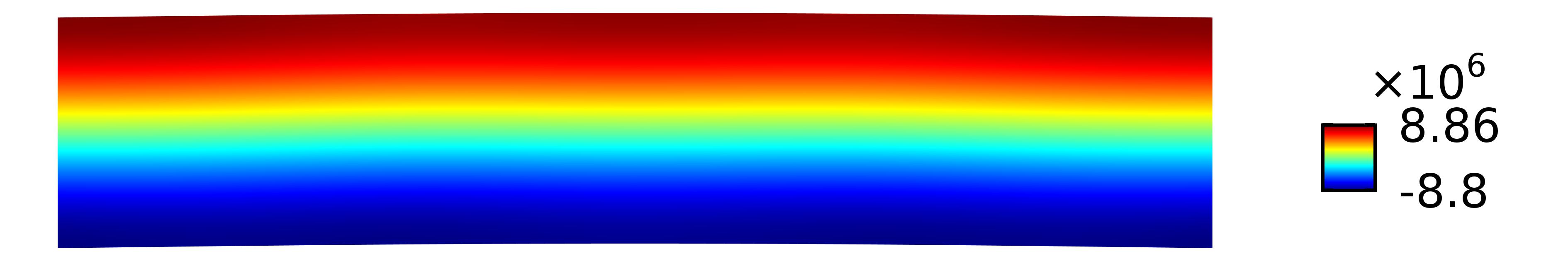}}&	
\subfigure[]{\includegraphics[width=80 mm]{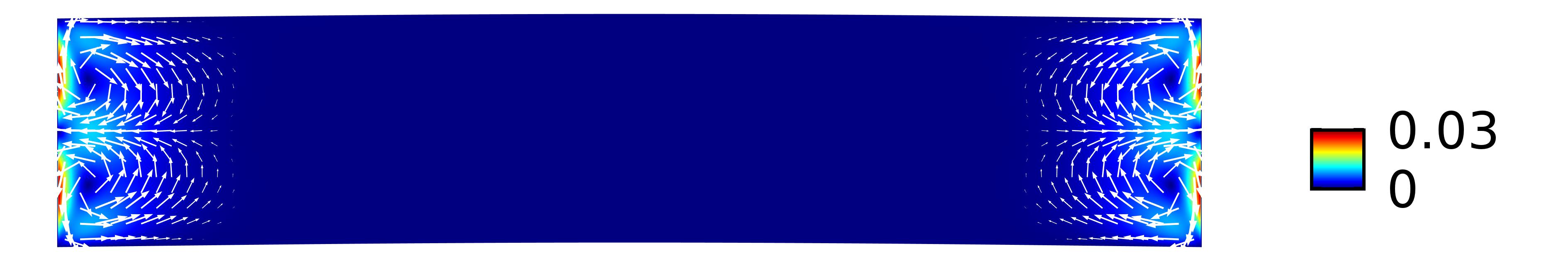}}\\
\end{tabular}
\caption{First order pressure fields (left column) and time-averaged second-order velocity fields (right column) in cross section of sinusoidal microchannels where geometrical wavelengths of symmetrical top and bottom walls are keeping fixed as $\lambda_g=2w$ and the microchannel's width to height ratio, $n$, varies from $1$ to $5$.}
\label {fig3} 
\end{figure}
\end{center}

\begin{figure*}[h!]
\begin{center} 
\begin{tabular}{l|ll}	
\;& $p_1$ & $\langle v_2 \rangle$ \\
\hline
$n=1$&	
\subfigure[]{\includegraphics[width=18 mm]{sym-wide100-p1-mod2}}&	
\subfigure[]{\includegraphics[width=18 mm]{sym-wide100-v2-mod2}}\\
$n=2$&	
\subfigure[]{\includegraphics[width=35 mm]{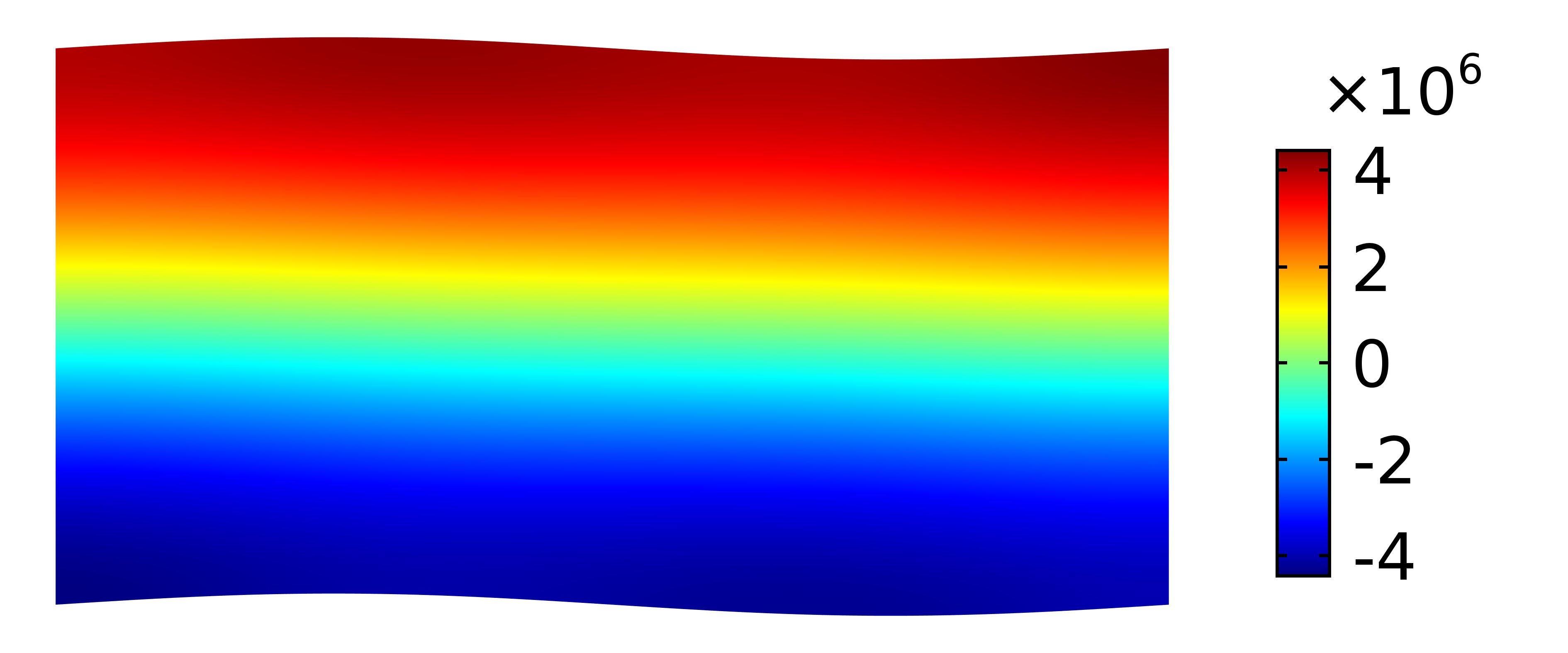}}&	
\subfigure[]{\includegraphics[width=35 mm]{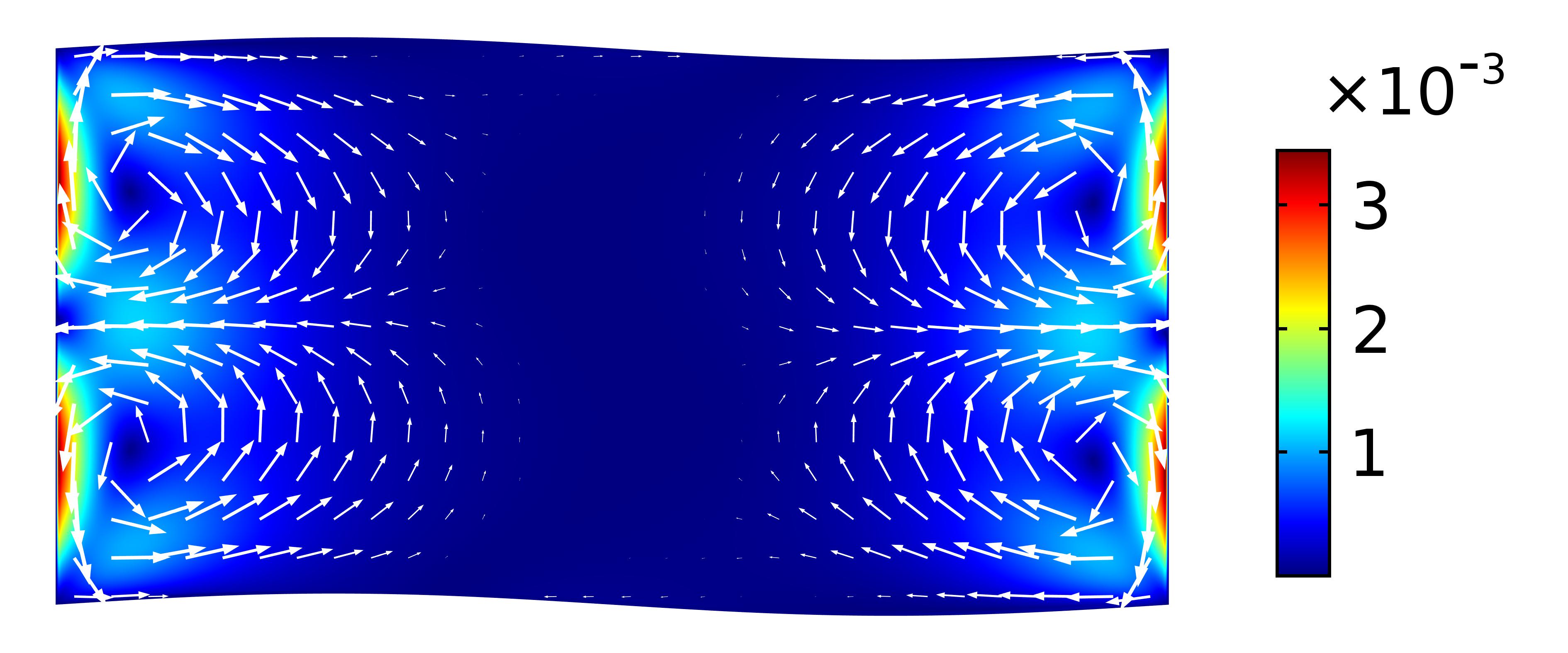}}\\
$n=3$&	
\subfigure[]{\includegraphics[width=55 mm]{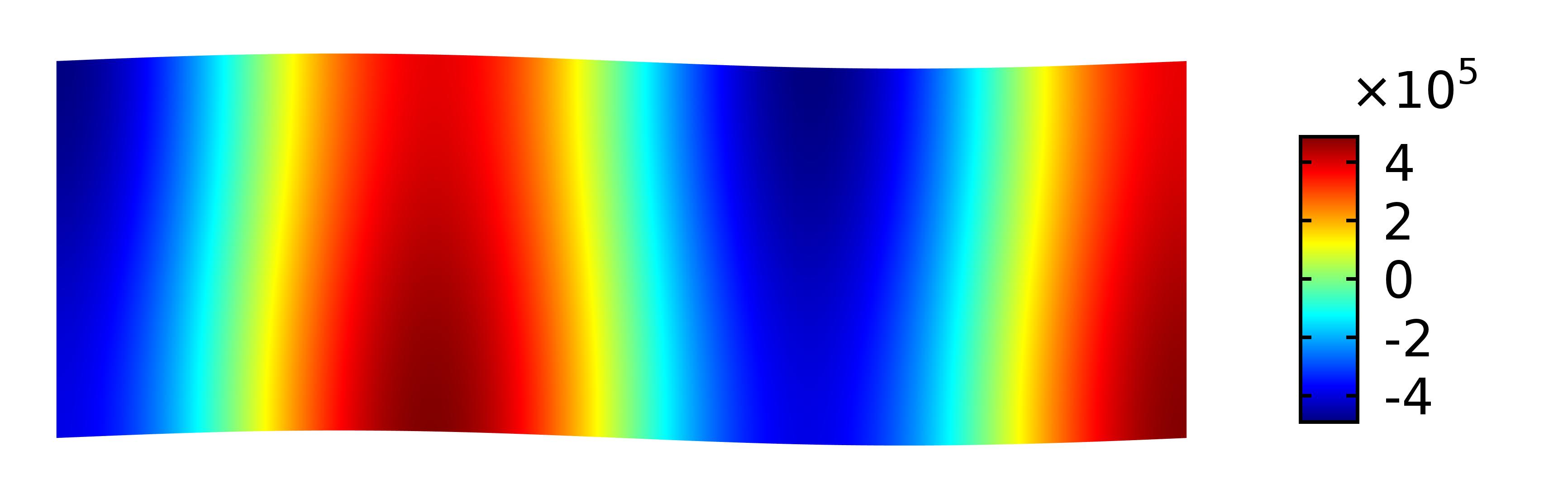}}&	
\subfigure[]{\includegraphics[width=55 mm]{sym-wide300-v2-mod2}}\\
$n=4$&	
\subfigure[]{\includegraphics[width=75 mm]{sym-wide400-p1-mod2}}&	
\subfigure[]{\includegraphics[width=75 mm]{sym-wide400-v2-mod2}}\\
$n=5$&	
\subfigure[]{\includegraphics[width=80 mm]{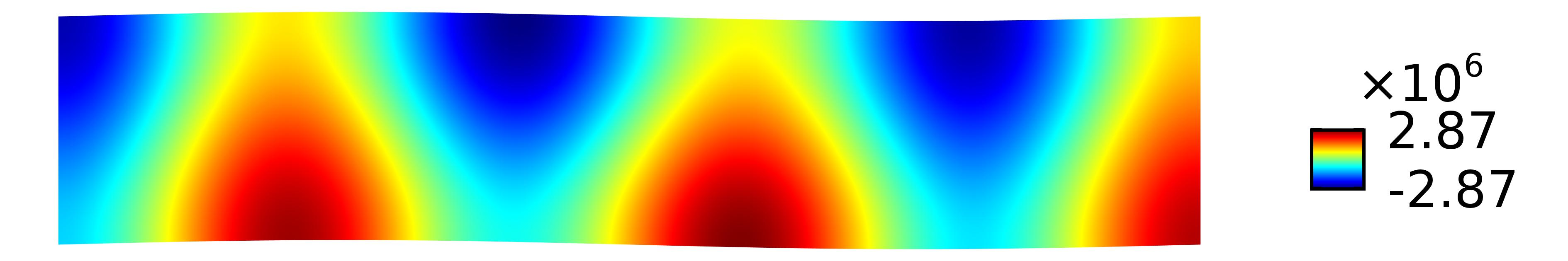}}&	
\subfigure[]{\includegraphics[width=80 mm]{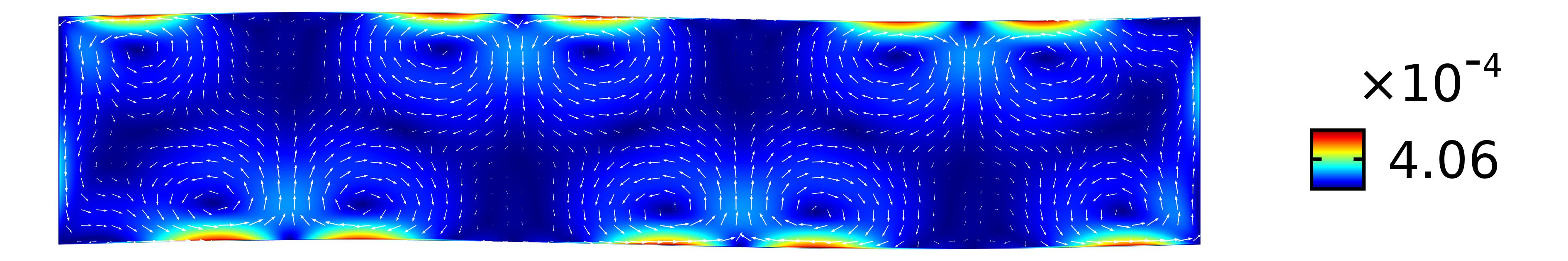}}
\end{tabular}
\caption{First order pressure fields (left column) and time-averaged second-order velocity fields (right column) in cross section of sinusoidal microchannels where geometrical wavelengths of symmetrical top and bottom walls are keeping fixed as $\lambda_g=w$ and the microchannel's width to height ratio, $n$, varies from $1$ to $5$.} 
\label {fig4}
\end{center}
\end{figure*}

Figures \ref{fig3} and \ref{fig4} illustrate examples for two fixed geometrical wavelengths, $\lambda_g=2w$ and $\lambda_g=w$, but different values of $n$ from 1 to 5 for each case. Results show that in some various microchannel's width to height ratios, the streaming patterns are completely different from the flat geometry streaming patterns.
Dominant repetitive vortices or twisting-like patterns are discoverable. For $\lambda_g=2w$ and odd numbers of $n$, streaming patterns tend to flat geometry. For $\lambda_g=w$ and even numbers of $n$ the same results are achieved. 

Acoustic streaming patterns for other geometrical wavelengths with variable ratios have been studied numerically and same pattern sequences have been extracted. Details will discuss in Sec. III to classify and formulate such appearing streaming patterns. 
 
\subsubsection{Effects of geometrical wavelength on acoustic streaming patterns} \label{Res-A-2}

Another investigation has carried out focusing on the different values of $\lambda_g$ but fixed $n$. Results show that $\lambda_g$ definitely affects on streaming patterns in some cases. Dominant repetitive vortices and twisting-like patterns are discoverable same as in cases with fixed $\lambda_g$ but changing $n$. In some repetitive cases, patterns resemble flat geometry pattern. Figure \ref{fig5} shows the results for $n=6$ and definitive magnitudes of $\lambda_g$. As depicted, acoustic standing waves rotate in some cases from vertical into horizontal. Noteworthy, the frequency of actuation and the direction of oscillations are vertically in all cases. More investigations show that the patterns in $n=600$ are achievable in other channel widths with classifiable values of $\lambda_g$. As a result, some formulations are proposed in Sec.III to make acoustic streaming patterns predictable as much as possible. 
	
\subsubsection{Effects of asymmetrical sinusoidal top and bottom walls} \label{Res-A-3}
Same simulations as above repeated for asymmetrical sinusoidal boundaries. Figure \ref{fig6} indicate cases with different values of $n$ but fixed $\lambda_g$ and Fig. \ref{fig7} shows selected simulations for different values of $\lambda_g$ but fixed $n$. Results declare that in asymmetrical cases previous patterns, as in symmetrical boundaries, are not achievable. The generating standing waves typically resemble flat geometry and streaming patterns carry less deviation from a rectangular case in compared to the symmetrical cases. However, some considerable patterns exist. At the geometrical wavelength of $\lambda_g=2w$ but high numbers of $n$ four additional boundary layers generate around the sinusoidal curved boundaries. As a result, four bulk streaming flows emerge in the bulk of the microchannel. In addition, at $\lambda_g=w$, some stretched fluid circles are generated as in Fig. \ref{fig7.d}. As such, fluid molecules or tiny particles transfer from one side to another side just through implying acoustical oscillations.

\subsection{Classification of streaming patterns in symmetrical sinusoidal microchannels}\label{RES-B}

This work aims to characterize the effect of sinusoidal boundaries on the acoustic streaming patterns in microfluidic systems. Focusing on natural values of the microchannel's width to height ratio from $1$ to $10$, and geometrical length-widths of $\lambda_g=\frac{4w}{2p}$, where $p=1,2,...,10$, leads to a $10\times10$ matrix which contains numerical results.
A classification is proposed to predict patterns with larger values of $\lambda_g$ and $n$ through an inductive reasoning. 

\begin{figure*}[ht]
\vspace{-1.5cm}
		\begin{center} 
		\begin{tabular}{l|ll}	
\;& $p_1$ & $\langle v_2 \rangle$ \\
\hline
flat&	
\subfigure[]{\includegraphics[height=14 mm]{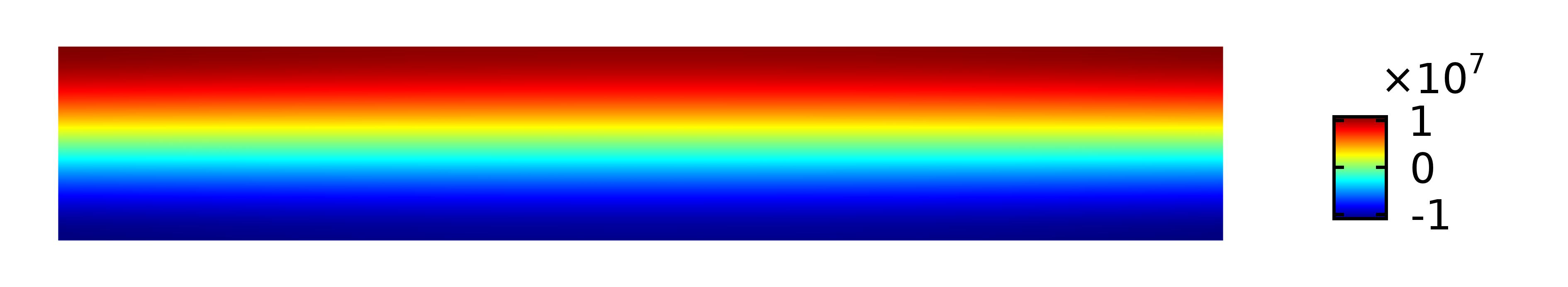}\label{fig5.a}}&
\subfigure[]{\includegraphics[height=13.9 mm]{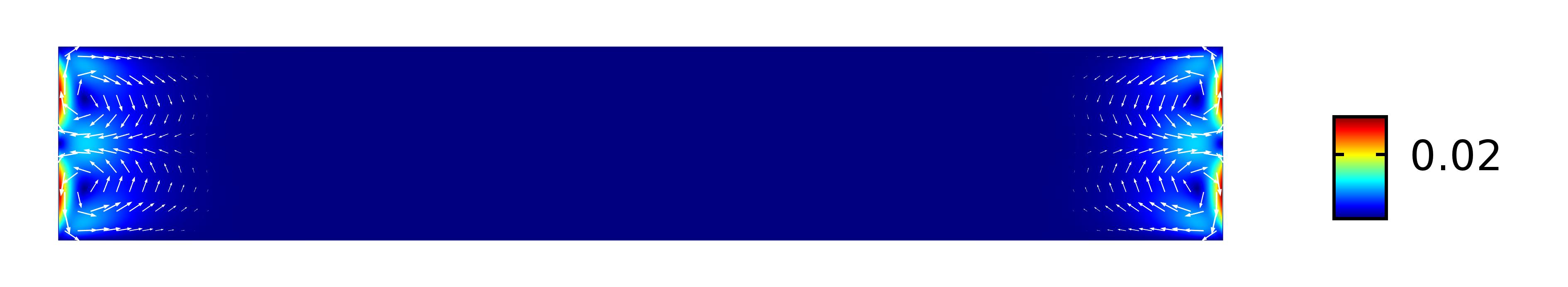}\label{fig5.b}}\\
$\lambda_g=2w$&	
\subfigure[]{\includegraphics[height=13.6 mm]{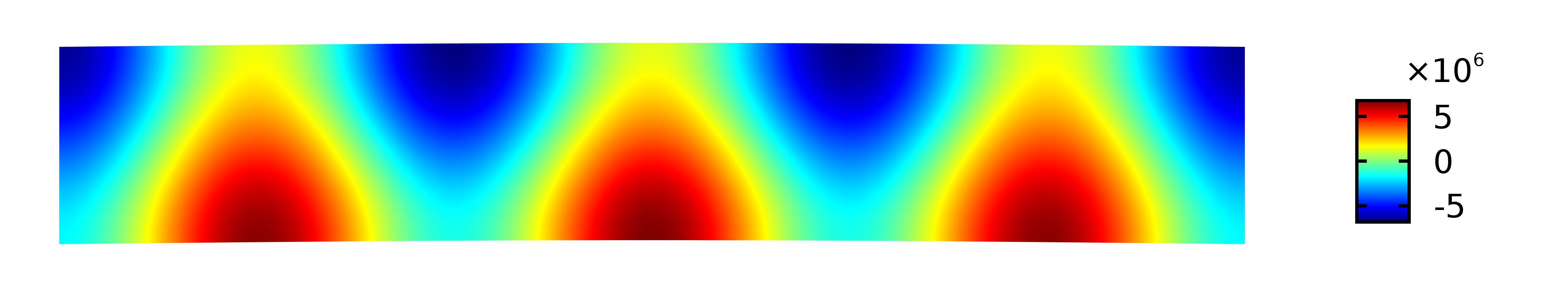}\label{fig5.c}}&
\subfigure[]{\includegraphics[height=15.8 mm]{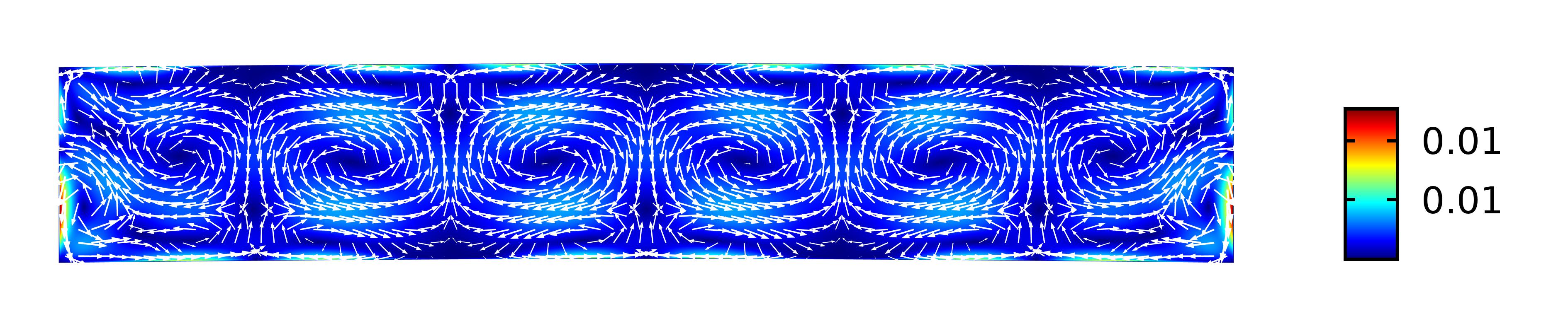}\label{fig5.d}}\\
$\lambda_g=w$&	
\subfigure[]{\includegraphics[height=13.6 mm]{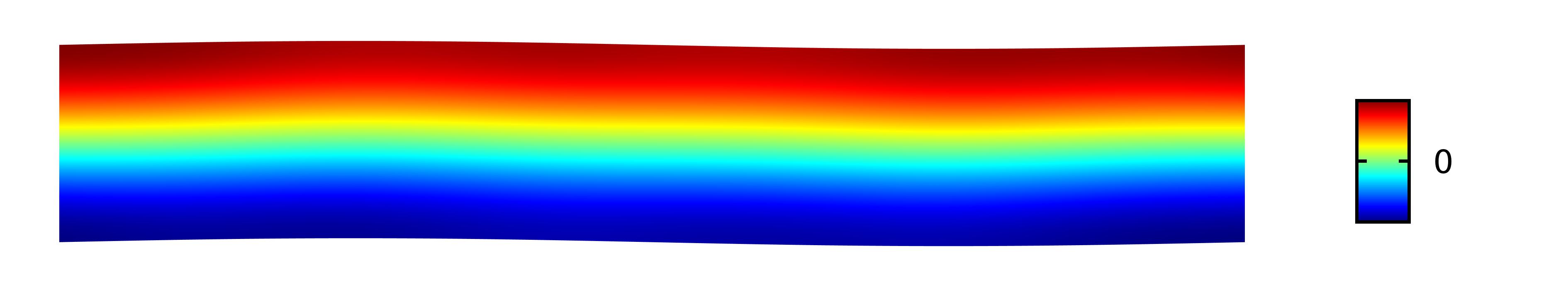}\label{fig5.e}}&
\subfigure[]{\includegraphics[height=15.5 mm]{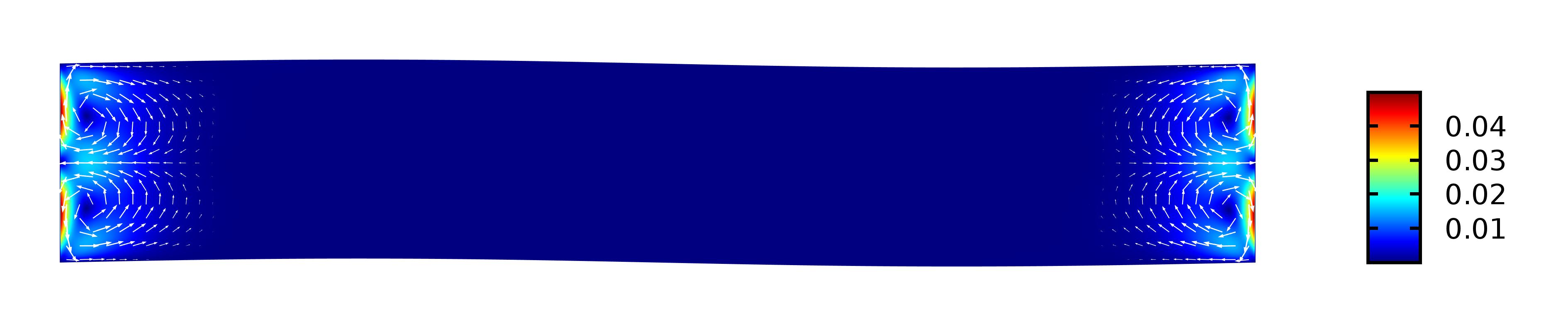}\label{fig5.f}}\\	
$\lambda_g=\frac{2}{3}w$&	
\subfigure[]{\includegraphics[height=12.7 mm]{sym-wide600-p1-mod3}\label{fig5.g}}&
\subfigure[]{\includegraphics[height=12.7 mm]{sym-wide600-v2-mod3}\label{fig5.h}}\\
$\lambda_g=\frac{1}{2}w$&	
\subfigure[]{\includegraphics[height=13.6 mm]{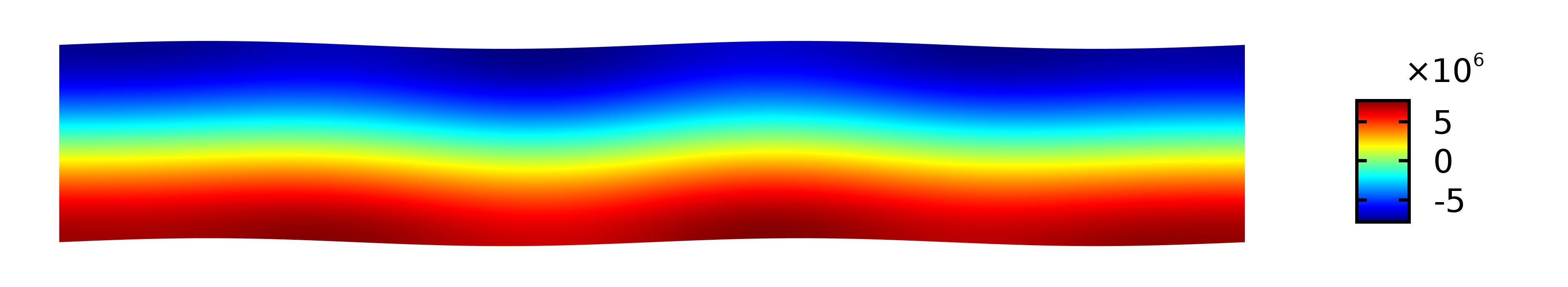}\label{fig5.i}}&
\subfigure[]{\includegraphics[height=15.5 mm]{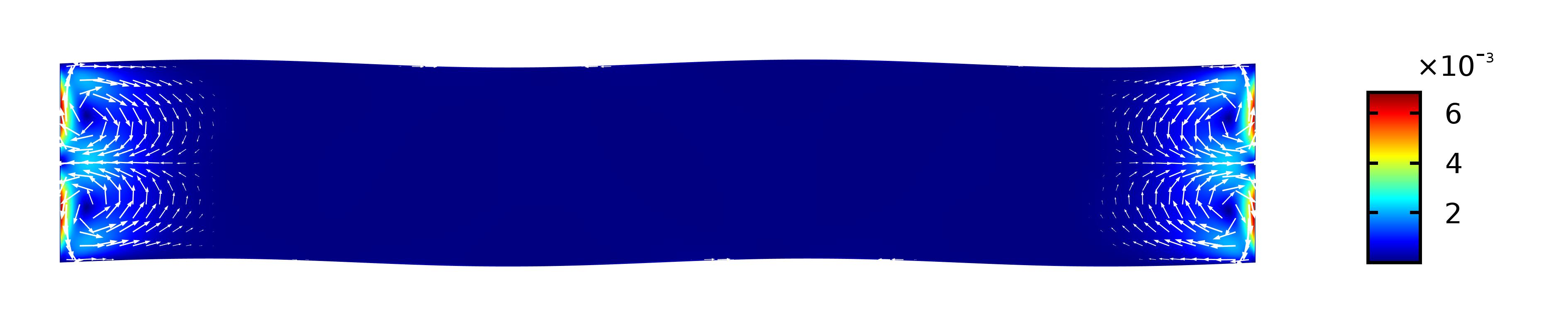}\label{fig5.j}}\\
$\lambda_g=\frac{2}{5}w$&
\subfigure[]{\includegraphics[height=15.5 mm]{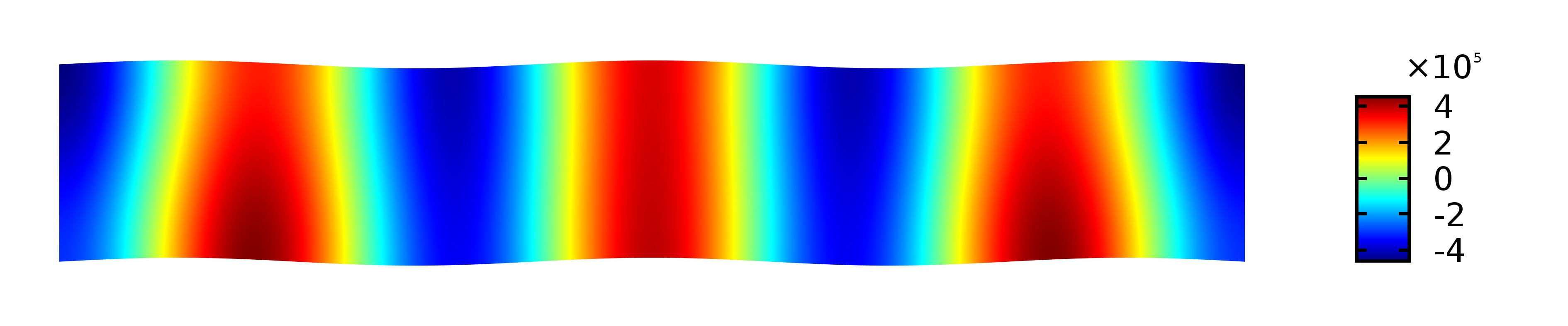}\label{fig5.k}}&
\subfigure[]{\includegraphics[height=16 mm]{sym-wide600-v2-mod5}\label{fig5.l}}\\
$\lambda_g=\frac{1}{3}w$&
\subfigure[]{\includegraphics[height=13.5 mm]{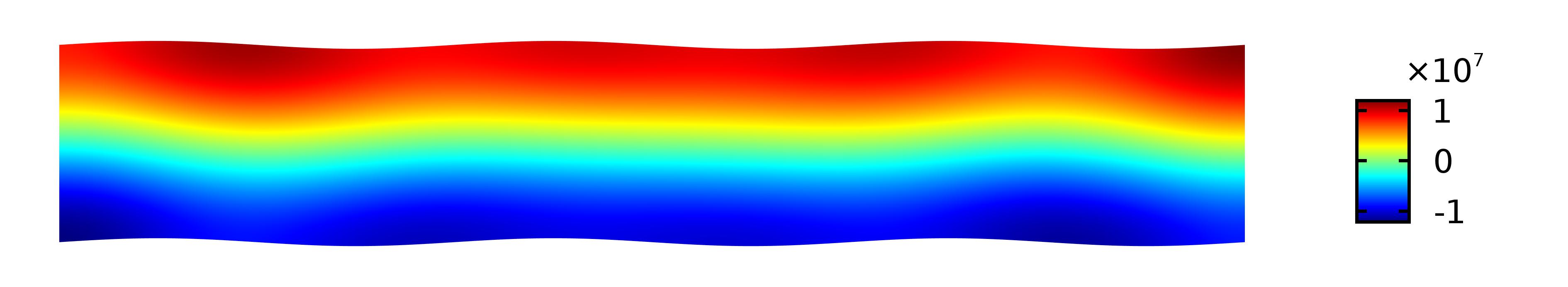}\label{fig5.m}}&
\subfigure[]{\includegraphics[height=15.5 mm]{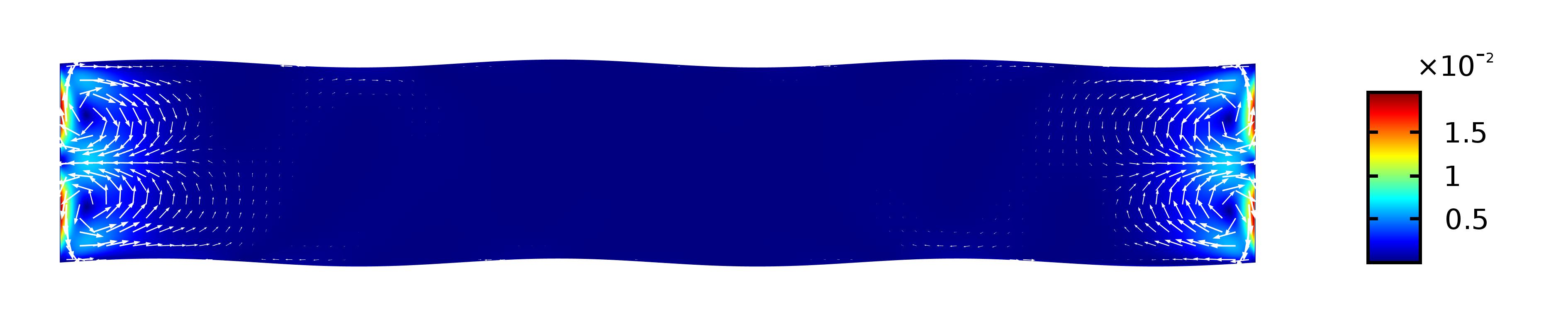}\label{fig5.n}}\\
$\lambda_g=\frac{2}{7}w$&
\subfigure[]{\includegraphics[height=15.5 mm]{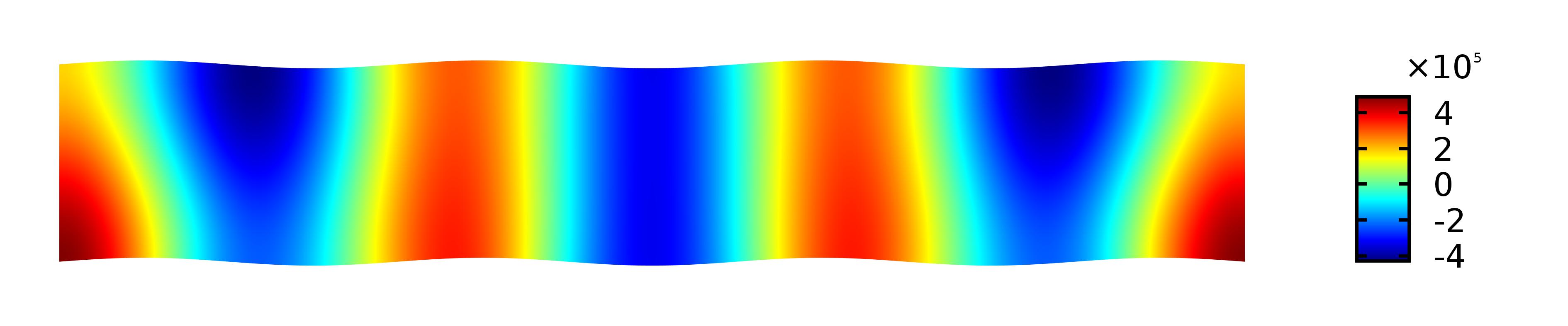}\label{fig5.o}}&
\subfigure[]{\includegraphics[height=16 mm]{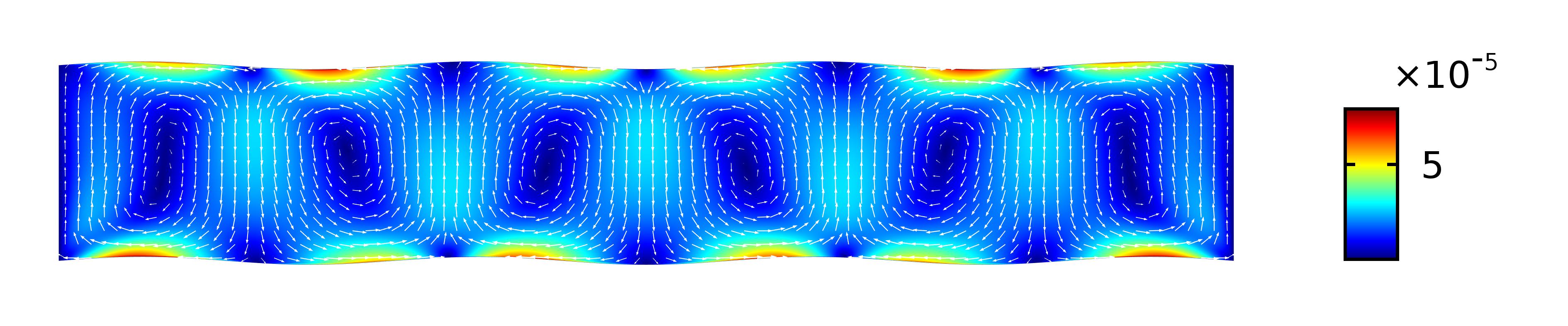}\label{fig5.p}}\\
$\lambda_g=\frac{1}{4}w$&	
\subfigure[]{\includegraphics[height=14 mm]{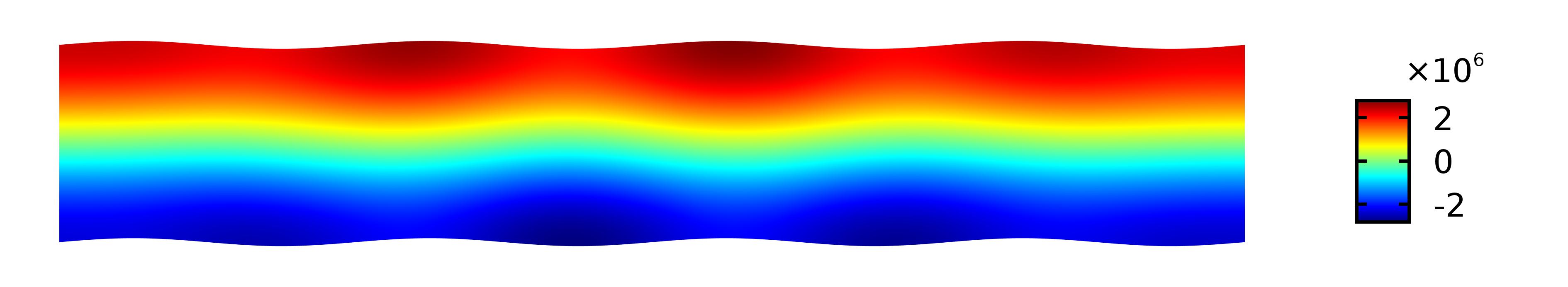}\label{fig5.q}}&
\subfigure[]{\includegraphics[height=16 mm]{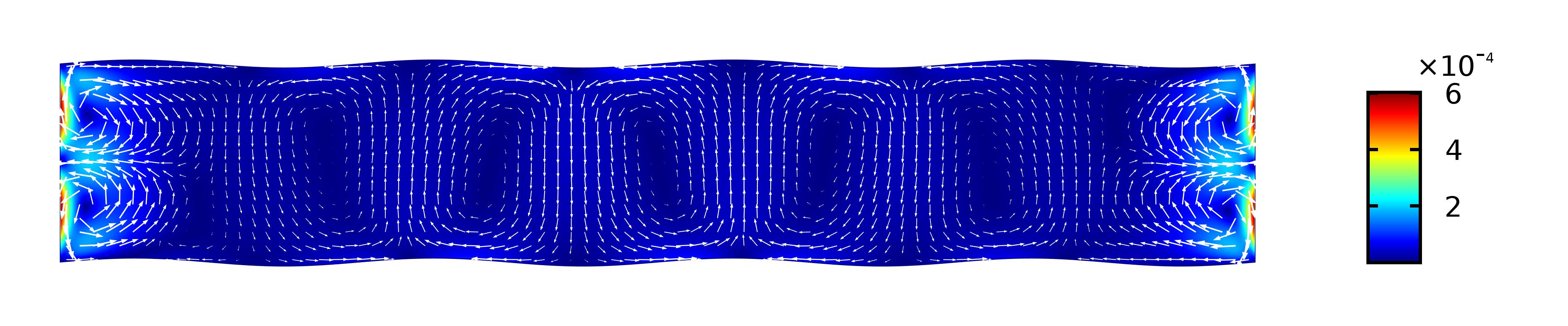}\label{fig5.r}}\\	
$\lambda_g=\frac{2}{9}w$&
\subfigure[]{\includegraphics[height=14 mm]{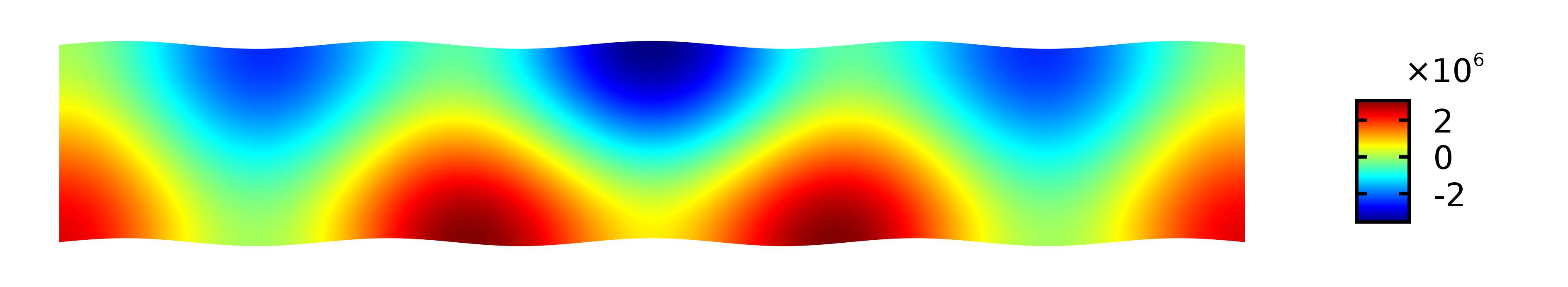}\label{fig5.s}}&
\subfigure[]{\includegraphics[height=16 mm]{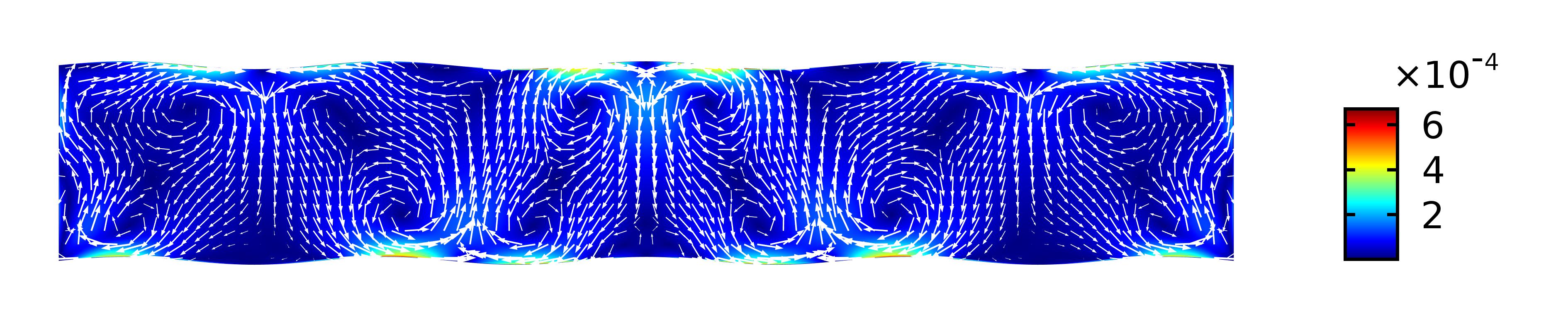}\label{fig5.t}}\\			
		\end{tabular}
\caption{First order pressure fields (left column) and time-averaged second-order velocity fields (right column) in cross section of sinusoidal microchannels where the microchannel's width to height ratio is keeping fixed as $n=6$ and geometrical wavelengths of symmetrical top and bottom walls are $\lambda_g=\frac{4}{2p}$ where $p=1,2,..., 7$.} 
			\label{fig5}
			\end{center}
	\end{figure*}

\begin{figure*}[ht]
	\begin{center} 
		\begin{tabular}{l|ll}	
			\;& $p_1$ & $\langle v_2 \rangle$ \\
			\hline
			$n=1$&	
			\subfigure[]{\includegraphics[width=23 mm]{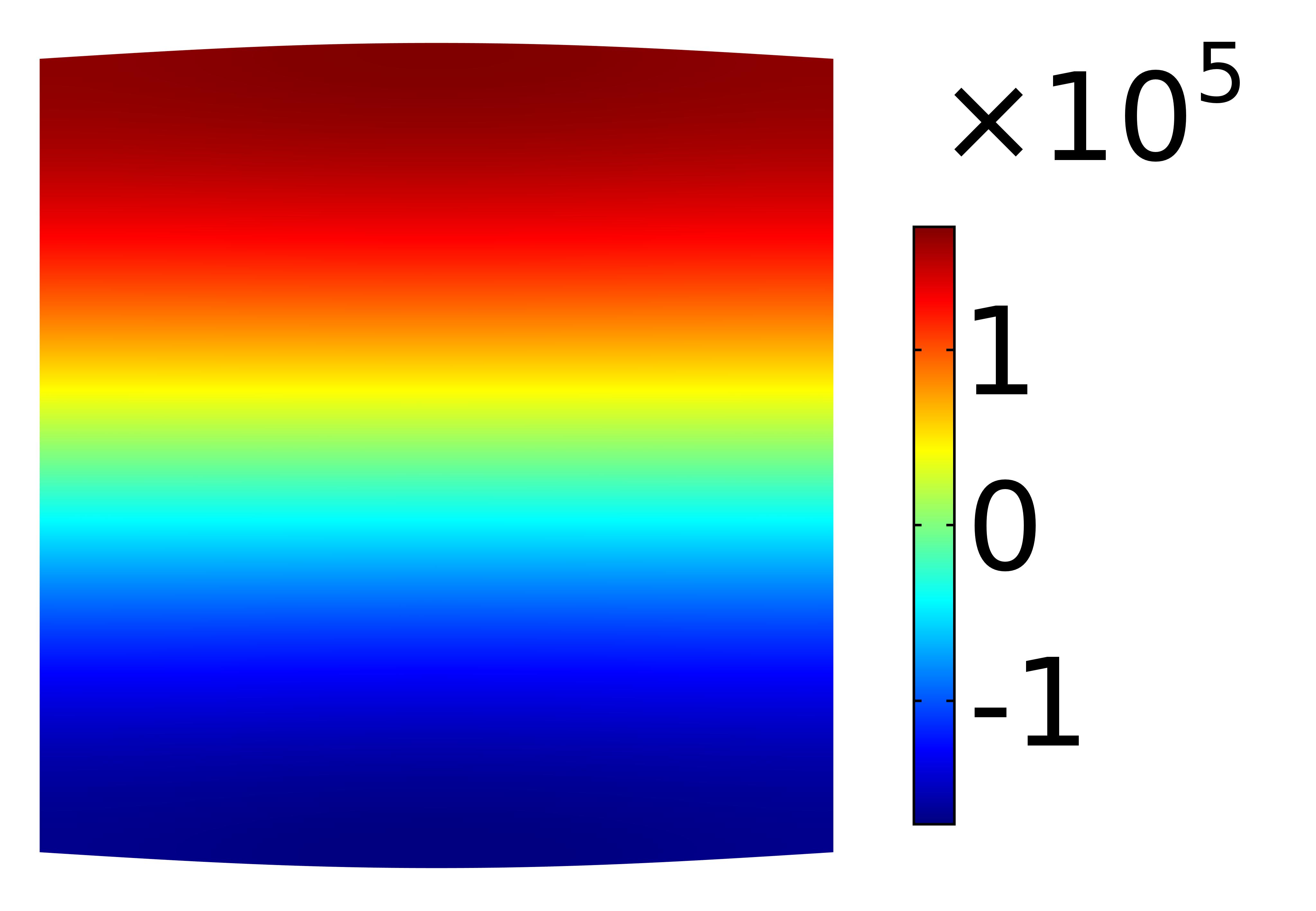}\label{fig6.a}}&	
			\subfigure[]{\includegraphics[width=23 mm]{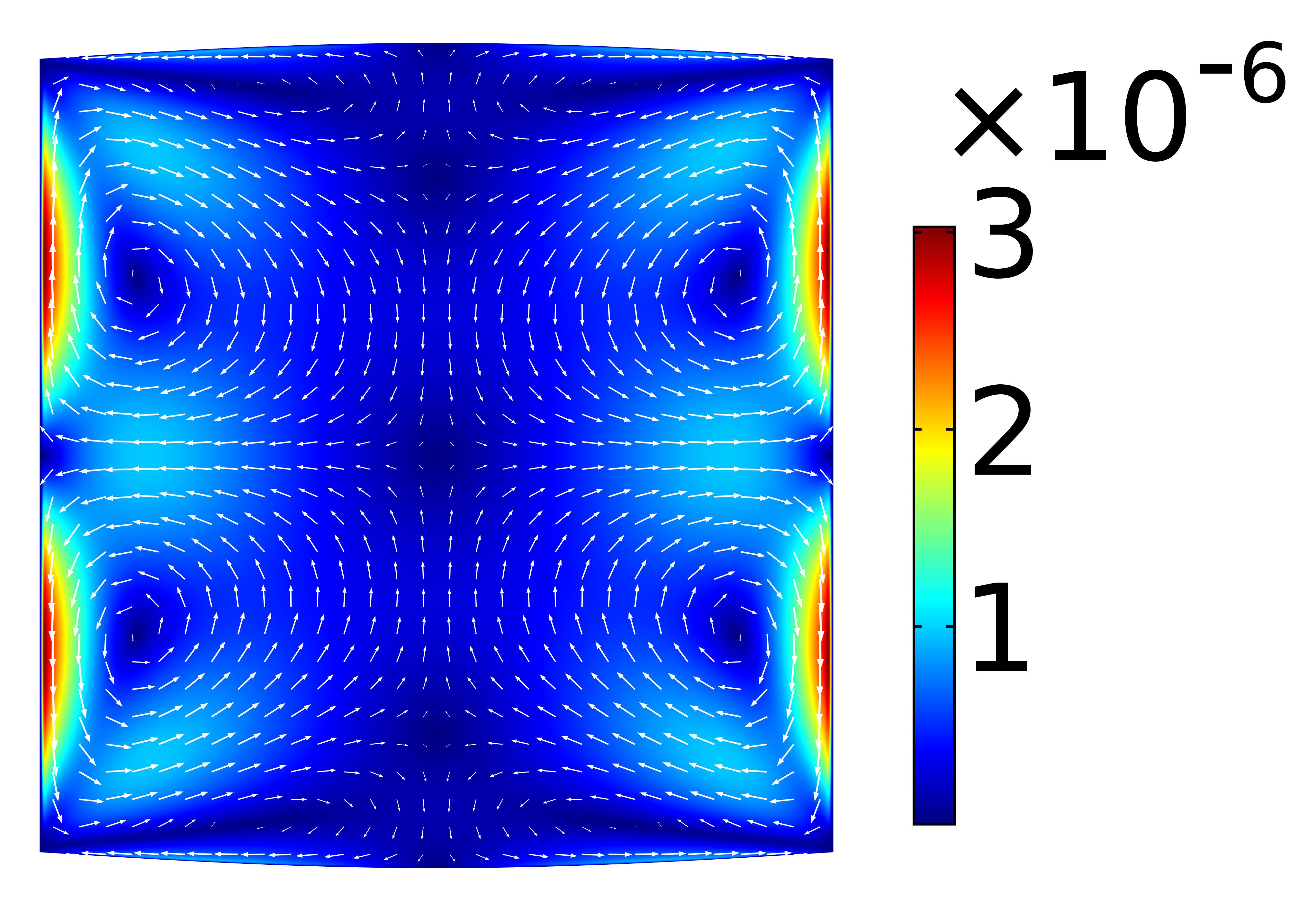}\label{fig6.b}}\\
			$n=2$&	
			\subfigure[]{\includegraphics[width=40 mm]{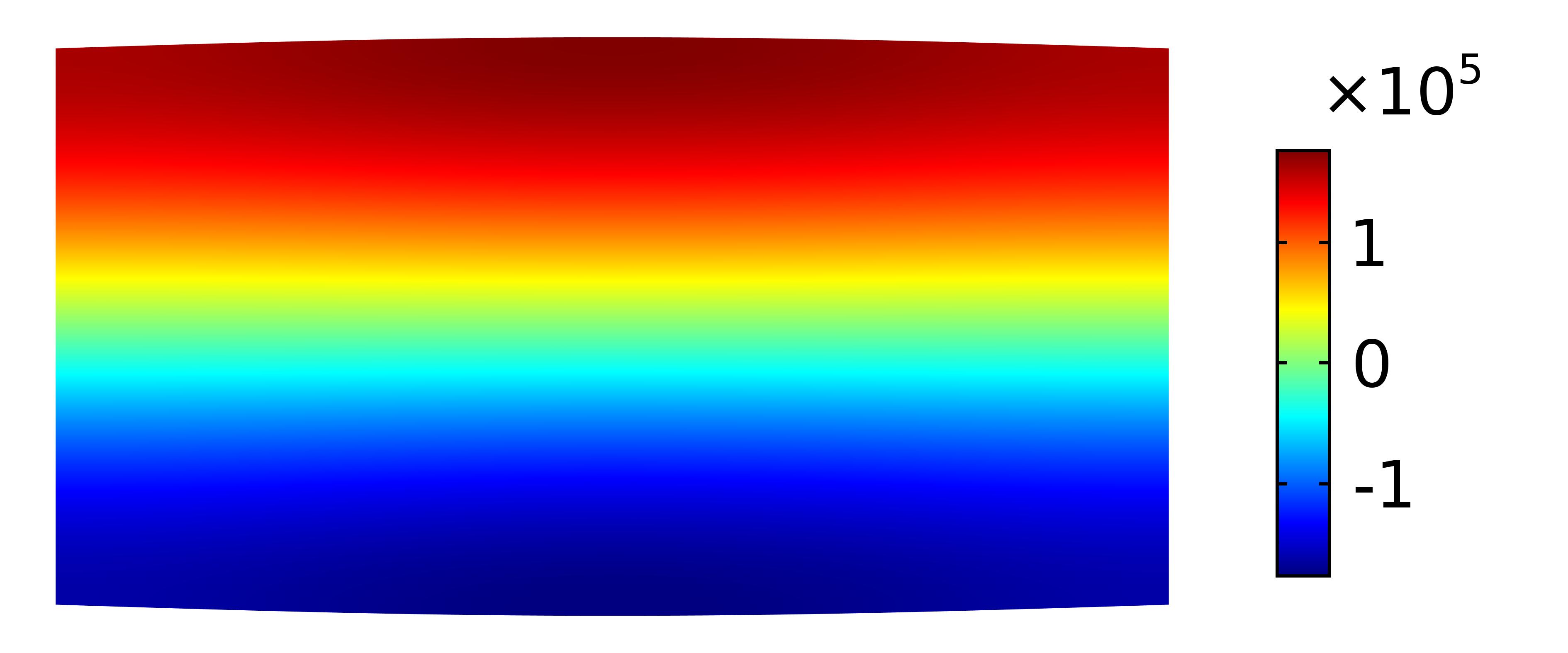}\label{fig6.c}}&	
			\subfigure[]{\includegraphics[width=40 mm]{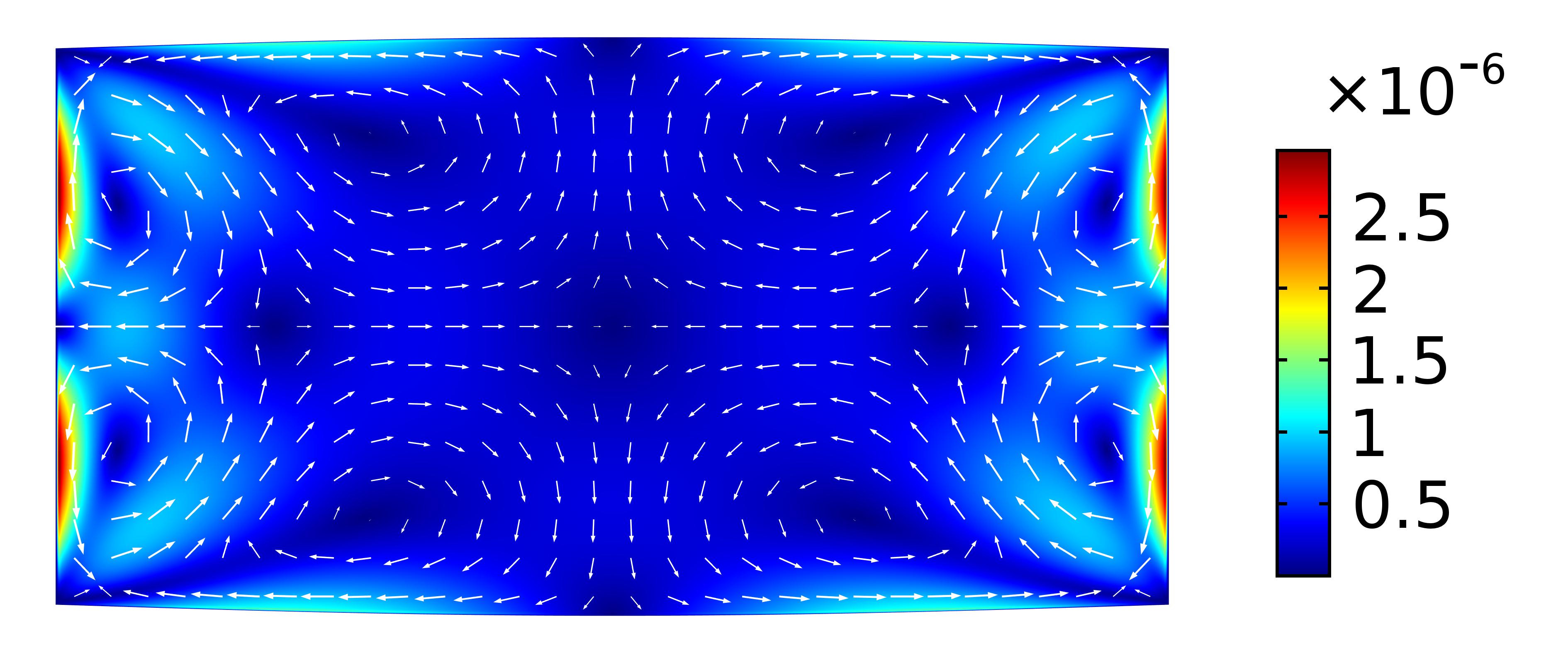}\label{fig6.d}}\\
			$n=3$&	
			\subfigure[]{\includegraphics[width=60 mm]{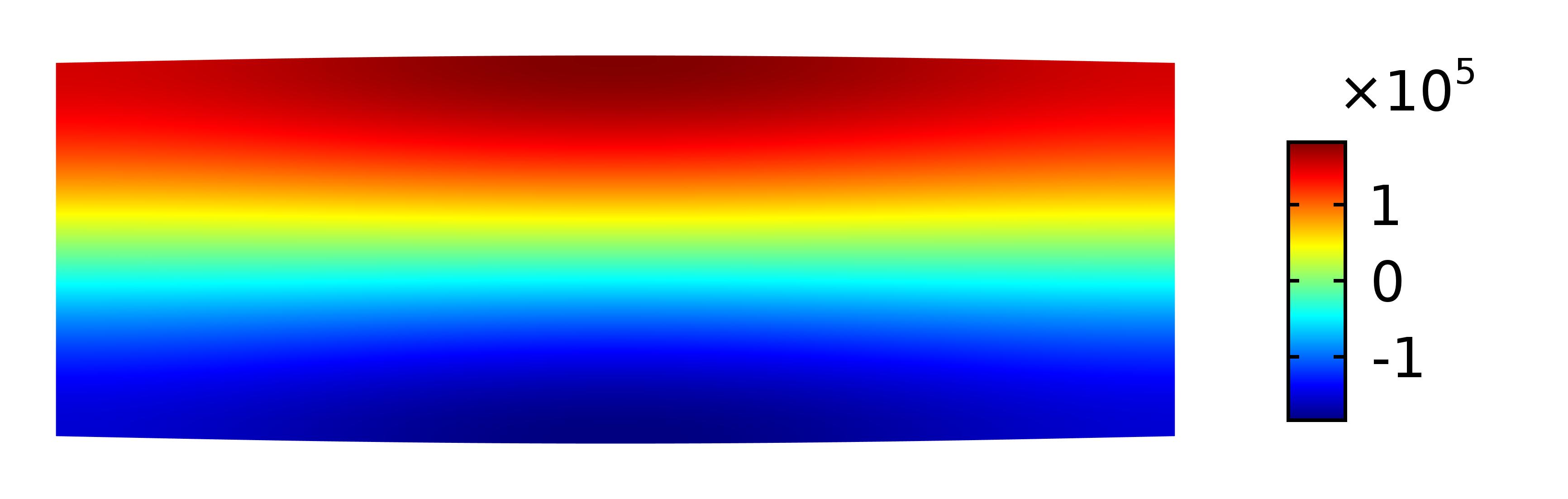}\label{fig6.e}}&	
			\subfigure[]{\includegraphics[width=60 mm]{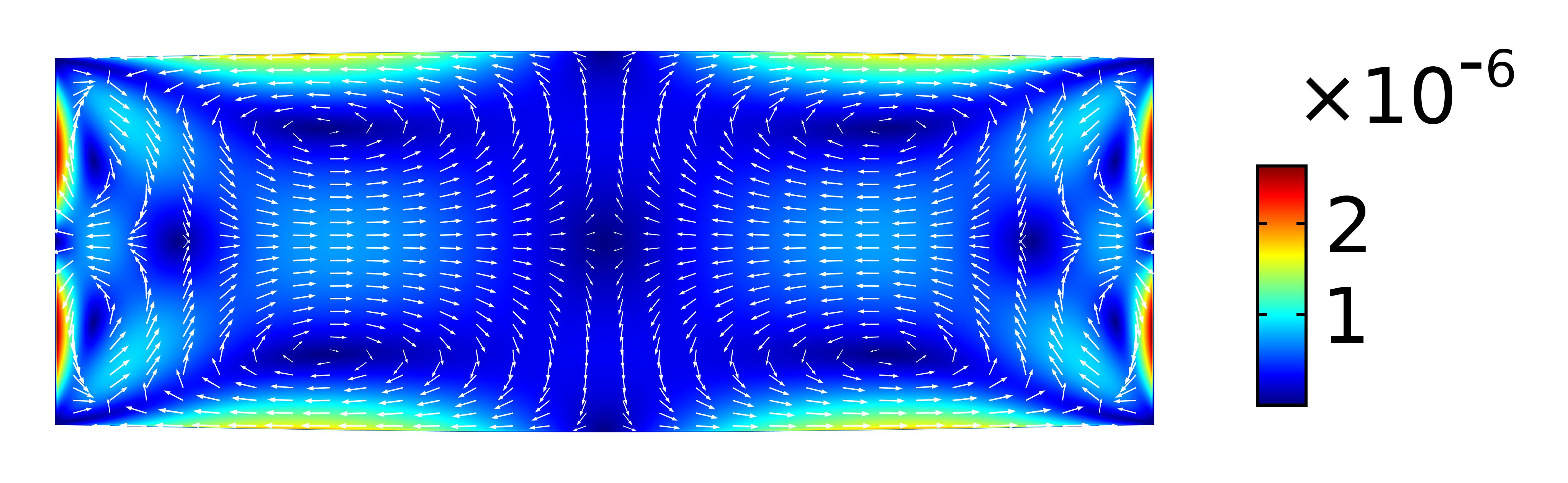}\label{fig6.f}}\\
			$n=4$&	
			\subfigure[]{\includegraphics[width=70 mm]{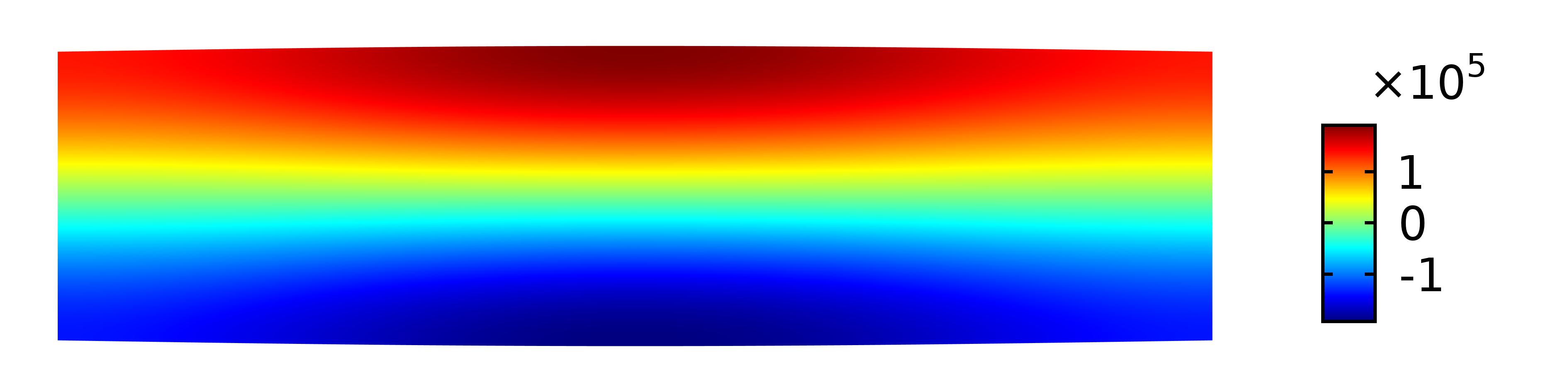}\label{fig6.g}}&	
			\subfigure[]{\includegraphics[width=70 mm]{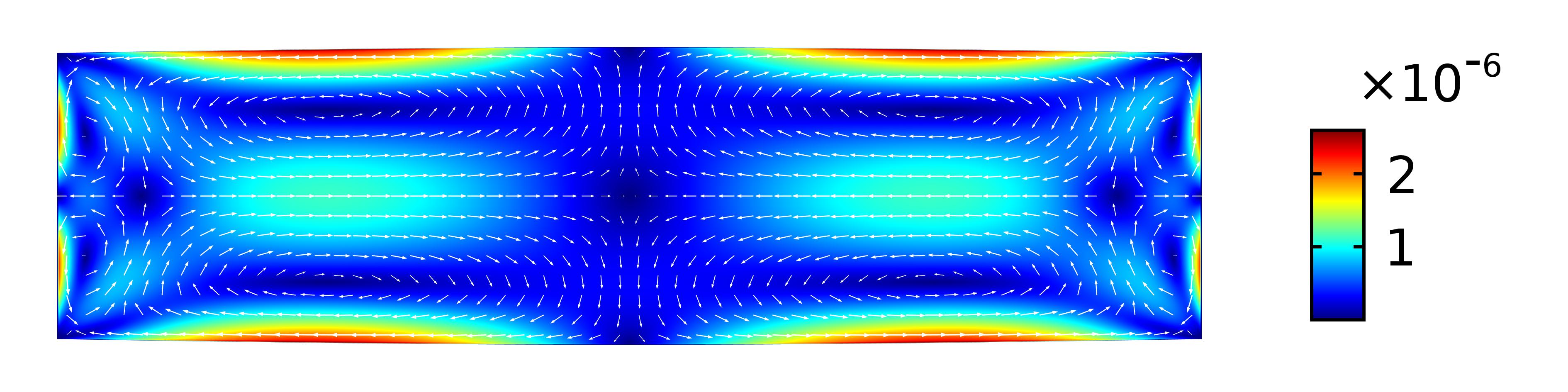}\label{fig6.h}}\\
			$n=5$&	
			\subfigure[]{\includegraphics[width=85 mm]{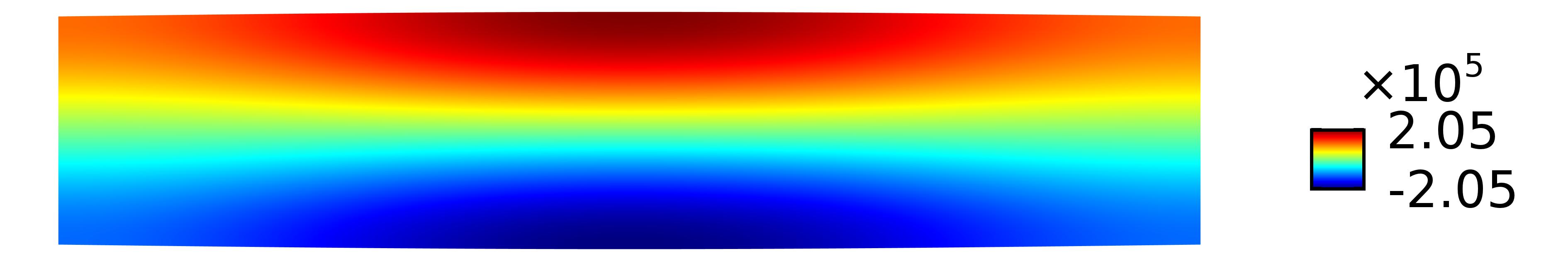}\label{fig6.i}}&	
			\subfigure[]{\includegraphics[width=85 mm]{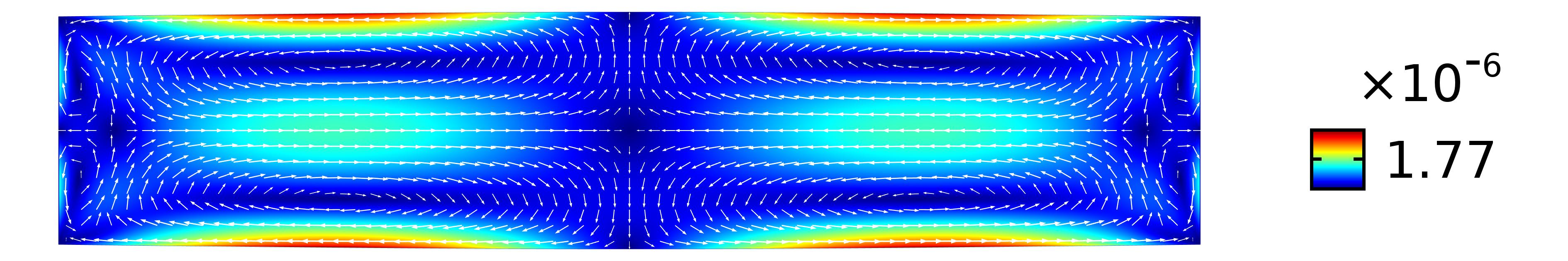}\label{fig6.j}}
		\end{tabular}
			\caption{First order pressure fields (left column) and time-averaged second-order velocity fields (right column) in cross section of sinusoidal microchannels where geometrical wavelengths of asymmetrical top and bottom walls are keeping fixed as $\lambda_g=2w$ and the microchannel's width to height ratio, $n$, varies from $1$ to $5$.} 
			\label{fig6}
	\end{center}
\end{figure*}

\begin{figure*}[ht] 
		\begin{tabular}{l|ll}	
\;& $p_1$ & $\langle v_2 \rangle$ \\
\hline
$\lambda_g=2w$&	
\subfigure[]{\includegraphics[height=12 mm]{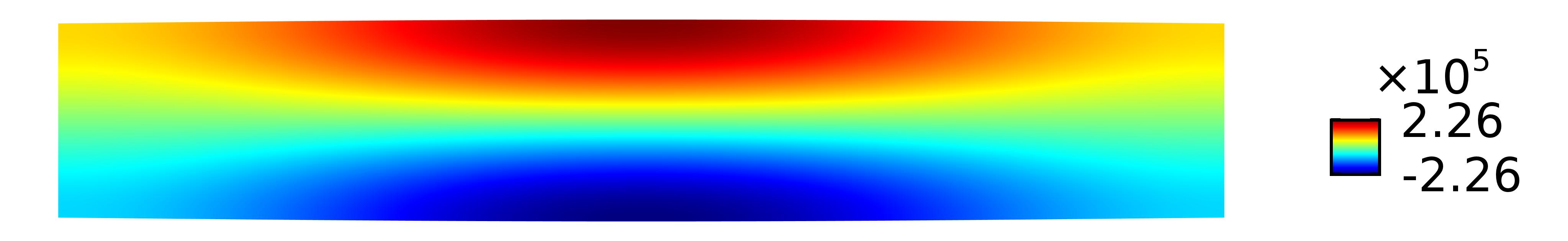}\label{fig7.a}}&
\hspace{-5mm}	
\subfigure[]{\includegraphics[height=12 mm]{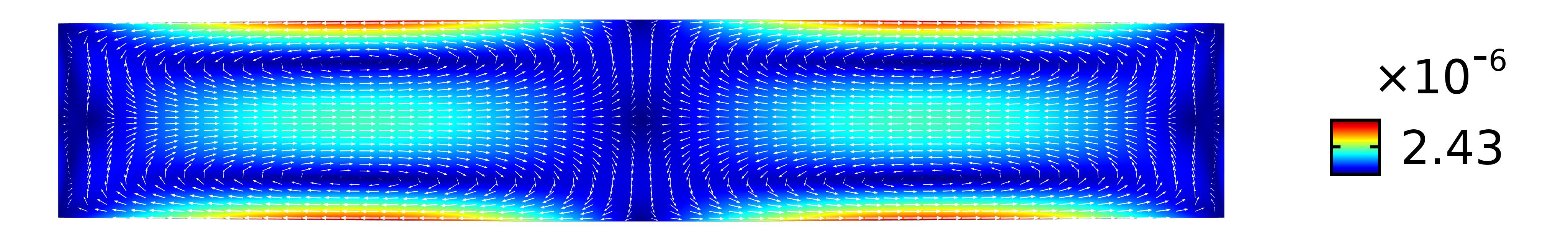}\label{fig7.b}}\\
$\lambda_g=w$&	
\subfigure[]{\includegraphics[height=12 mm]{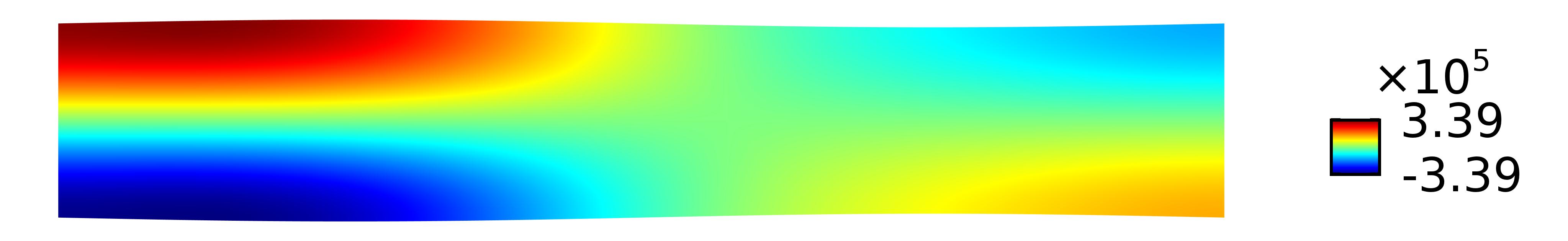}\label{fig7.c}}&
\hspace{-5mm}
\subfigure[]{\includegraphics[height=12 mm]{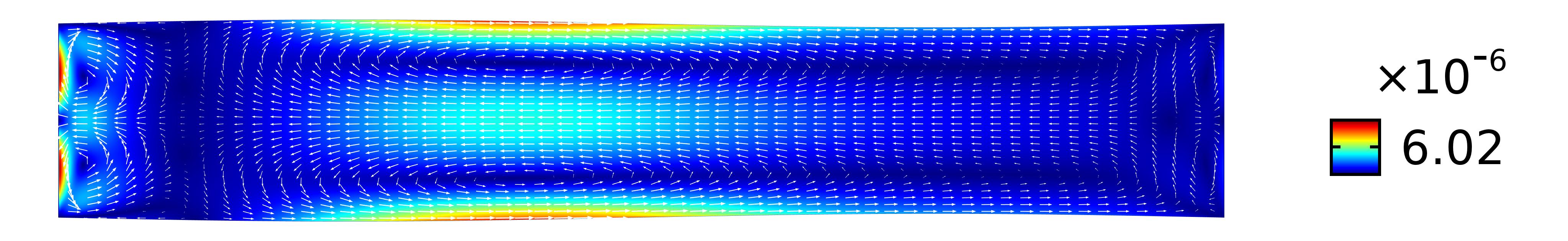}\label{fig7.d}}\\
$\lambda_g=\frac{2}{3}w$&	
\subfigure[]{\includegraphics[height=12 mm]{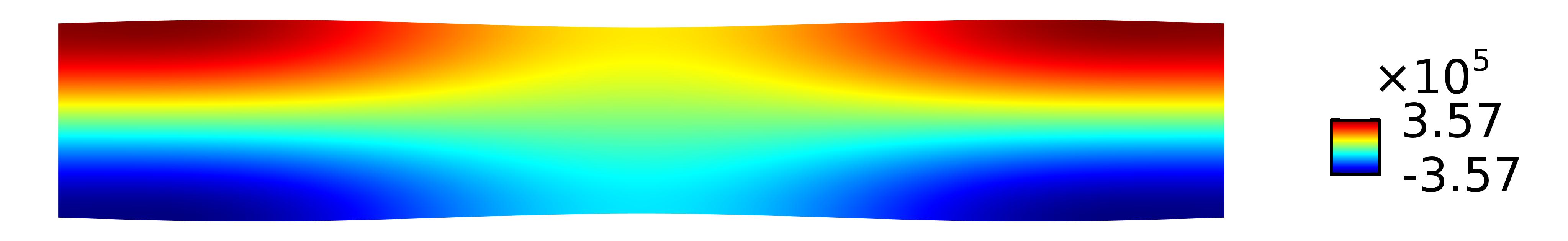}\label{fig7.e}}&
\hspace{-5mm}
\subfigure[]{\includegraphics[height=12 mm]{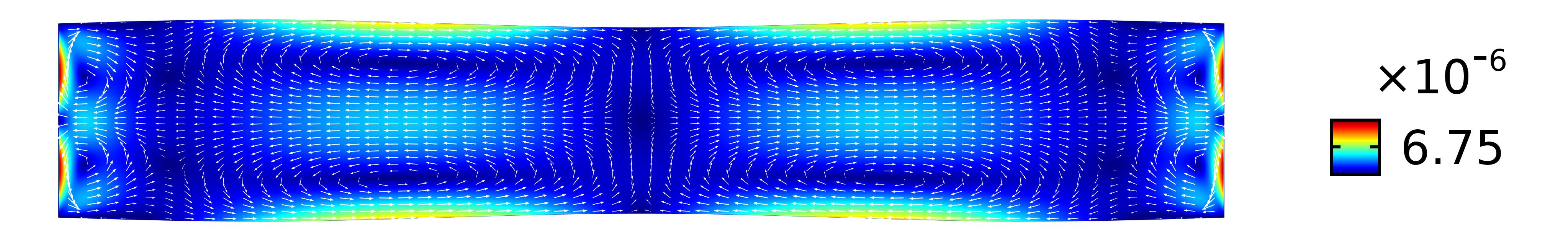}\label{fig7.f}}\\	
$\lambda_g=\frac{1}{2}w$&	
\subfigure[]{\includegraphics[height=12 mm]{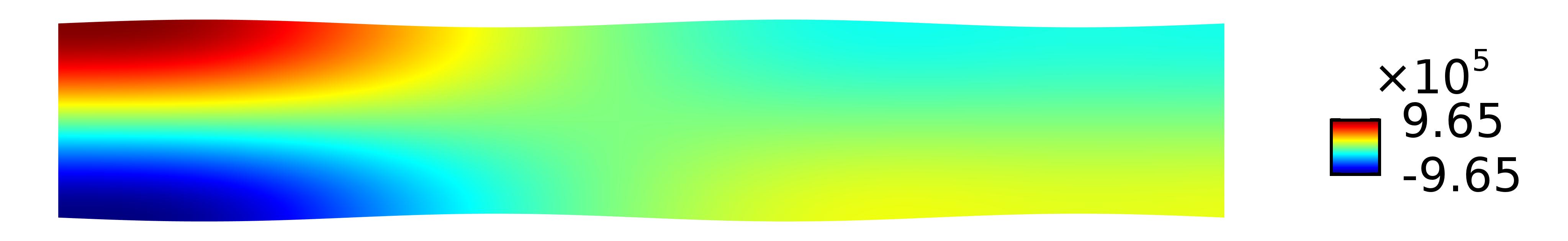}\label{fig7.g}}&
\hspace{-5mm}
\subfigure[]{\includegraphics[height=12 mm]{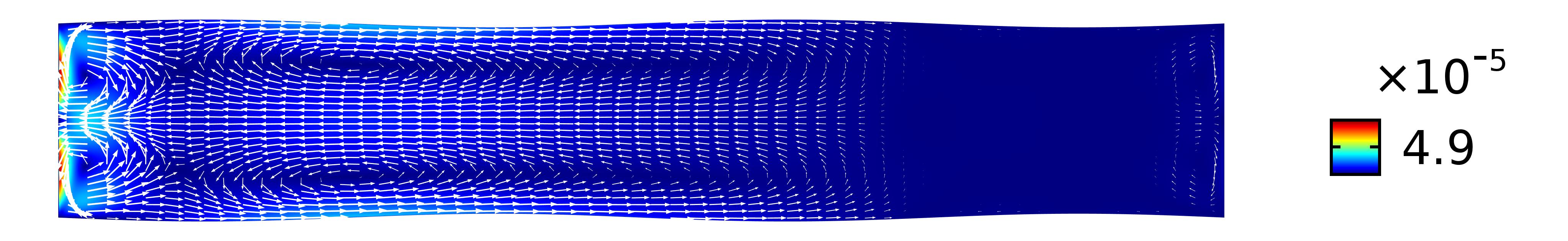}\label{fig7.h}}\\
$\lambda_g=\frac{2}{5}w$&	
\subfigure[]{\includegraphics[height=12 mm]{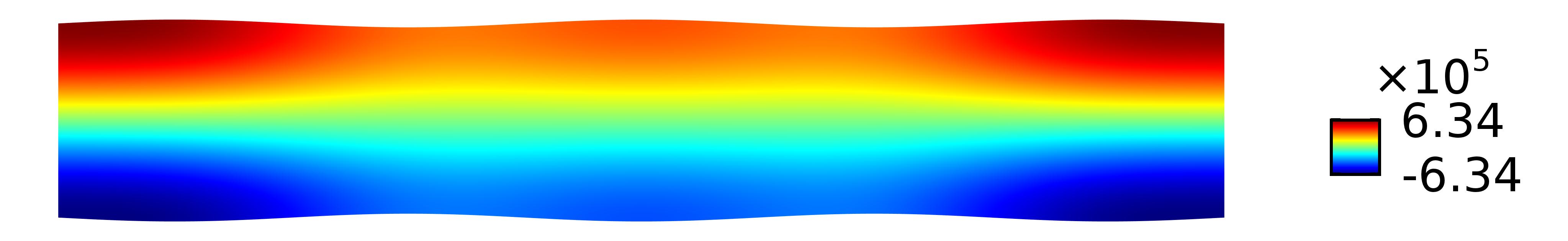}\label{fig7.i}}&
\hspace{-5mm}
\subfigure[]{\includegraphics[height=12 mm]{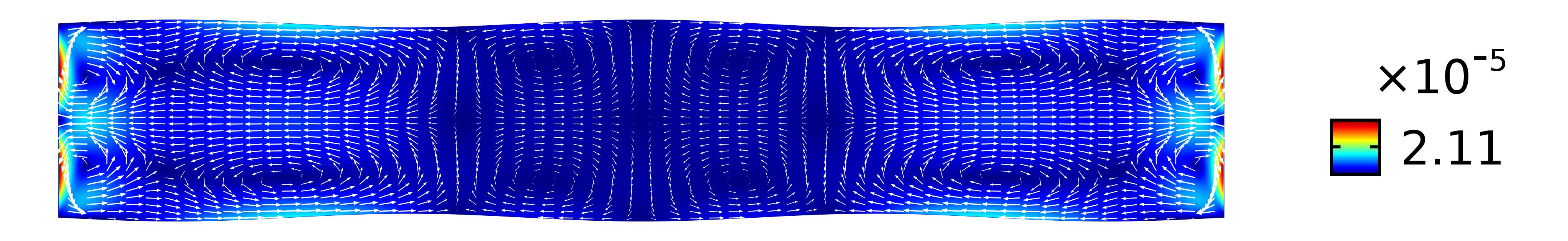}\label{fig7.j}}\\
$\lambda_g=\frac{1}{3}w$&
\subfigure[]{\includegraphics[height=12 mm]{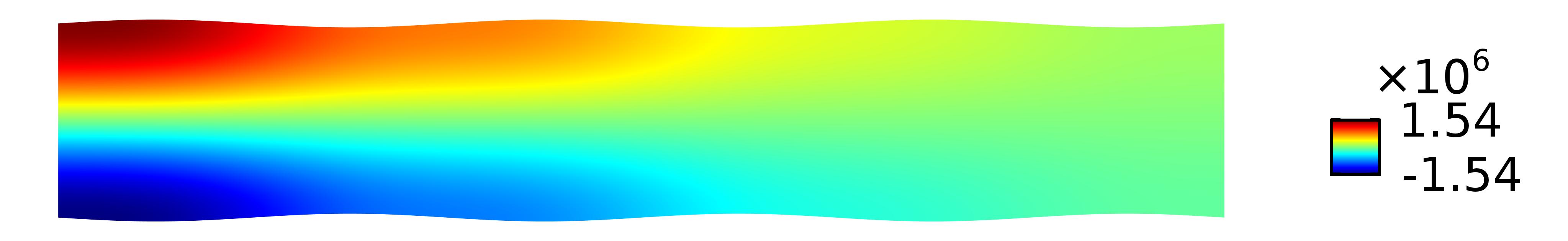}\label{fig7.k}}&
\hspace{-5mm}
\subfigure[]{\includegraphics[height=12 mm]{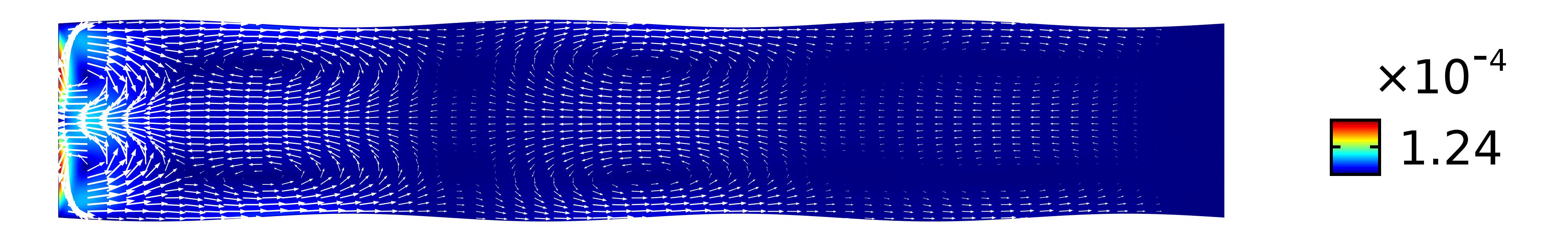}\label{fig7.l}}\\
$\lambda_g=\frac{2}{7}w$&
\subfigure[]{\includegraphics[height=12 mm]{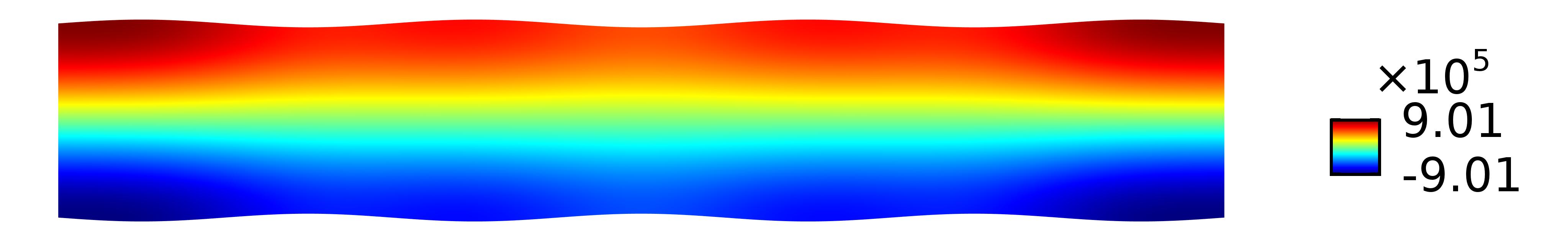}\label{fig7.m}}&
\hspace{-5mm}
\subfigure[]{\includegraphics[height=12 mm]{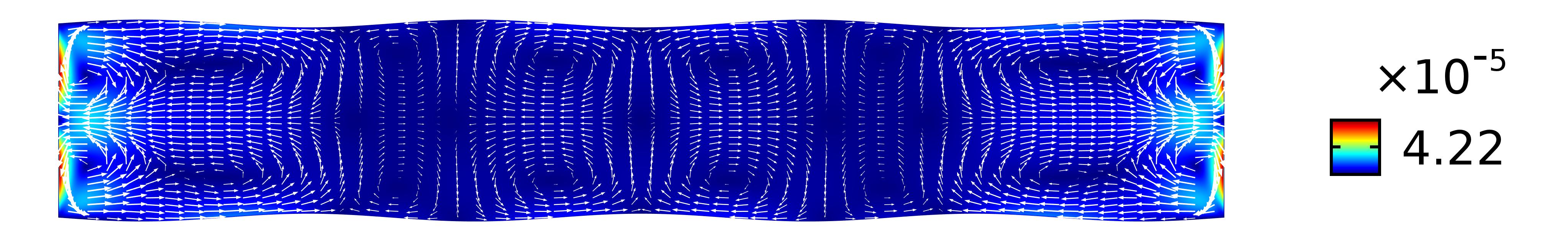}\label{fig7.n}}\\
$\lambda_g=\frac{1}{4}w$&
\subfigure[]{\includegraphics[height=12 mm]{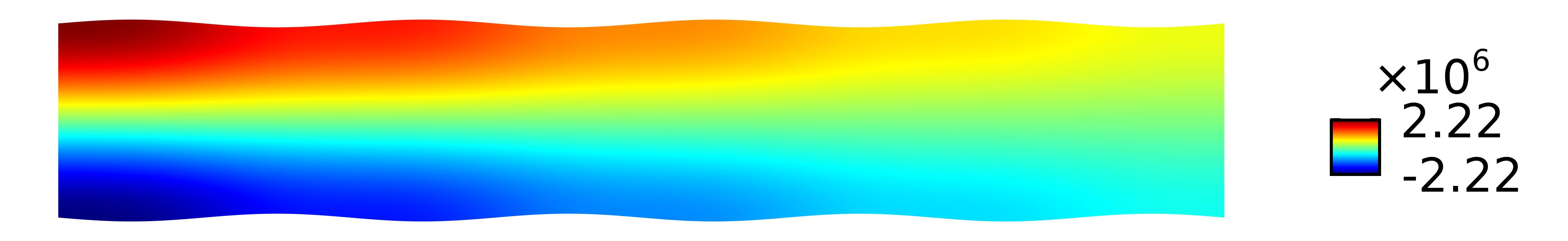}\label{fig7.o}}&
\hspace{-5mm}
\subfigure[]{\includegraphics[height=12 mm]{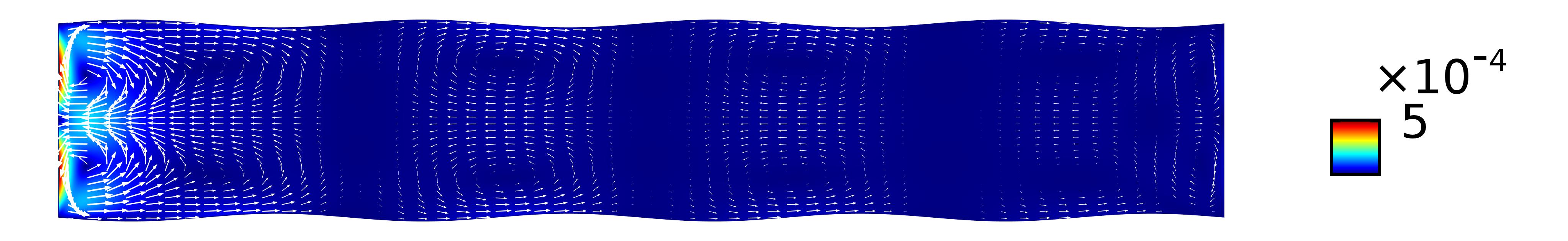}\label{fig7.p}}			
		\end{tabular}
			\caption{First order pressure fields (left column) and time-averaged second-order velocity fields (right column) in cross section of sinusoidal microchannels where the microchannel's width to height ratio is keeping fixed as $n=6$ and geometrical wavelengths of assymetrical top and bottom walls are $\lambda_g=\frac{4}{2p}w$ where $p=1,2,..., 8$.} 
			\label{fig7}
	\end{figure*}

First formula is declared for fast streaming patterns like Fig. \ref{fig5.d}. It resembles to a chain with $n$ numbers of rings.
\begin{equation}
\lambda_g=\frac{4}{2n-10}w, \qquad\textrm{where} \; n\geq 6 \in R.
\label{eq16}
\end{equation} 

Next formula is proposed for the patterns like Fig. \ref{fig5.l} that $n$ numbers of dominant vortices are recognizable.   
\begin{equation}
\lambda_g=\frac{4}{2n-2}w, \qquad\textrm{where} \; n\geq 2 \in R.
\label{eq17}
\end{equation} 

Similar patterns, with a shift of one vortex forward, exist (As shown in Fig. \ref{fig5.p}). The formulation is 
\begin{equation}
\lambda_g=\frac{4}{2n+2}w, \qquad\textrm{where} \; n\geq 1 \in R.
\label{eq18}
\end{equation} 
 
Another series of similar patterns could be defined considering that two corresponding vortices are generated instead of one dominant vortex as in two previous classes. 
\begin{equation}
\lambda_g=\frac{4}{2n-6}w, \qquad\textrm{where} \; n\geq 4 \in R.
\label{eq19}
\end{equation} 

Next formula is defined for identical patterns imposing a shift of one vortex forward (See Fig. \ref{fig5.t}).
\begin{equation}
\lambda_g=\frac{4}{2n+6}w, \qquad\textrm{where} \; n\geq 1 \in R.
\label{eq20}
\end{equation} 

For the cases very similar to flat geometry, a general relation can be proposed as 
 
\begin{subequations}
	\begin{align}
&\lambda_g=\frac{4}{2p-1}w, &\qquad\textrm{where} \; p\geq 1 \in R \qquad\textrm{for odd numbers of $n$}\\
&\lambda_g=\frac{4}{2p}w, &\qquad\textrm{where} \; p\geq 1 \in R \qquad\textrm{for even numbers of $n$}
	\end{align}
\label{eq21}
\end{subequations} 
  
\begin{figure*}[ht]
\vspace{-1cm}
	\begin{center} 
		\begin{tabular}{l|l} 
			$\lambda_g$ , $n$ & $\langle v_2 \rangle$ \\
			\hline
			$4w$ , 1 &	 
			\subfigure {\includegraphics[height=19 mm] {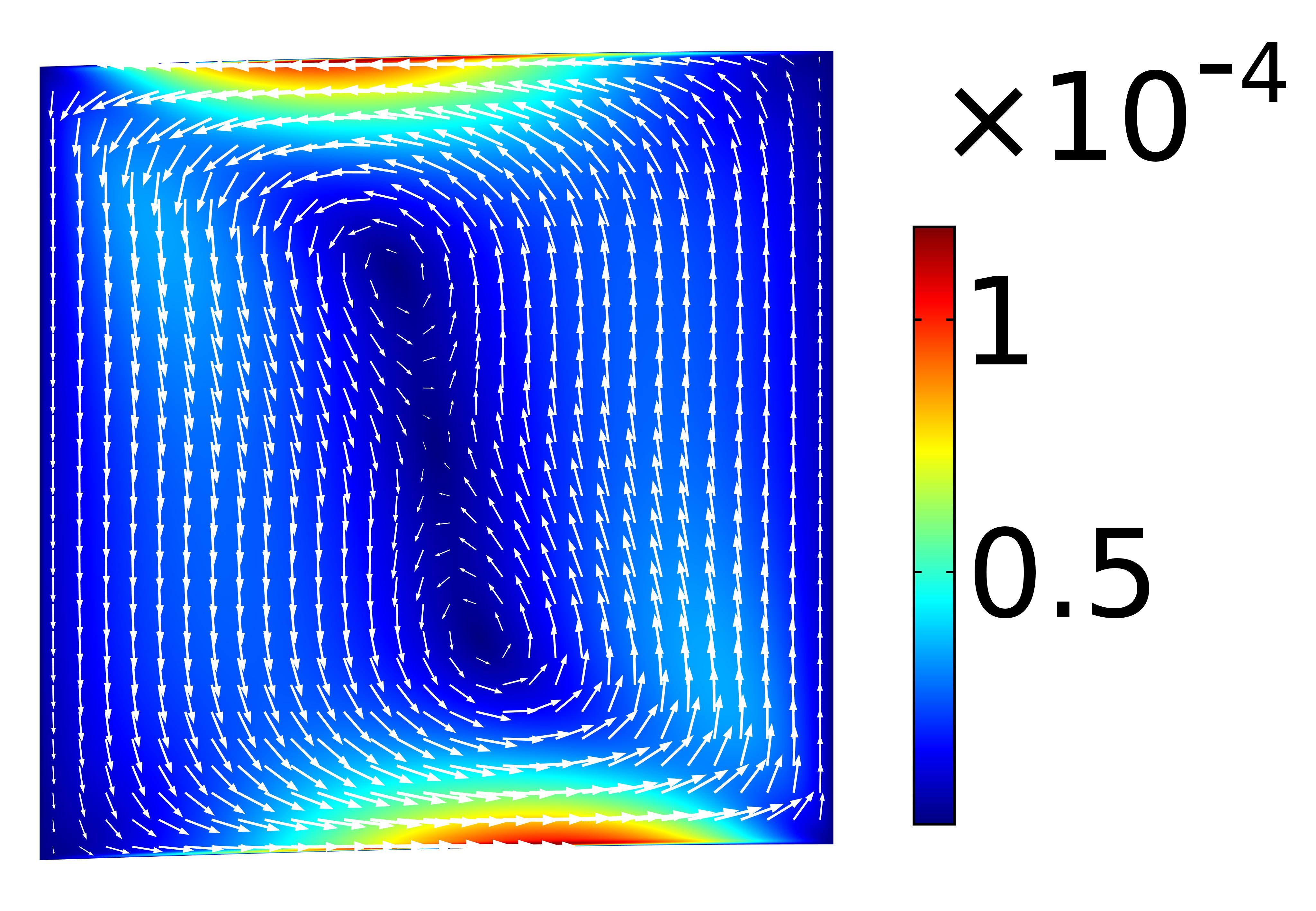}}\\  
			$2w$ , 2 &	
			\subfigure {\includegraphics[height=19 mm] {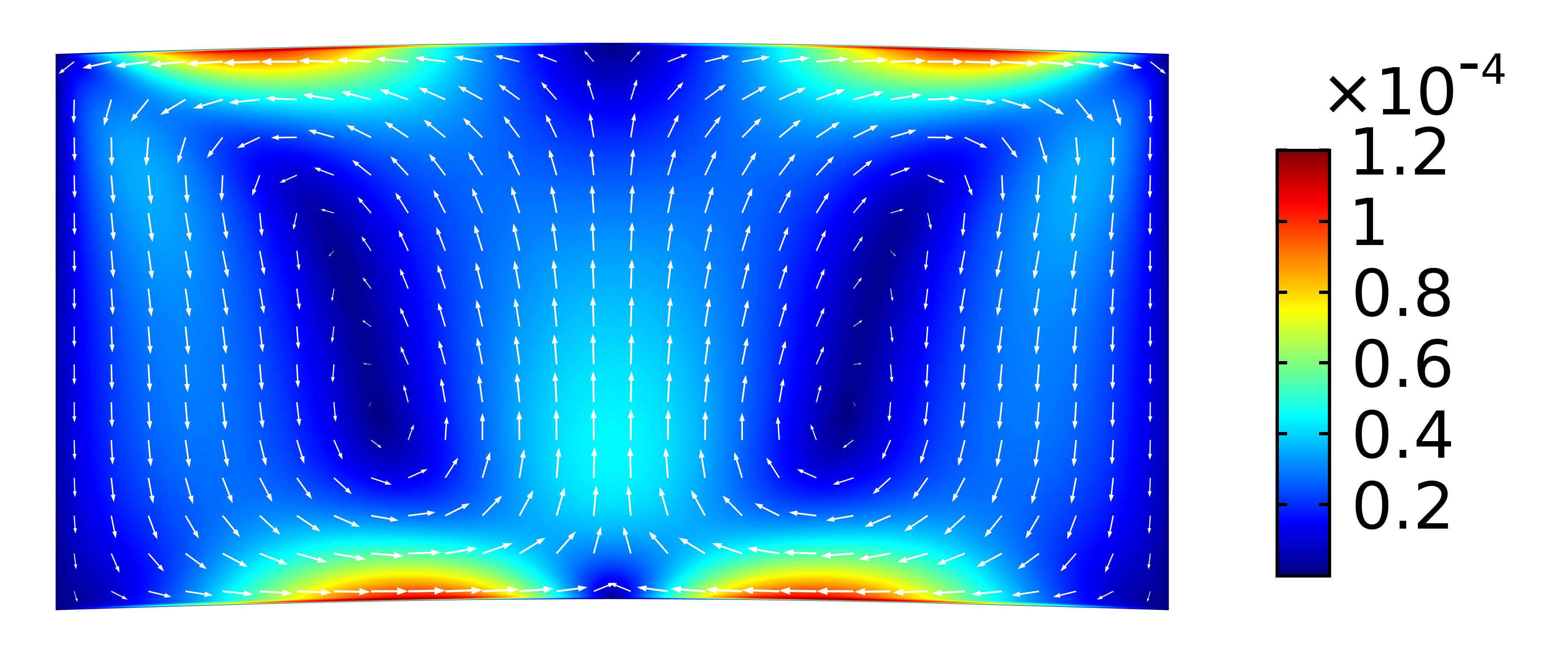}}\\  
			$w$ , 3 &	 
			\subfigure {\includegraphics[height=21 mm] {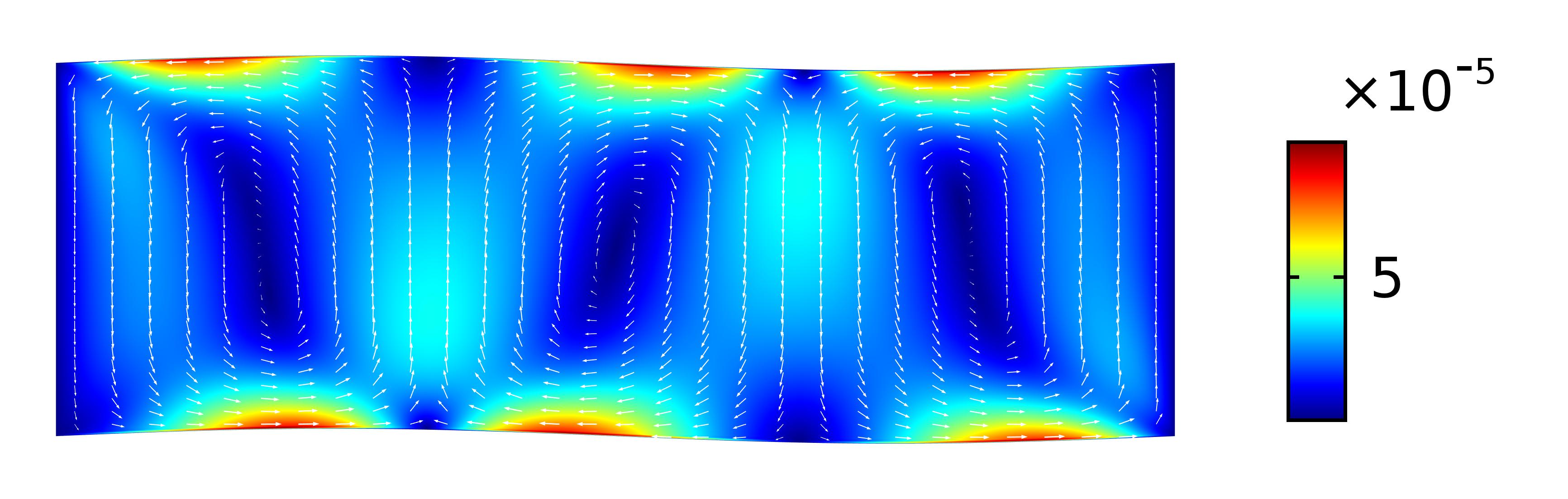}}\\  
			$\frac{2}{3} w$ , 4 &
			\subfigure {\includegraphics[height=21 mm] {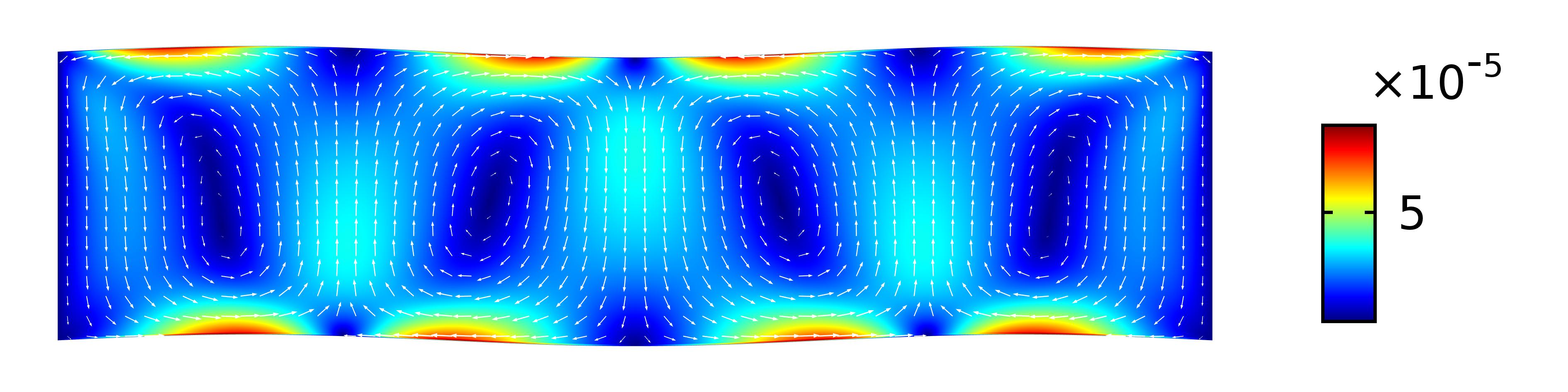}}\\
			$\frac{1}{2} w$ , 5 &
			\subfigure {\includegraphics[height=18 mm] {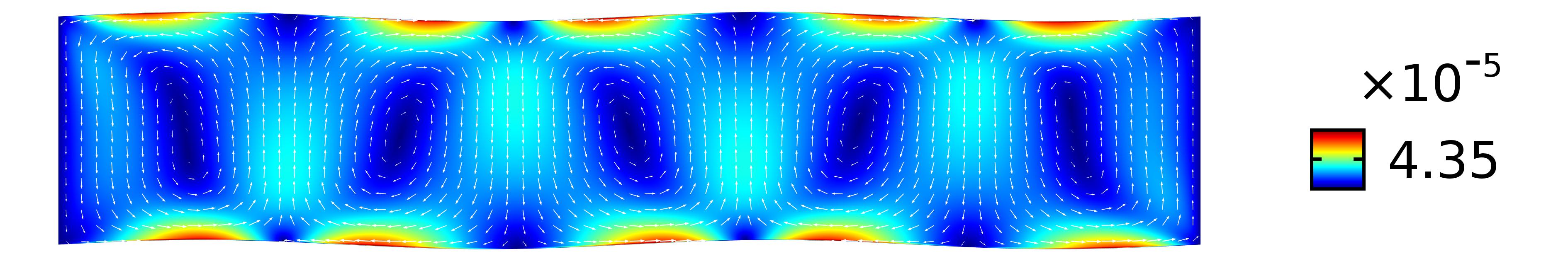}}\\
			$\frac{2}{5} w$ , 6 &
			\subfigure {\includegraphics[height=25 mm] {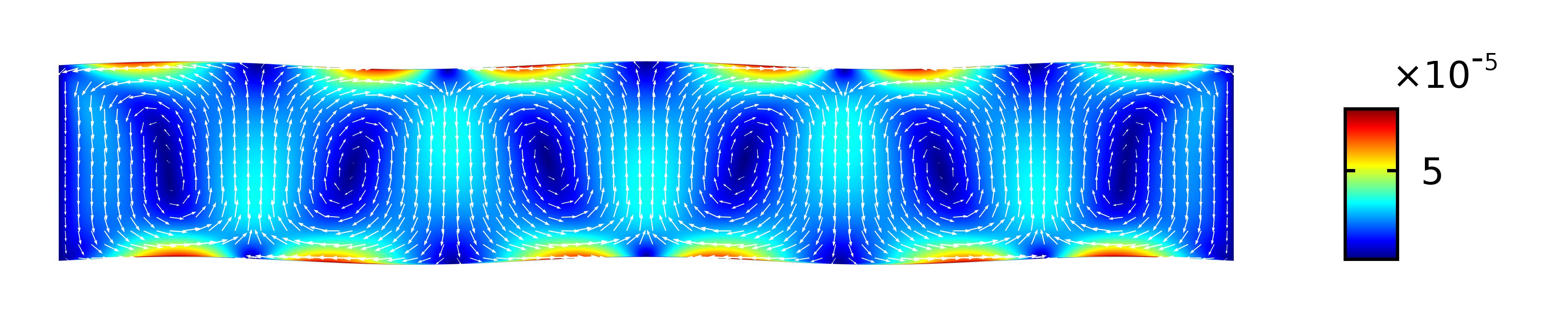}}\\
			$\frac{1}{3} w$ , 7 & 
			\subfigure {\includegraphics[height=24 mm] {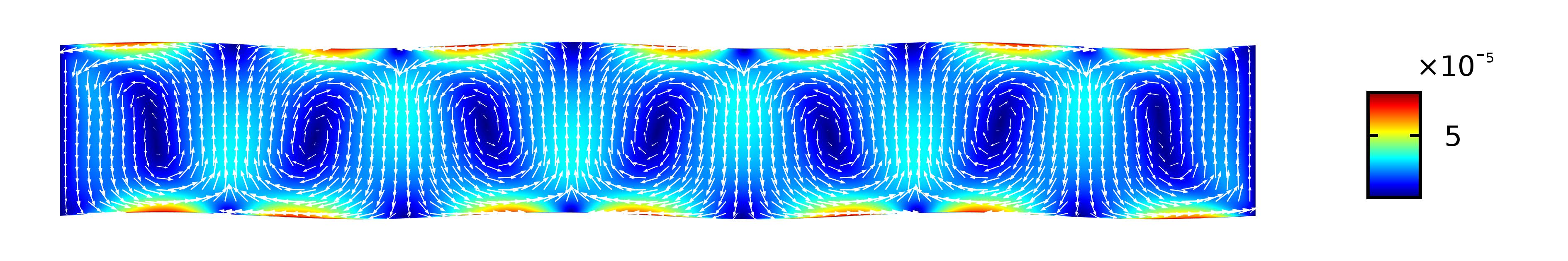}}\\
			$\frac{2}{7} w$ , 8 & 
			\subfigure {\includegraphics[height=24 mm] {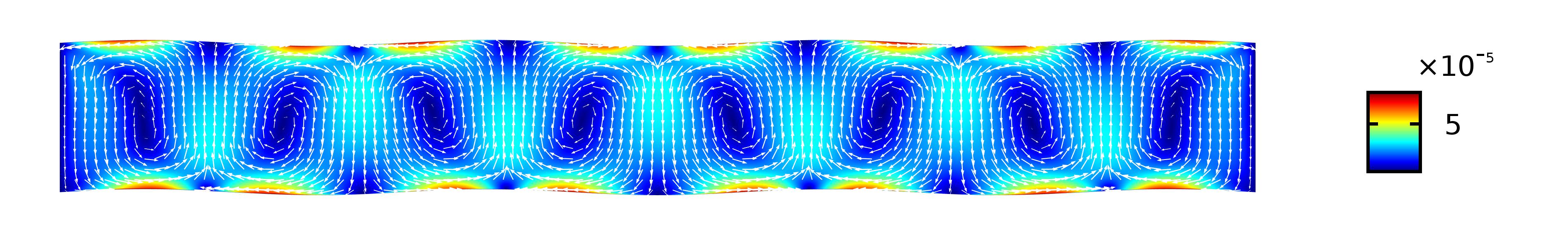}}
			\end{tabular} 
		\caption{An example of repetitive acoustic streaming patterns inside sinusoidal microchannels when the relation between geometrical wavelengths $\lambda_g$ and the microchannel's width to height ratio is defined as $\lambda_g=\frac{4}{2n-2}w$ where $n\geq 2 \in R$.}
		\label{fig8} 
		\end{center} 
	\end{figure*}

Figure \ref{fig8} validates Eq. \ref{eq15} of acoustic streaming classified patterns. A single vortex is accessible when $n=1$ and $\lambda_g=4w$ in addition to the proposed formula. 

Noteworthy, other cases with half numbers of $n$ or geometrical wavelengths of $\lambda_g=\frac{4w}{2p-1}$ with $p=1,2,...,10$, could be investigated. The same classification could be defined but be ignored in this study.

\subsection{Investigation of microchannel's building blocks}\label{RES-C}

\begin{figure*}[ht]
	\begin{center} 
		\begin{tabular}{l|cccc} 
 &	Block1 & Block2  & Block3 & Block4 \\
 \hline
			$p_1$ &	
			\subfigure {\includegraphics[height=22 mm] {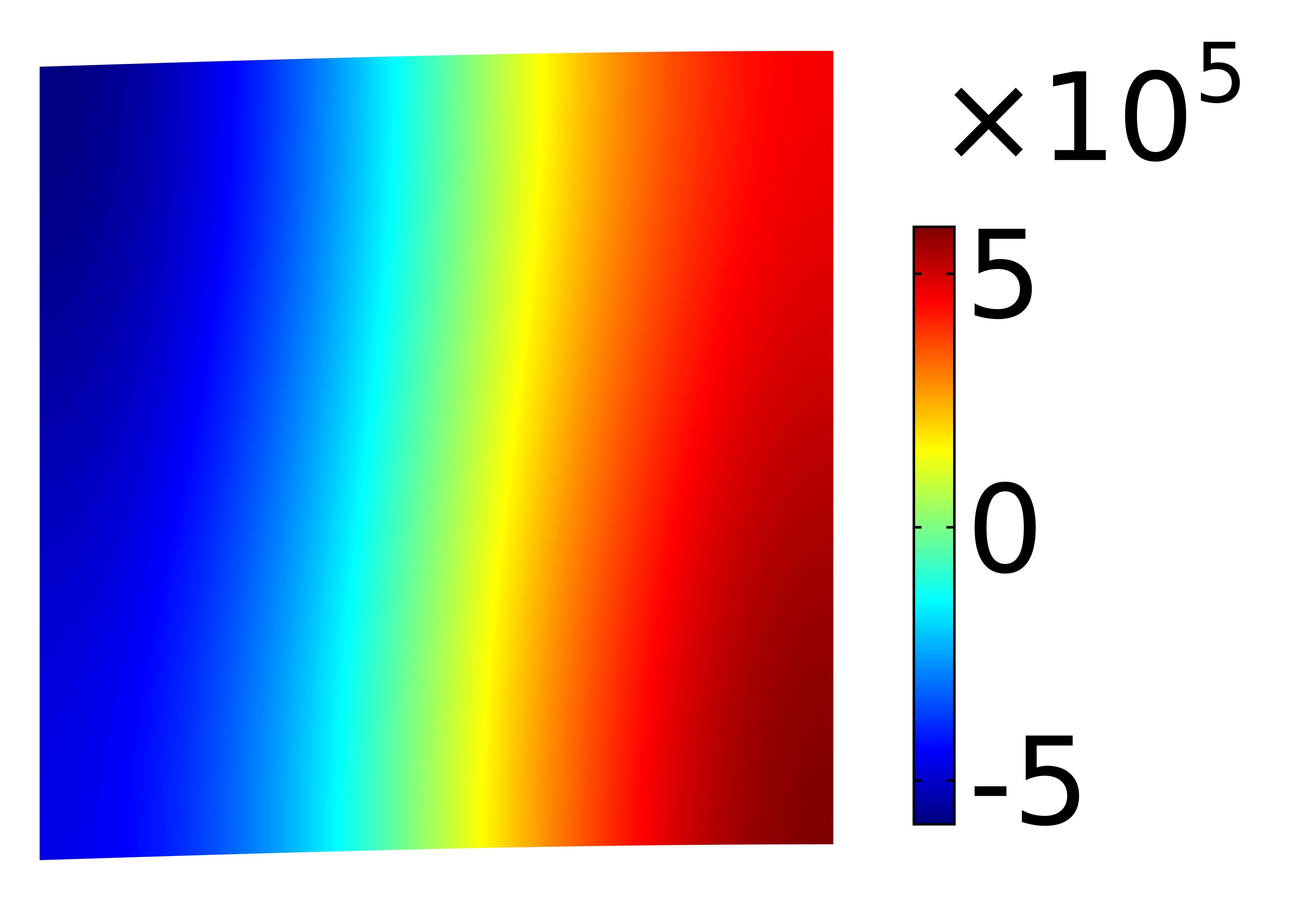}}& 
			\subfigure {\includegraphics[height=22 mm] {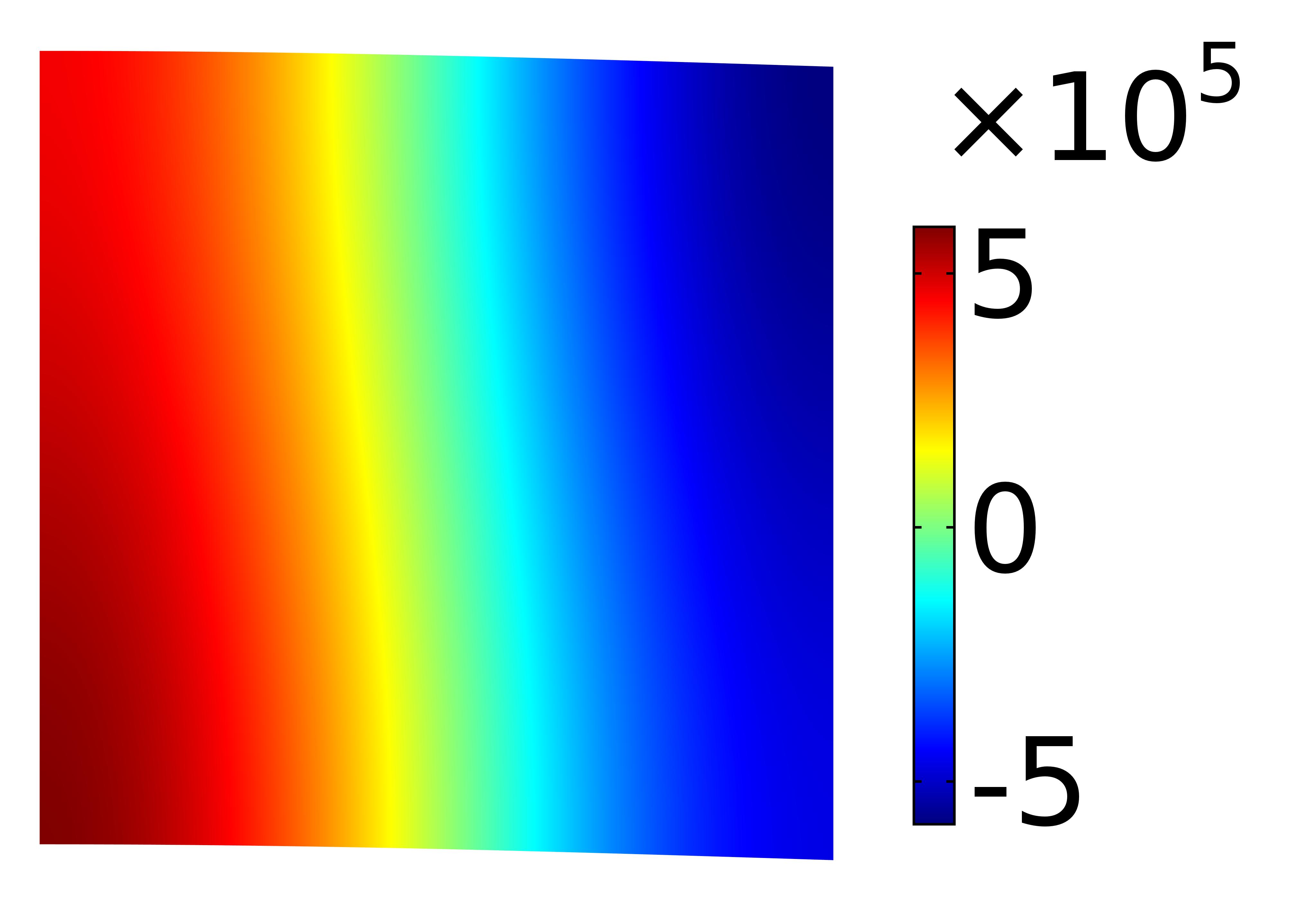}}& 
			\subfigure {\includegraphics[height=22 mm] {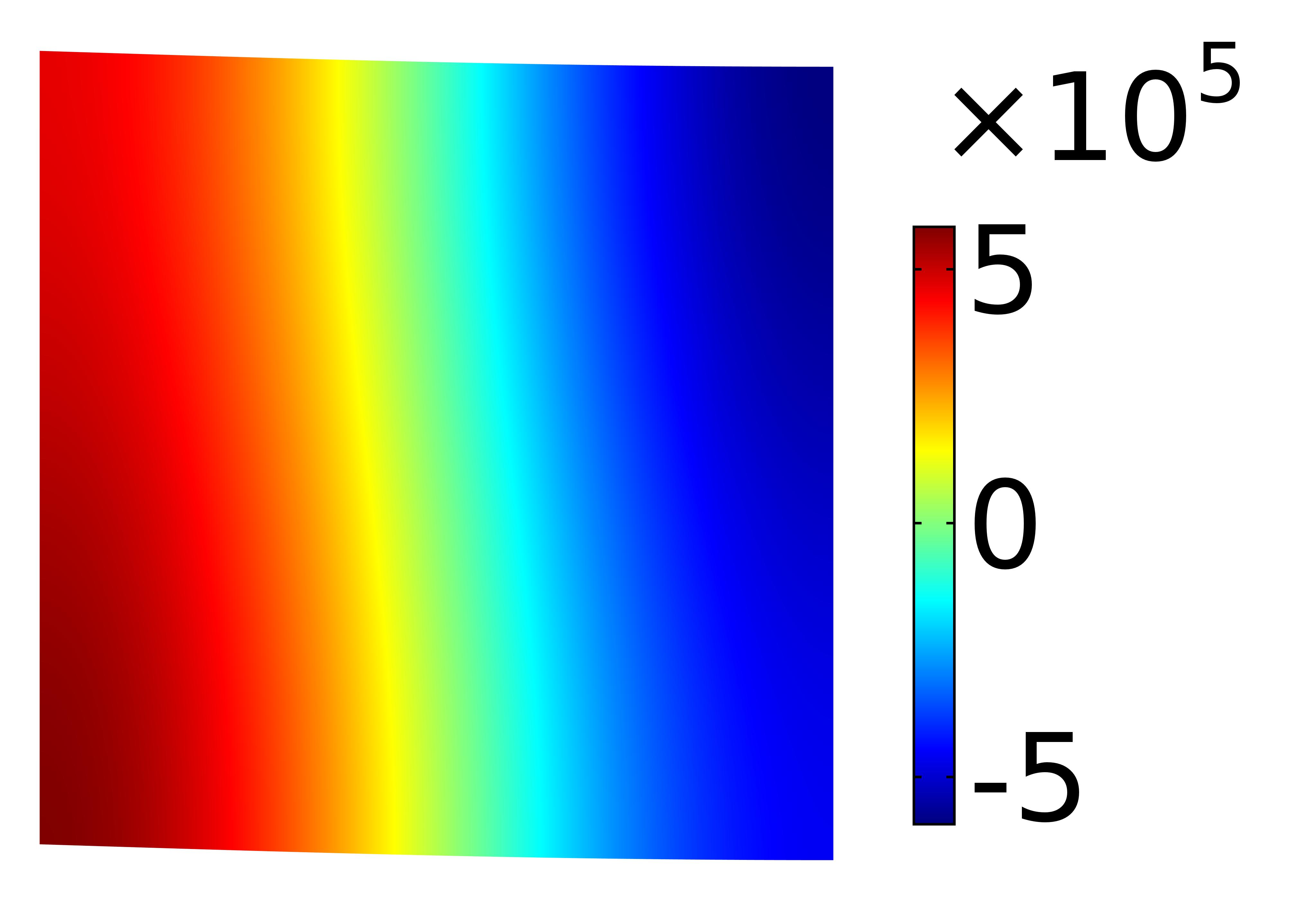}}&
			\subfigure {\includegraphics[height=22 mm] {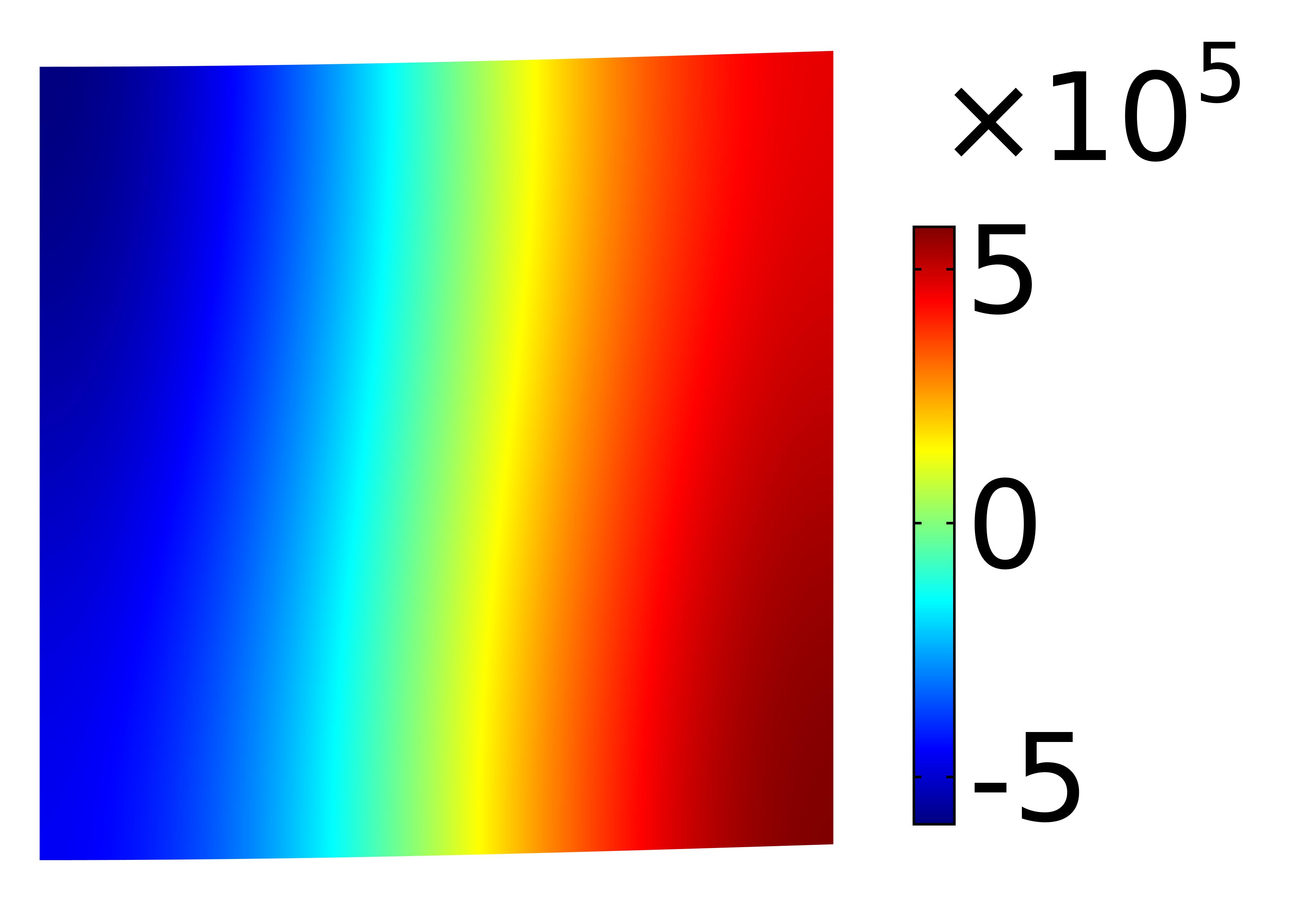}}\\ 
			$\langle v_2 \rangle$ &
			\subfigure {\includegraphics[height=22 mm] {sym-wide100-v2-mod05-block1.jpg}}& 
			\subfigure {\includegraphics[height=22 mm] {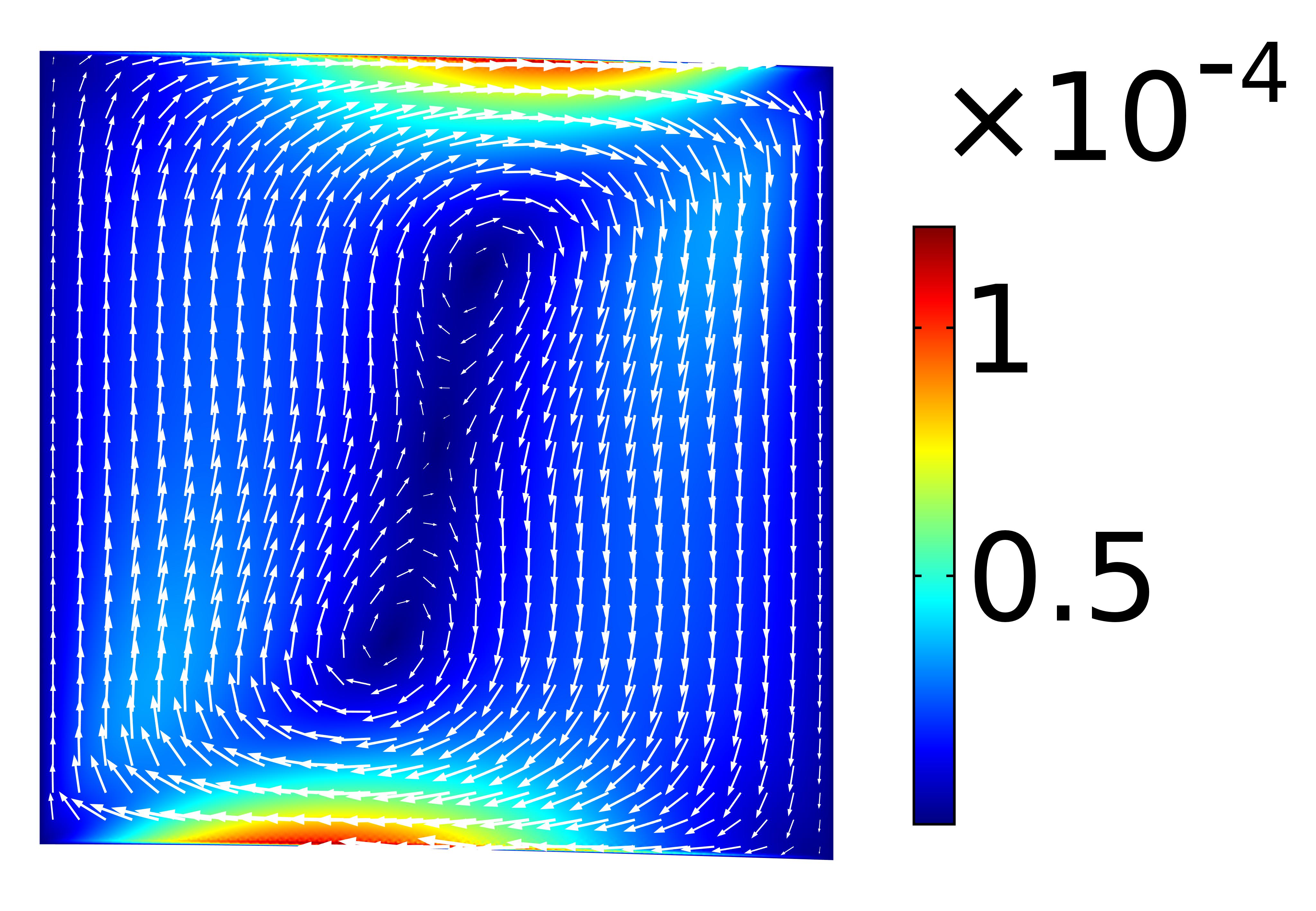}}& 
			\subfigure {\includegraphics[height=22 mm] {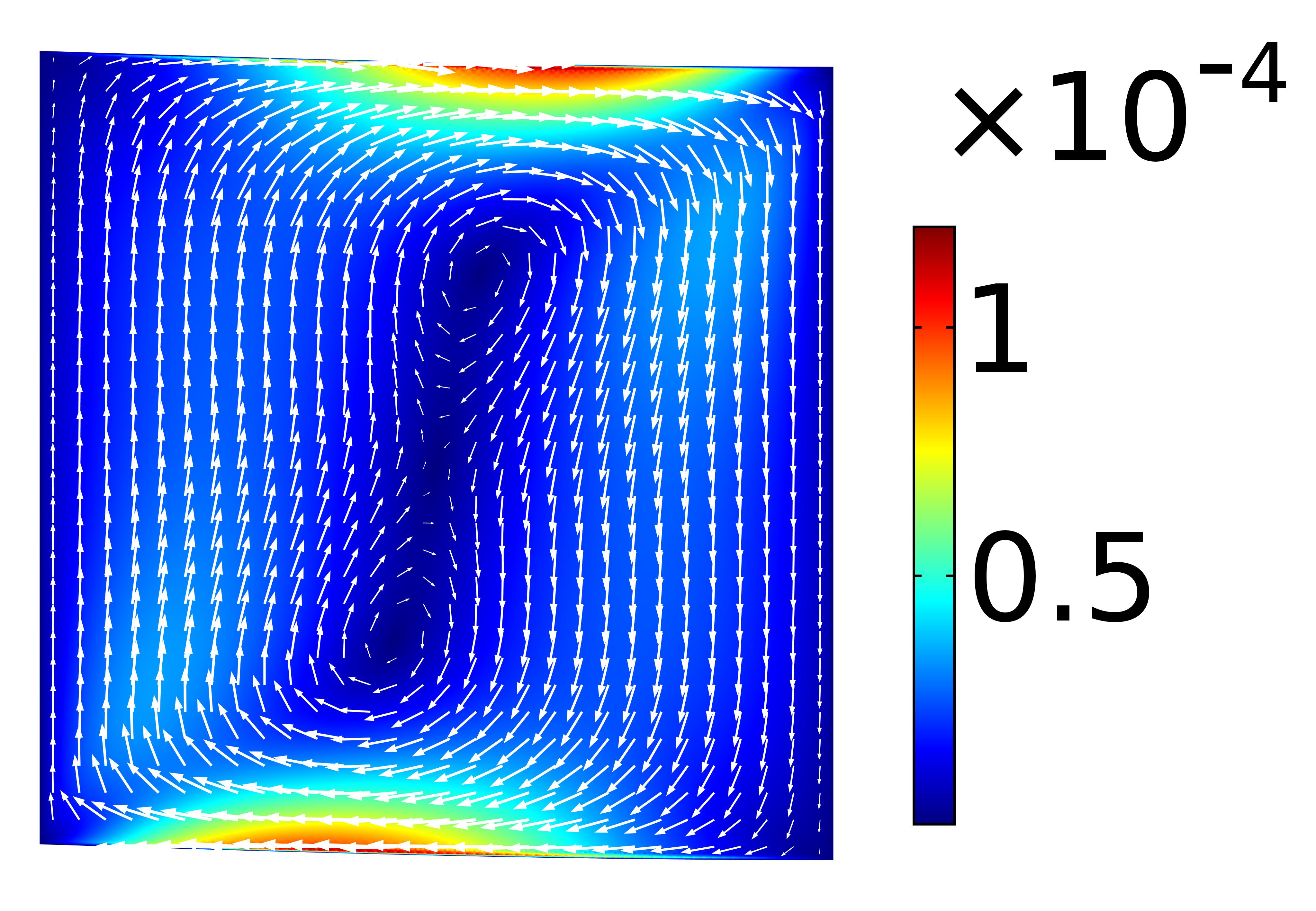}}&
			\subfigure {\includegraphics[height=22 mm] {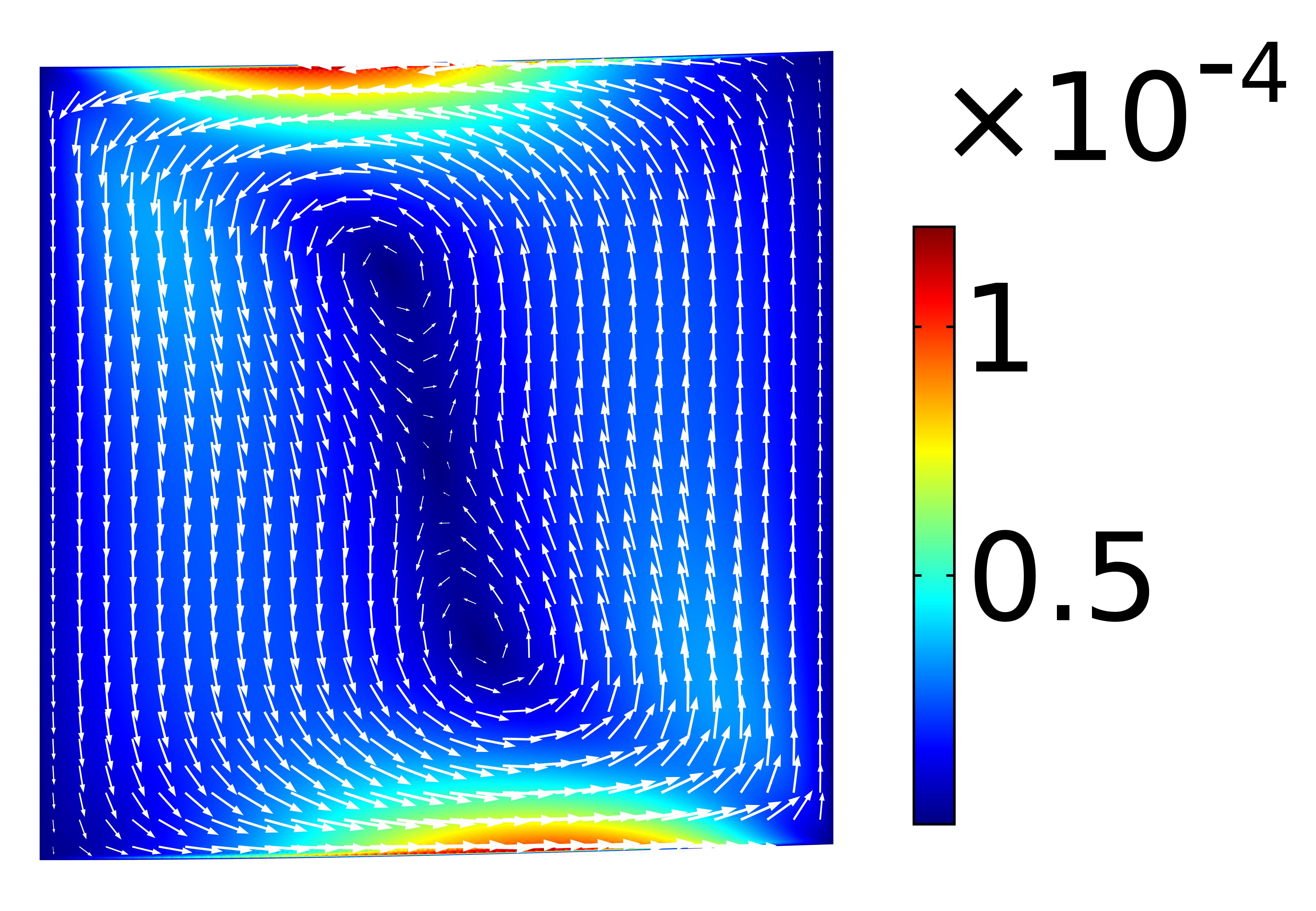}}\\
			\end{tabular} 
		\caption{First order pressure field, $p_1$, and acoustic streaming velocity fields ,$\langle v_2 \rangle$. for building blocks of a symmetrical sinusoidal microchannel with $n=4$ and $\lambda_g=w$. Each block has microchannel's width to height ratio of $n=1$ and geometrical wavelength of $\lambda_g=4w$. Geometrical phase for blocks1 to block4 are $\phi_g=0,\frac{\pi}{2}, \pi, \frac{3\pi}{2}$, respectively.}
		\label{fig9} 
		\end{center} 
	\end{figure*}

It might be supposed that repetitive patterns could be generated when some building blocks with a single-vortex pattern are joined to make a long sinusoidal microchannel. In this section, it is clarified that the repetition of acoustic streaming patterns is not in a trivial manner.
As in Fig. \ref{fig9} a microchannel with $n=4$ and $\lambda_g=w$ is separated into four microchannels with $n=1$ and $\lambda_g=4w$. 

\begin{figure*}[ht]
	\begin{center}		
		\begin{tabular}{l|cc} 
$\;$ &	Block 1 and 2 & Block 3 and 4 \\
 \hline
			$p_1$ &	  
			\subfigure {\includegraphics[height=18 mm] {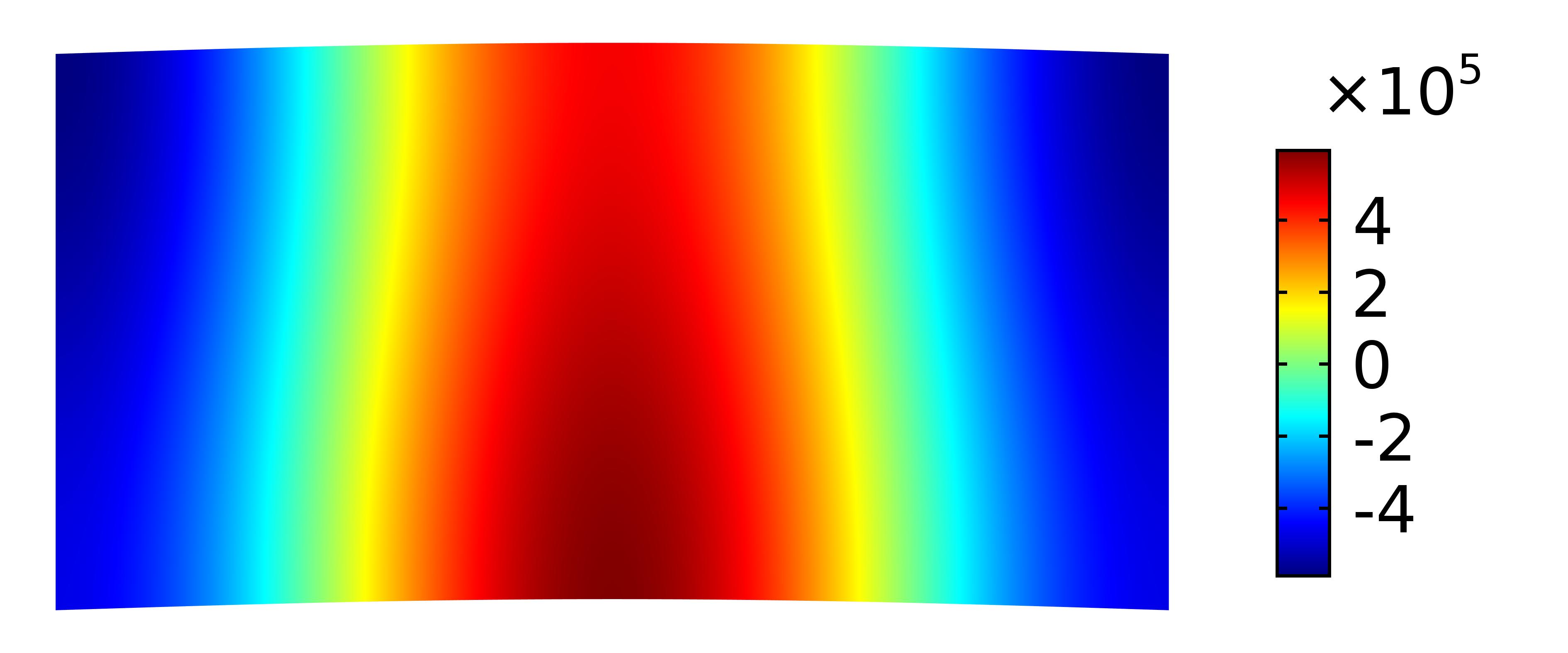}}& 
			\subfigure {\includegraphics[height=18 mm] {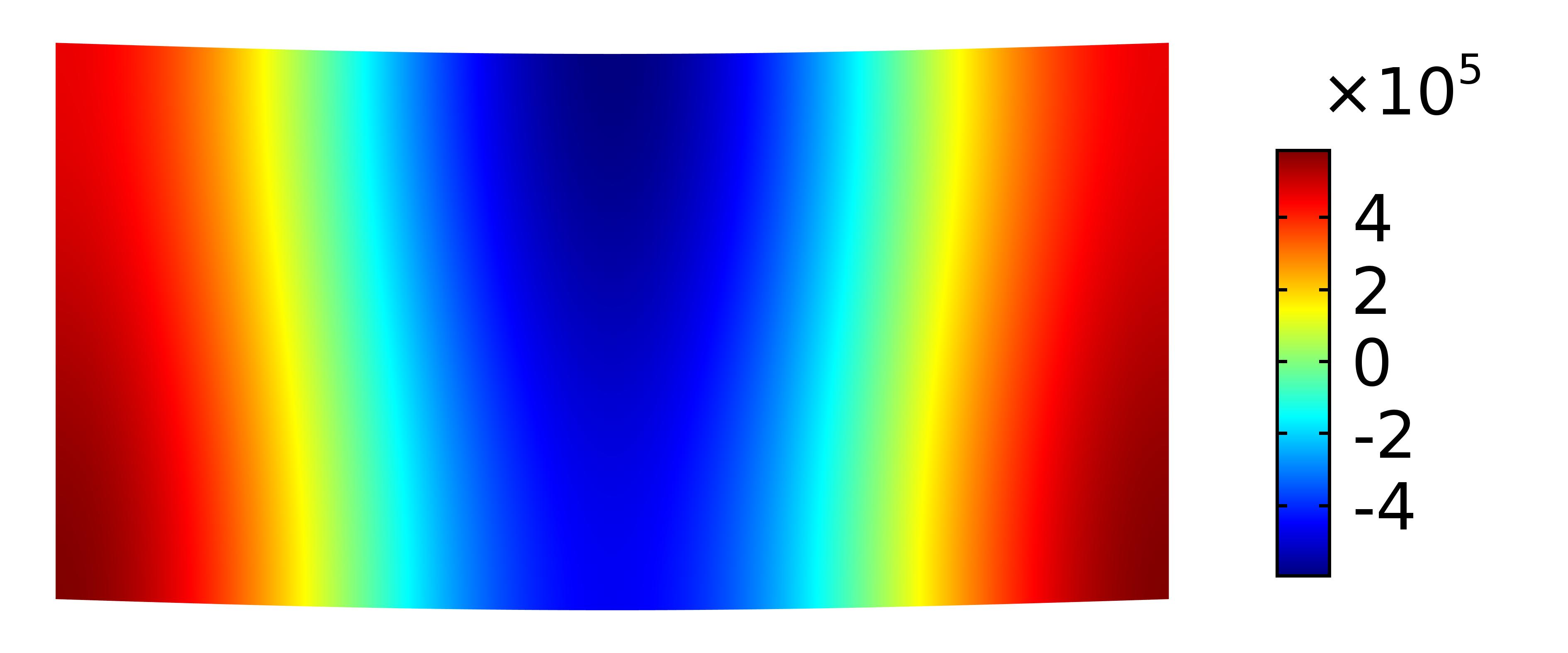}}\\  
			$\langle v_2 \rangle$ &
			\subfigure {\includegraphics[height=18 mm] {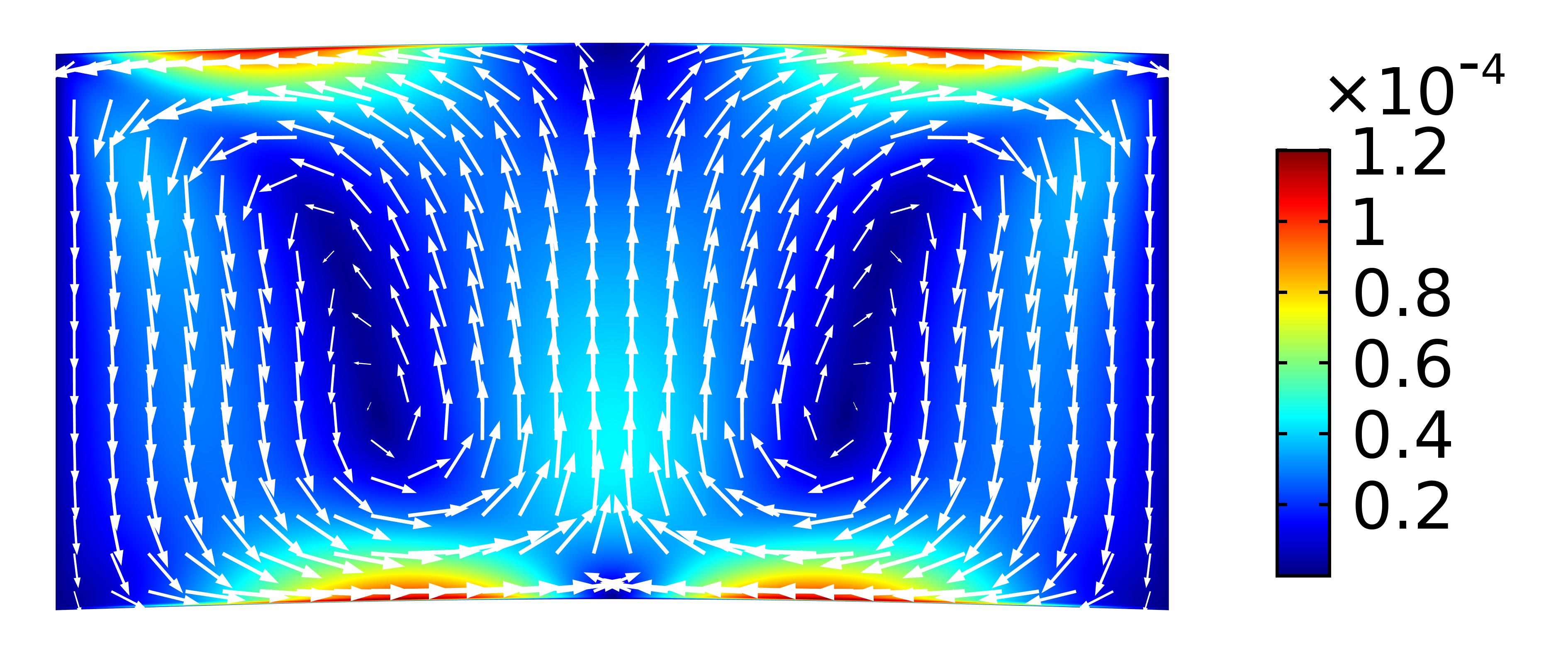}}& 
			\subfigure {\includegraphics[height=18 mm] {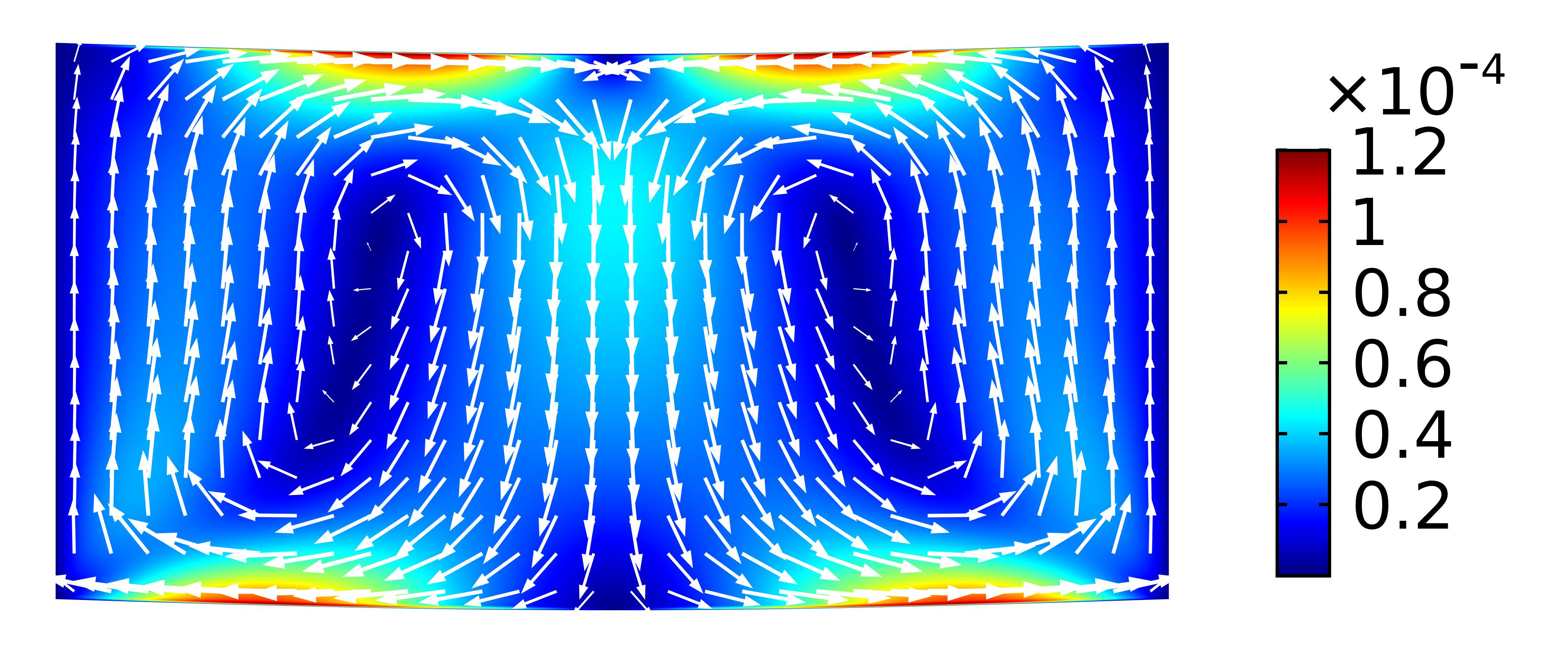}}\\
			\end{tabular} 
		\caption{First order pressure field, $p_1$, and acoustic streaming velocity field ,$\langle v_2 \rangle$, for two adjacent building blocks of a symmetrical sinusoidal microchannel. Final blocks have microchannel's width to height ratio of $n=2$ and geometrical wavelength of $\lambda_g=2w$.}
		\label{fig10} 
		\end{center} 
	\end{figure*}

\begin{figure*}[ht]
	\begin{center}
		\begin{tabular}{l|ll} 
			\; &	$p_1$ & $\langle v_2 \rangle$ \\
\hline
			$n=1$ &	
			\subfigure {\includegraphics[height=10 mm] {sym-wide100-p1-mod05-block1.jpg}}& 
			\subfigure {\includegraphics[height=10 mm] {sym-wide100-v2-mod05-block1.jpg}}\\  
			$n=2$ &	
			\subfigure {\includegraphics[height=10 mm] {sym-wide200-p1-mod1-block1.jpg}}&
			\subfigure {\includegraphics[height=10 mm] {sym-wide200-v2-mod1-block1.jpg}}\\  
			$n=3$ &	
			\subfigure {\includegraphics[height=10 mm] {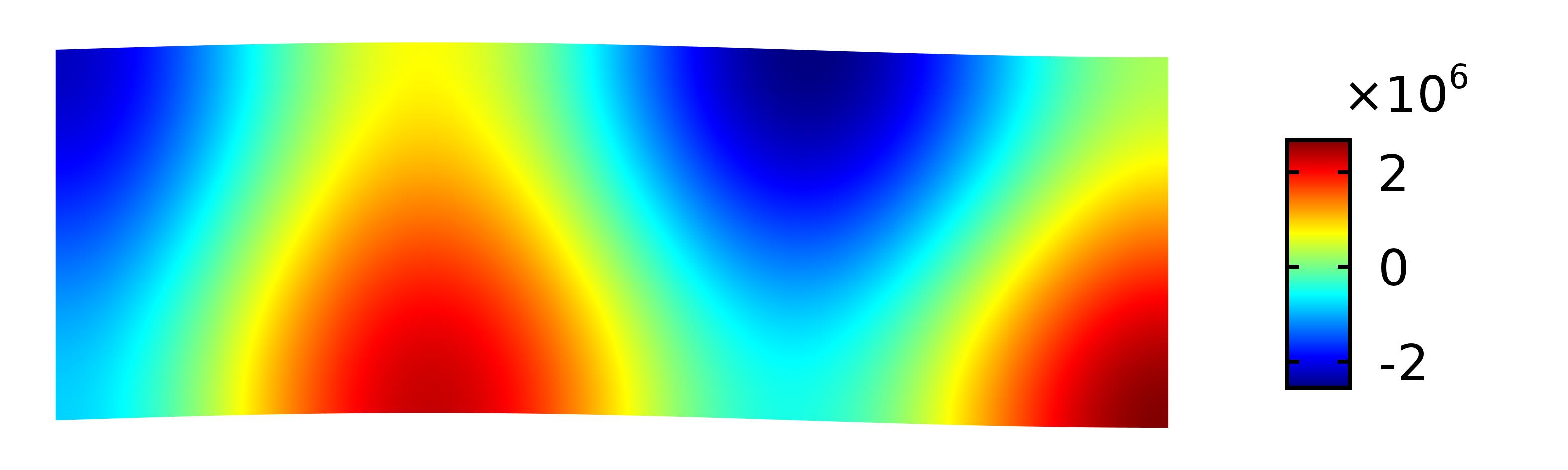}}& 
			\subfigure {\includegraphics[height=10 mm] {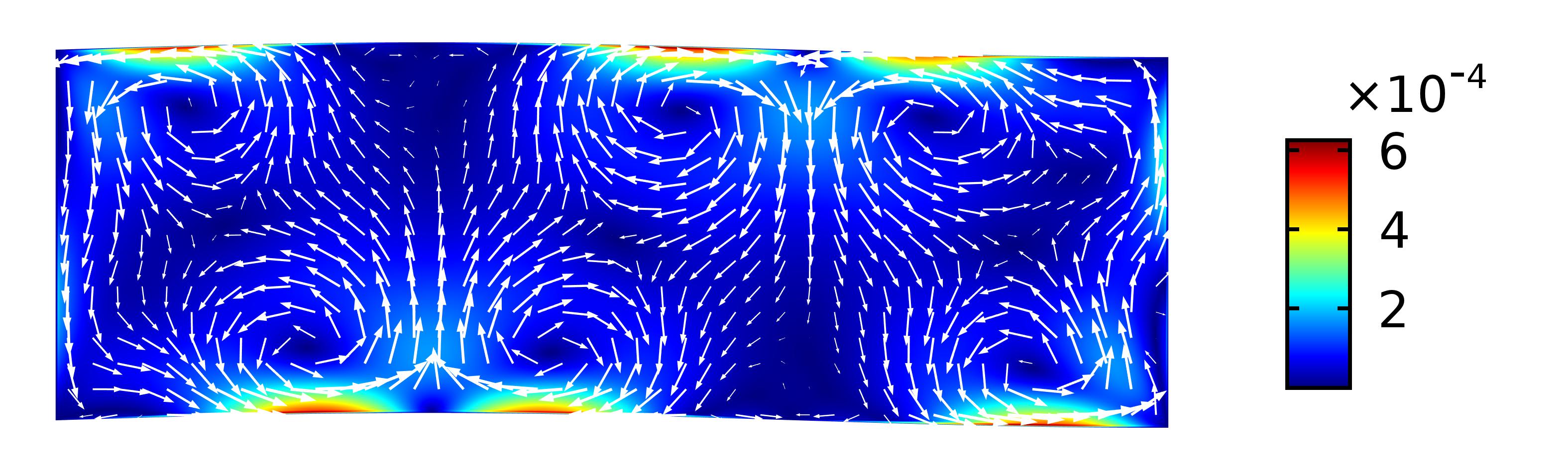}}\\  
			$n=4$ &
			\subfigure {\includegraphics[height=11 mm] {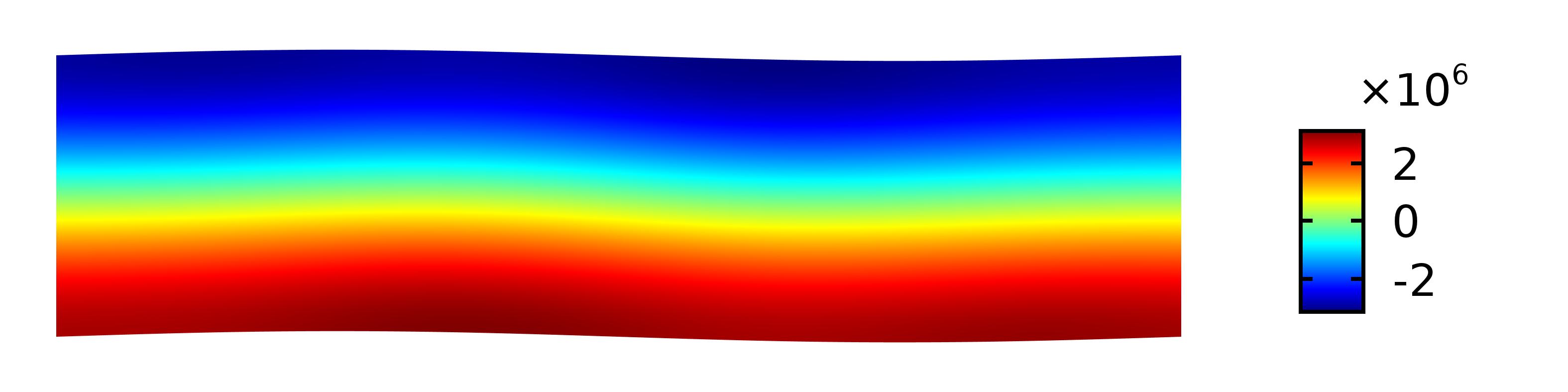}}&
			\subfigure {\includegraphics[height=11 mm] {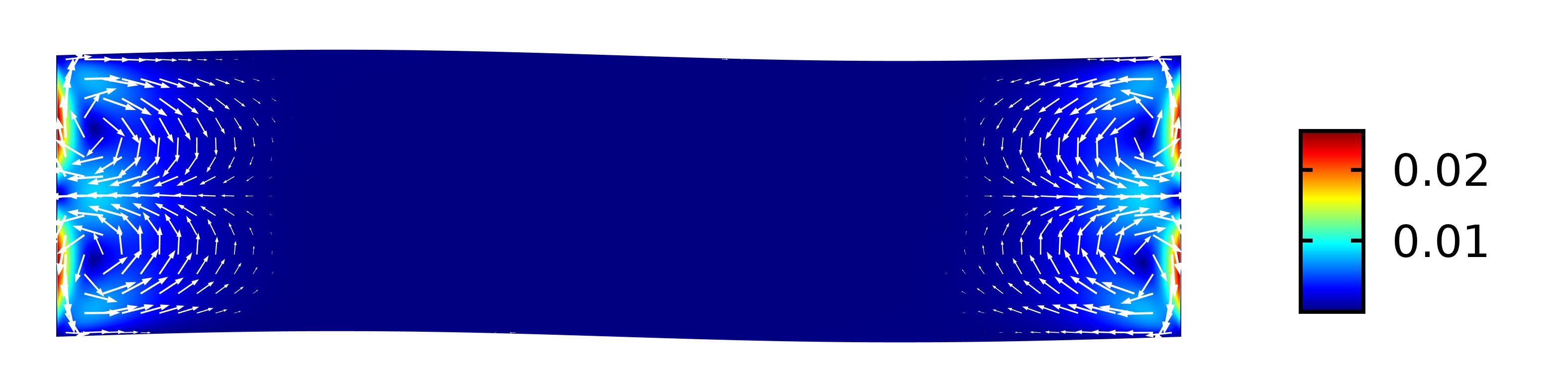}}\\
			$n=5$ &
			\subfigure {\includegraphics[height=10 mm] {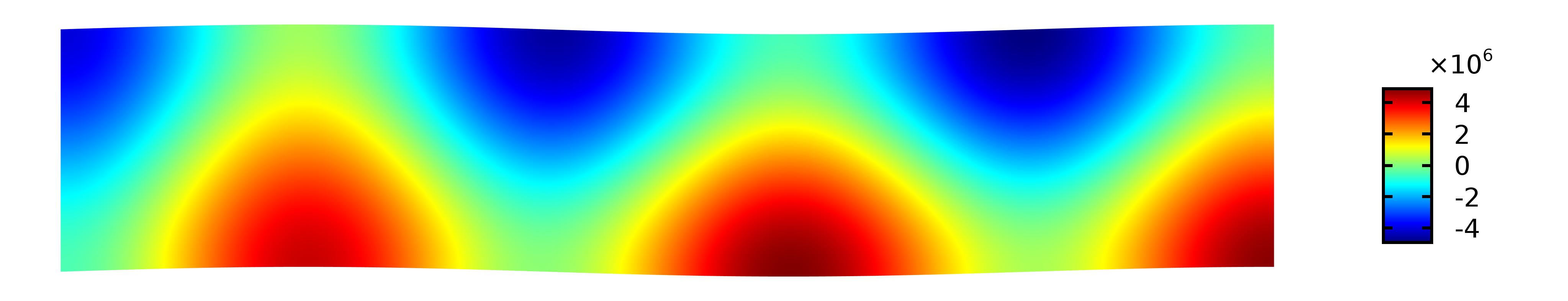}}&
			\subfigure {\includegraphics[height=10 mm] {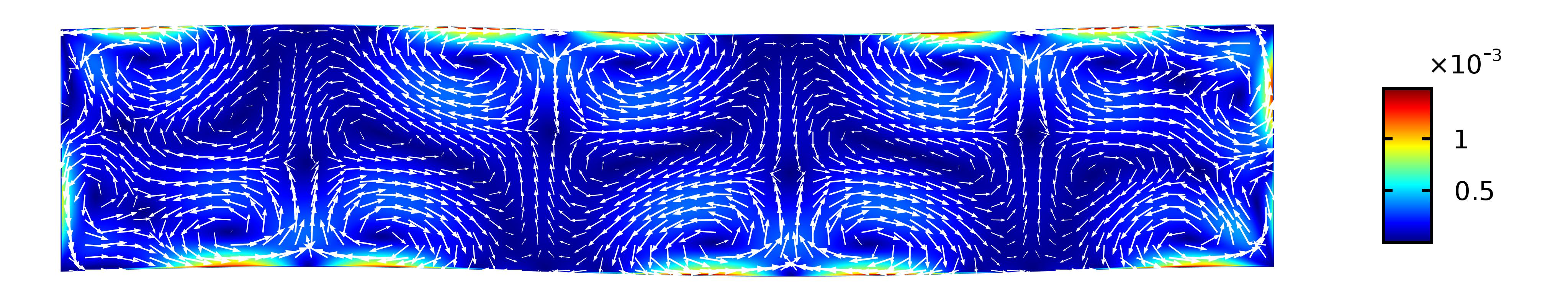}}\\
			$n=6$ &
			\subfigure {\includegraphics[height=10.5 mm] {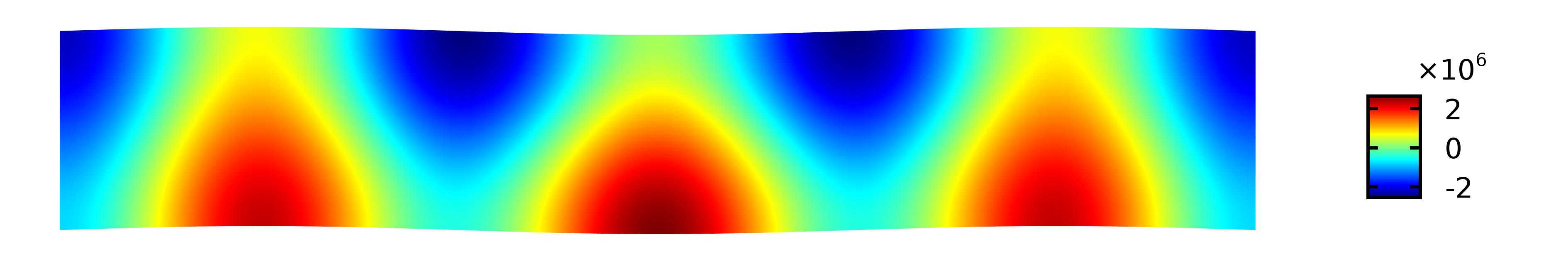}}&
			\subfigure {\includegraphics[height=10.5 mm] {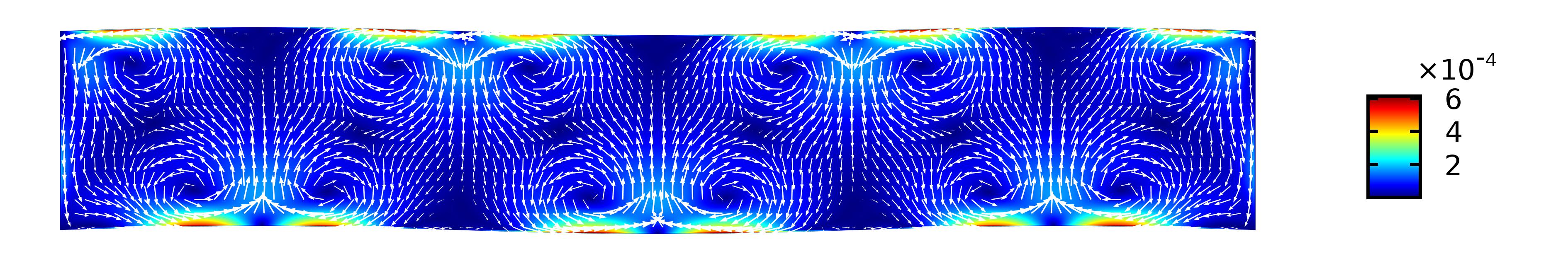}}\\
			$n=7$ &
			\subfigure {\includegraphics[height=12.5 mm] {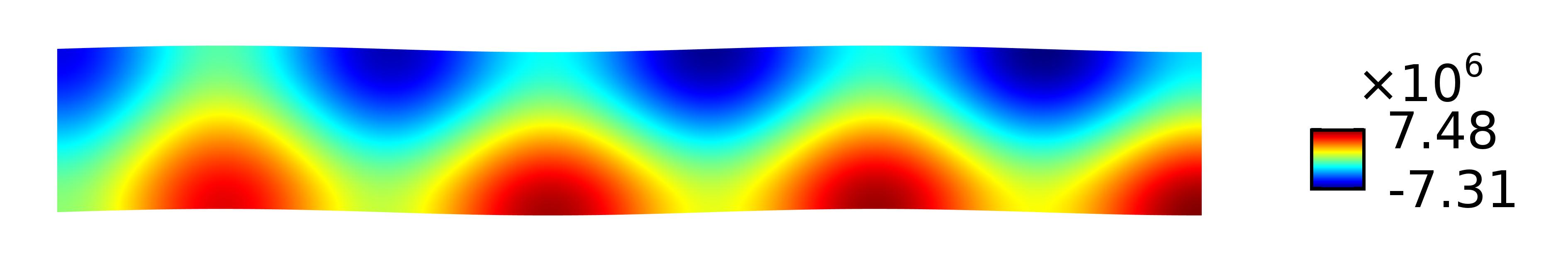}}&
			\subfigure {\includegraphics[height=12.5 mm] {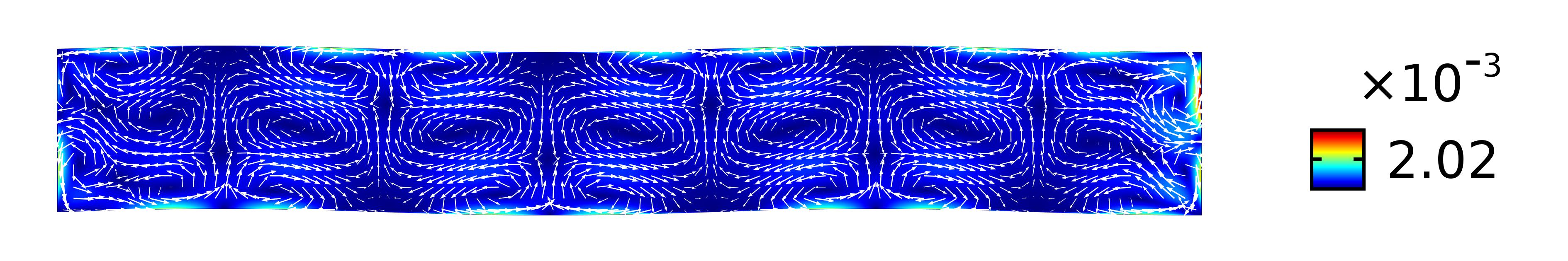}}\\
			$n=8$ &
			\subfigure {\includegraphics[height=10 mm] {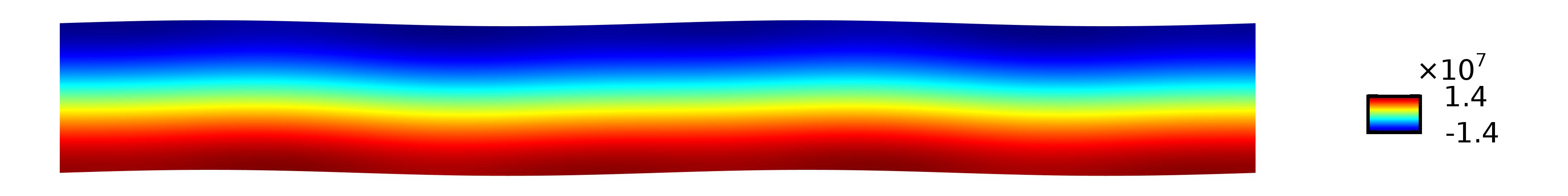}}&
			\subfigure {\includegraphics[height=10 mm] {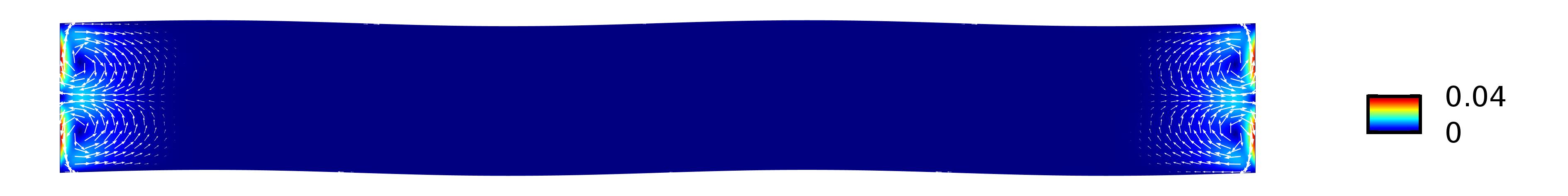}}
			\end{tabular} 
		\caption{Examples of $p_1$ and $\langle v_2 \rangle$ when adding building blocks of Fig. \ref{fig9} one by one from $n=1$ to $n=8$. Each block has geometrical wavelength of $\lambda_g=4w$}
		\label{fig11} 
		\end{center} 
	\end{figure*}

Figure \ref{fig10} shows the results when two blocks are joined. Patterns are satisfying in this case and two dominant vortices are functioned normally with the same order of the magnitude. As the third block is added, each dominant vortex separates into two vortices. Adding the fourth block destructs all the patterns and the resulting pattern becomes the same as the rectangular case. Surprisingly, adding the fifth block, make bulk vortices emerge again (See Fig. \ref{fig11}). 

\begin{figure*}[h] 
\begin{center}
\begin{tabular}{l|ll} 
\; & $p_1$ & $\langle v_2 \rangle$ \\
\hline
$n=1$ & 
\subfigure {\includegraphics[height=15 mm] {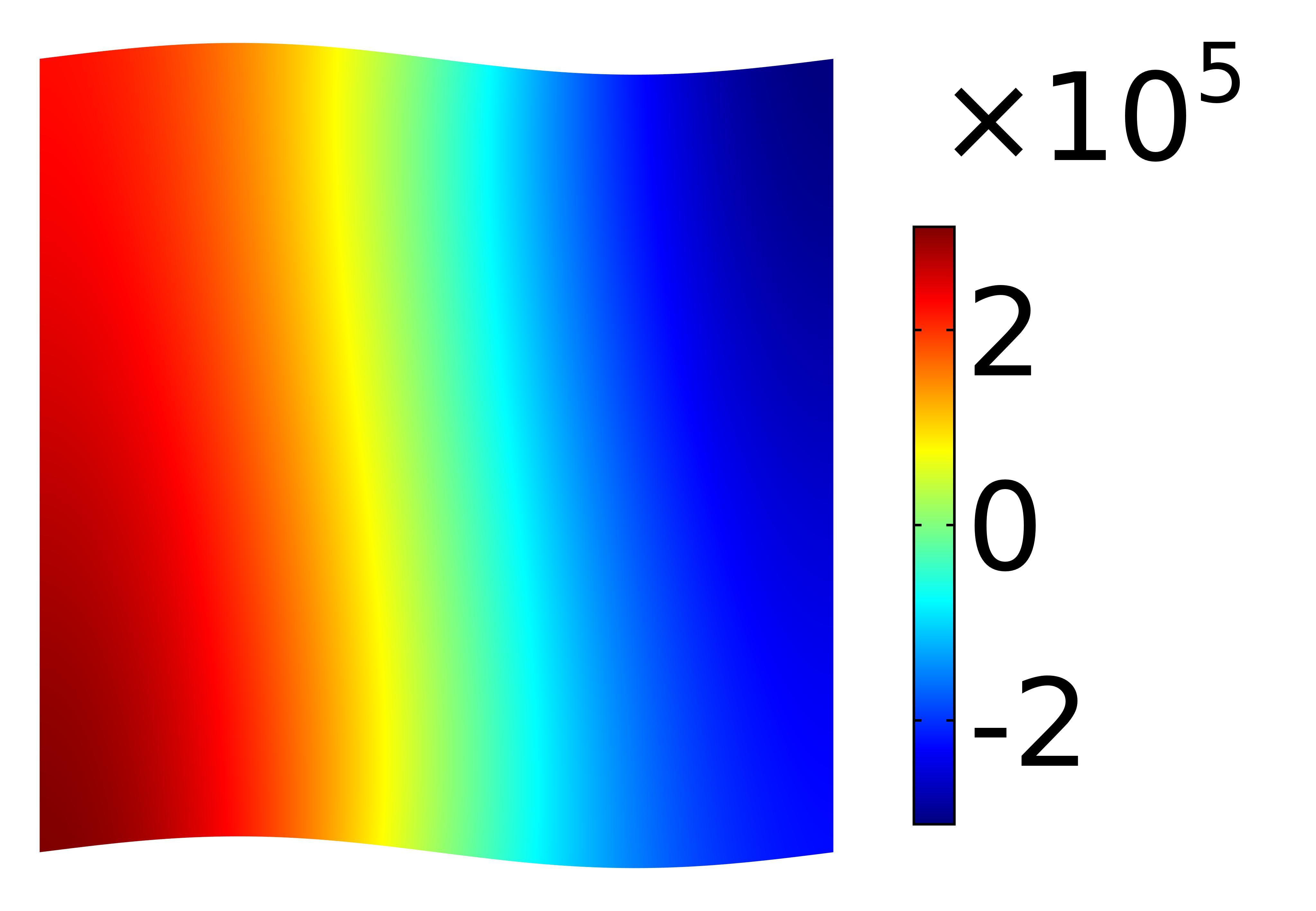}}& 
\subfigure {\includegraphics[height=15 mm] {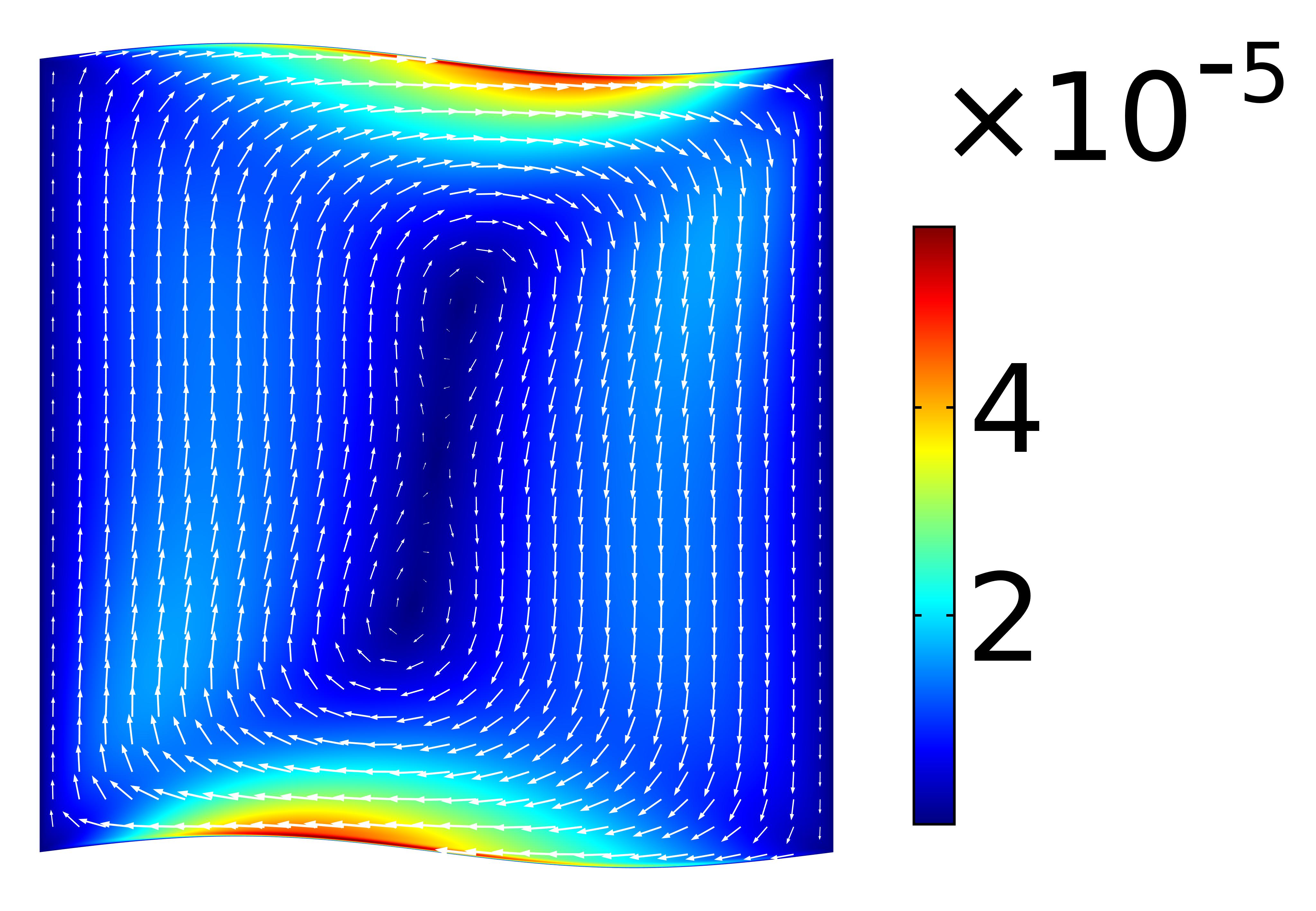}}\\ 
$n=2$ & 
\subfigure {\includegraphics[height=15 mm] {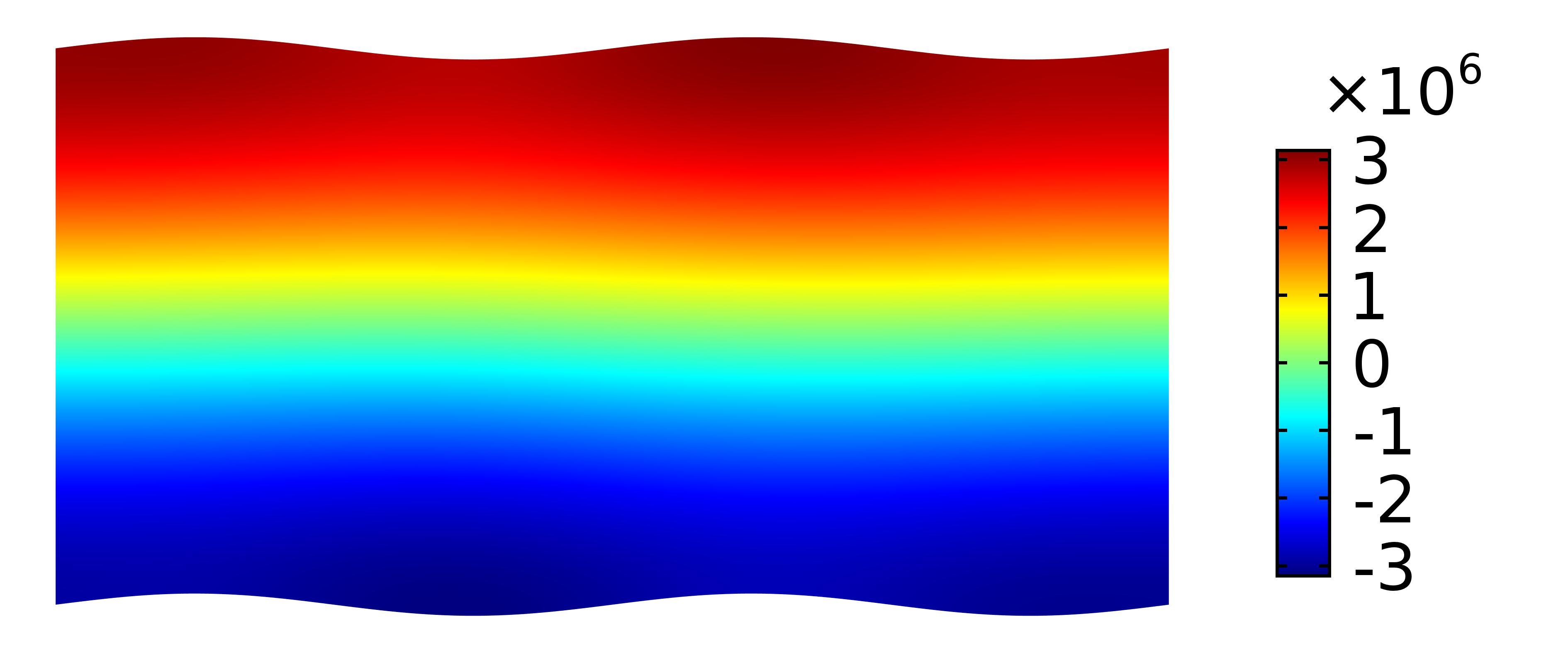}}&
\subfigure {\includegraphics[height=15 mm] {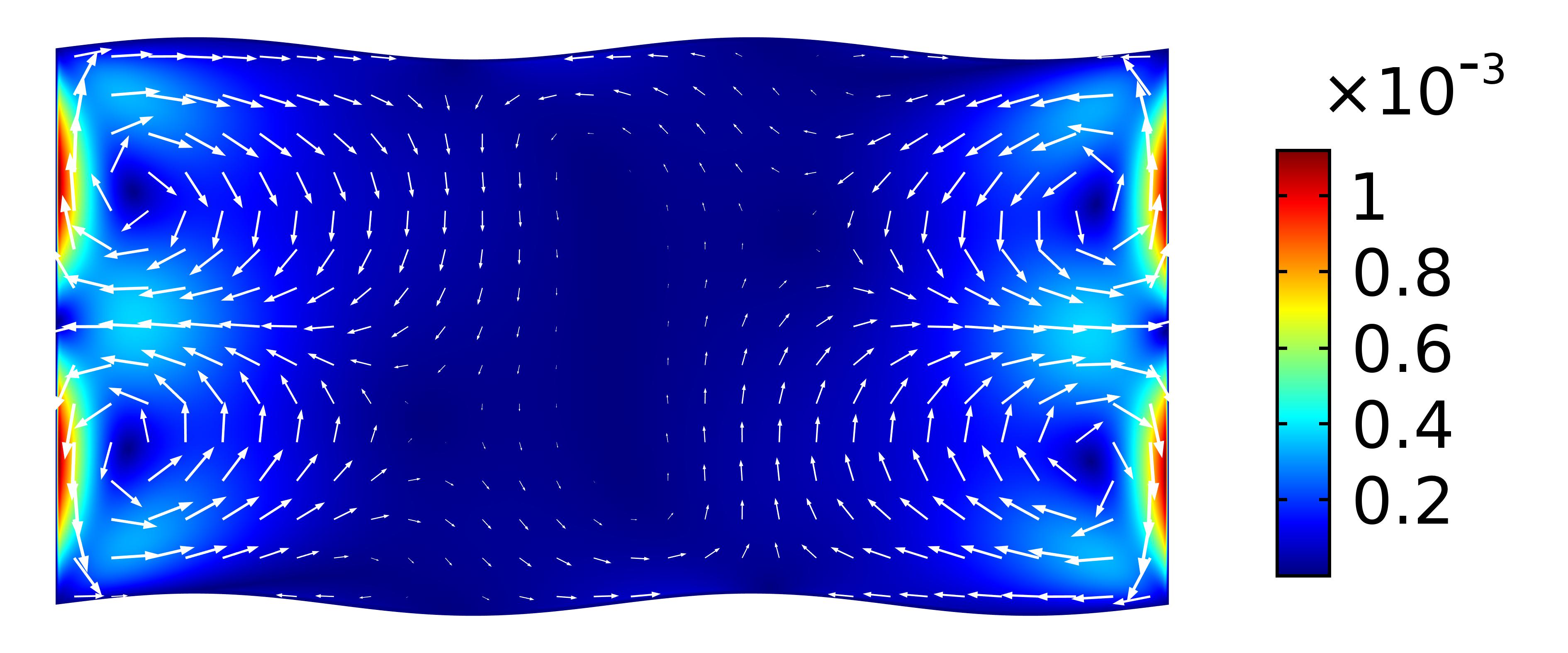}}\\ 
$n=3$ & 
\subfigure {\includegraphics[height=16 mm] {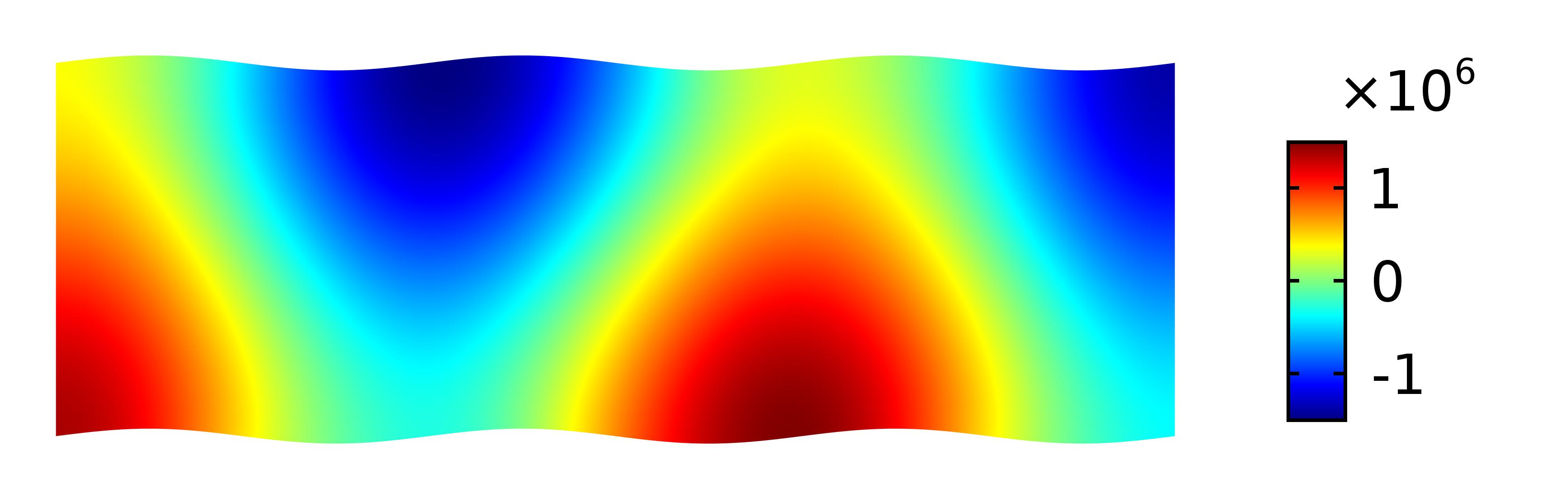}}& 
\subfigure {\includegraphics[height=16 mm] {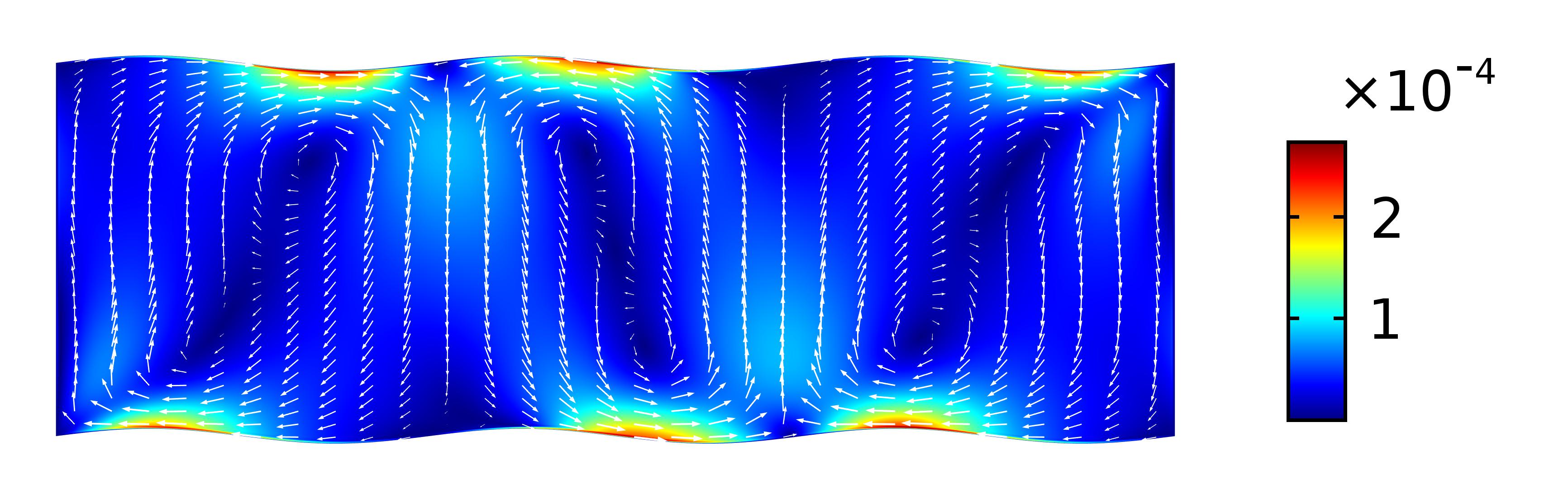}}\\ 
$n=4$ &
\subfigure {\includegraphics[height=17 mm] {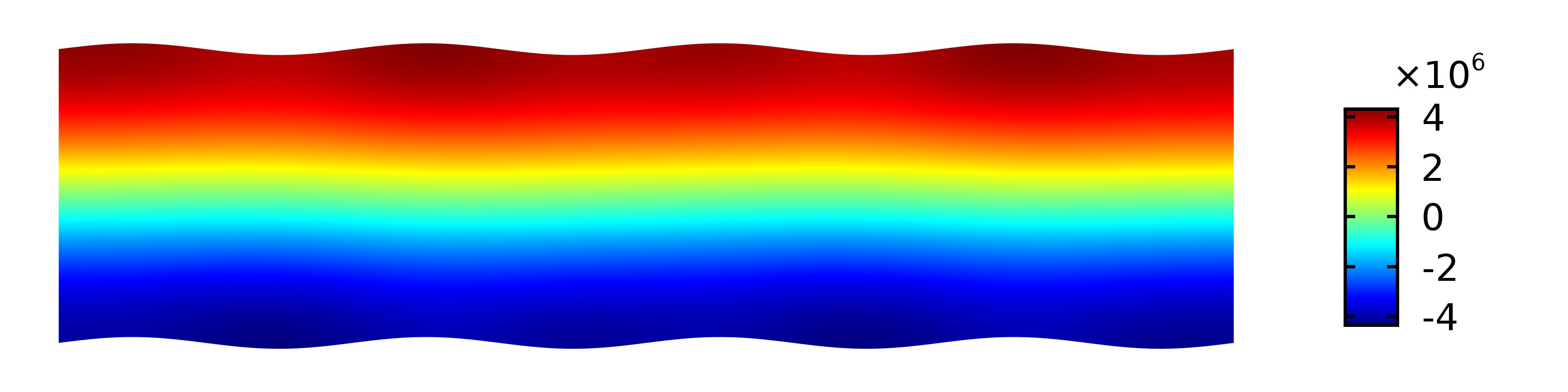}}&
\subfigure {\includegraphics[height=17 mm] {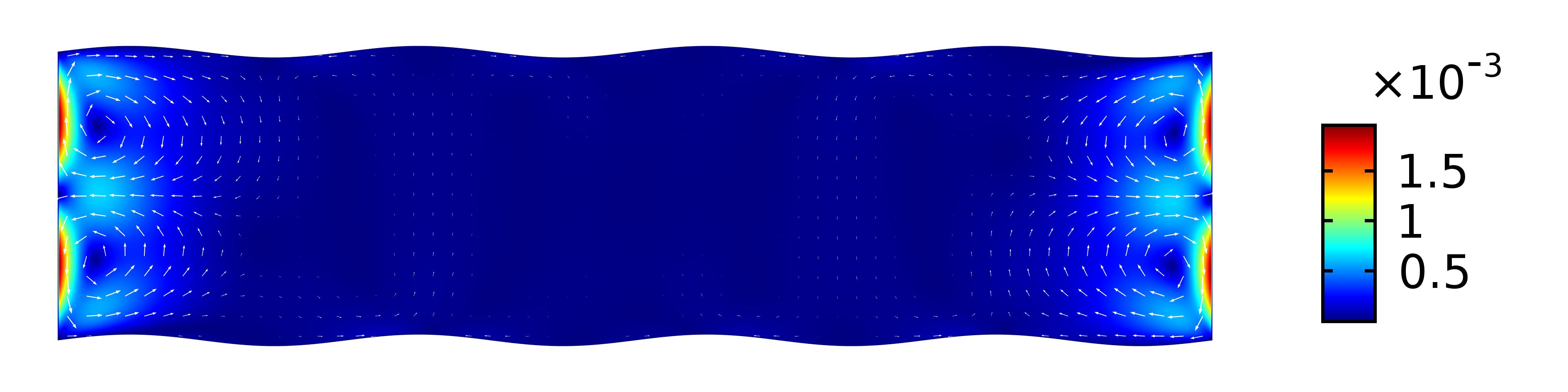}}\\
$n=5$ &
\subfigure {\includegraphics[height=14 mm] {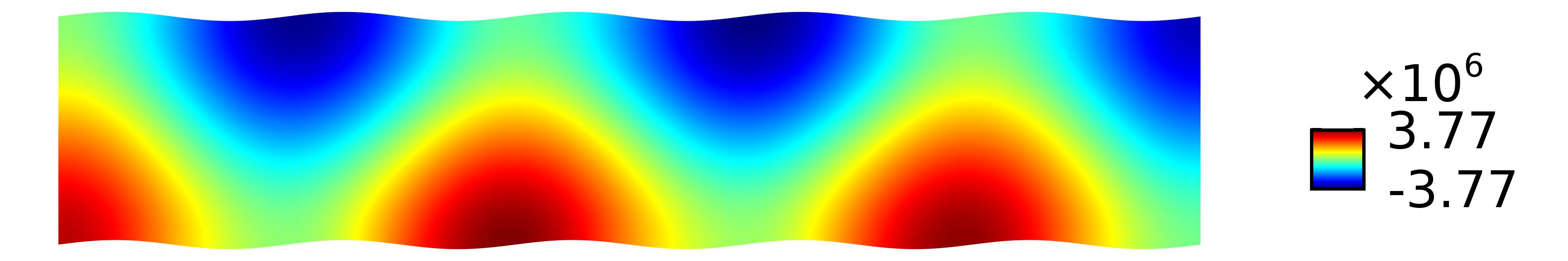}}&
\subfigure {\includegraphics[height=14 mm] {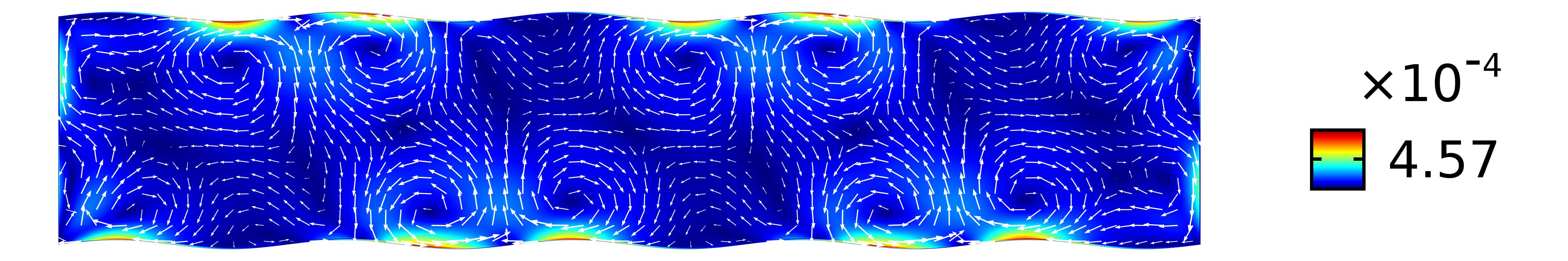}}\\
\end{tabular} 
\caption{Examples of $p_1$ and $\langle v_2 \rangle$ when adding building blocks one by one from $n=1$ to $n=5$. Each block has geometrical wavelength of $\lambda_g=w$.}
\label{fig12} 
\end{center} 
\end{figure*}

One other example for different behavior of joined building blocks in comapred with separated ones is shown in Fig. \ref{fig12} where each block have $n=1$ and $\lambda_g=w$.

\subsection{Trapping submicrometer particles inside symmetrical sinusoidal microchannels}\label{RES-D}

Novel acoustic streaming repetitive patterns have introduced numerically earlier in this paper. Patterns with repetitive single dominant vortices are capable of trapping sub-micron particles inside a sinusoidal microchannel in tweezing points. The most important finding is that such trapping now is possible through modification of the geometry of the boundaries instead of adding more oscillating boundaries. Several attempts have been made to create such patterns using multiple actuators with flat geometry \cite{Antfolk2014,bernassau2013interactive}.

\begin{figure*}[ht]
	\begin{center} 
		\begin{tabular}{l}
			\subfigure[\;$p_1$]{\includegraphics[width=170 mm]{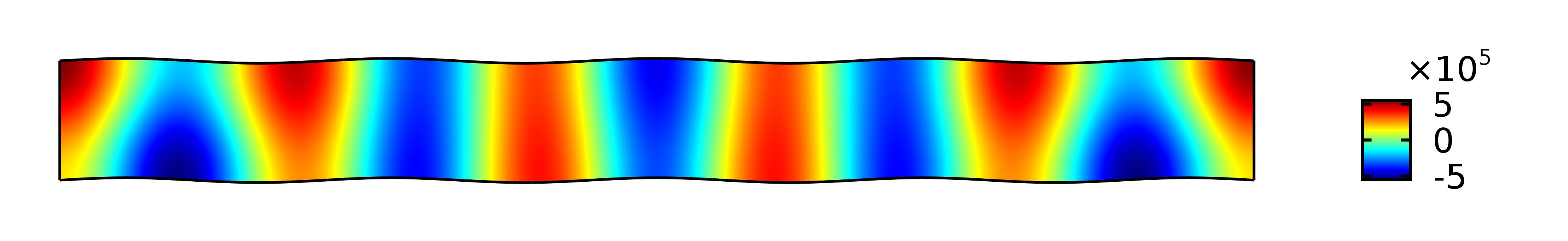}\label{fig13.a}}\\
			\subfigure[\;$\langle v_2\rangle$]{\includegraphics[width=170 mm]{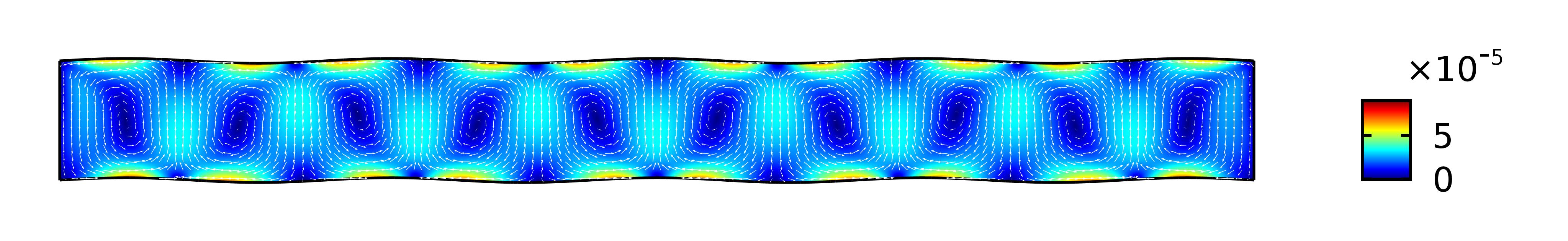}\label{fig13.b}}\\	
			\subfigure[\; t=0 s]{\includegraphics[width=143 mm]{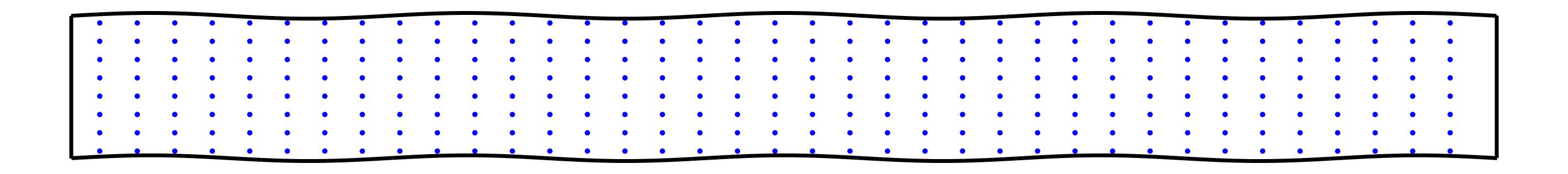}\label{fig13.c}}\\	
			\subfigure[\; t=10 s]{\includegraphics[width=143 mm]{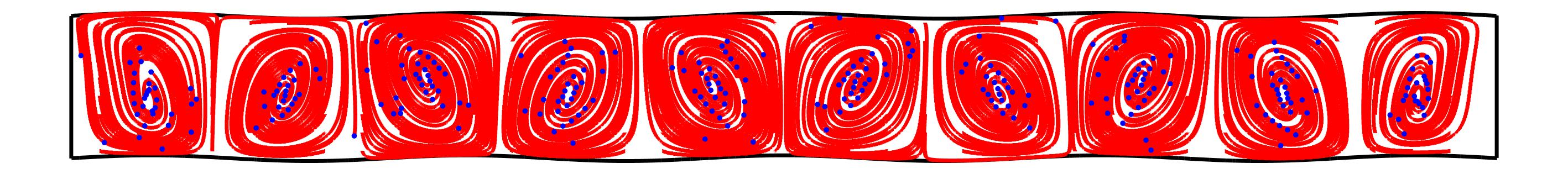}\label{fig13.d}}
		\end{tabular}
		\caption{(a) First order pressure field, (b) time-averaged second-order velocity field, (c) snapshot of movement of particles with the radius of $a=0.25 \mu$m at $t=0$ and (d) the snapshot at $t=10$s for a symmetrical sinusoidal microchannel with geometrical wavelength of  $\lambda_g=\frac{2}{9}$ and microchannel's width to height ratio of $n=10$. Trapping of tiny particles become possible by manipulation of boundary geometries in a defined manner.}
		\label {fig13} 
	\end{center}
\end{figure*}

Two-dimensional cross section of a sinusoidal microchannel with $1$mm width, and geometrical wavelength of $\frac{2}{9}$ is shown in Fig. \ref{fig13}. Top and bottom sinusoidal walls are actuated at the frequency of $f_v=\frac{c}{2h}$. 

The snapshots after 10 seconds for simulation of particles with radius of $a=0.25\;\mu$m inside the microchannel indicates that sub-micrometer particles tend to focus being affected by acoustic streaming fluid flows (See Fig.\ref{fig13.d}).

\section{Conclusion}\label{Conc}
In this paper, we aimed at numerically characterizing two-dimensional acoustic streaming patterns generated in the fluid inside the microchannels with acoustically oscillating sinusoidal boundaries.
Concerning the fact that the acoustic streaming patterns are extremely sensitive to the geometry, some geometrical parameters have been investigated that was the microchannel's width to height ratio, symmetry or asymmetry of sinusoidal walls and geometrical wavelength of sinusoidal boundaries. 
Results indicated that while the top and bottom sinusoidal boundaries had vertically been actuated at the resonance frequency of basic hypothetical rectangular microchannel, some repetitive acoustic streaming patterns generated. Such patterns could have been never produced in rectangular geometry with flat boundaries with only one directionally oscillation of boundaries. Relations between geometrical parameters and emerging acoustic streaming patterns led us to suggest some formulas in order to predict more cases. Results and formulations were not trivial at a glance. Consequently, an application has been proposed numerically to trap sub-micron particles inside a sinusoidal microchannel in some tweezing points. All conclusions stated in this paper can be leading points to optimize the performance of acoustofluidic devices.




\newpage
\bibliographystyle{apsrev4-1}
\bibliography{main-Arxiv}

\begin{thebibliography}{30}%
\makeatletter
\providecommand \@ifxundefined [1]{%
 \@ifx{#1\undefined}
}%
\providecommand \@ifnum [1]{%
 \ifnum #1\expandafter \@firstoftwo
 \else \expandafter \@secondoftwo
 \fi
}%
\providecommand \@ifx [1]{%
 \ifx #1\expandafter \@firstoftwo
 \else \expandafter \@secondoftwo
 \fi
}%
\providecommand \natexlab [1]{#1}%
\providecommand \enquote  [1]{``#1''}%
\providecommand \bibnamefont  [1]{#1}%
\providecommand \bibfnamefont [1]{#1}%
\providecommand \citenamefont [1]{#1}%
\providecommand \href@noop [0]{\@secondoftwo}%
\providecommand \href [0]{\begingroup \@sanitize@url \@href}%
\providecommand \@href[1]{\@@startlink{#1}\@@href}%
\providecommand \@@href[1]{\endgroup#1\@@endlink}%
\providecommand \@sanitize@url [0]{\catcode `\\12\catcode `\$12\catcode
  `\&12\catcode `\#12\catcode `\^12\catcode `\_12\catcode `\%12\relax}%
\providecommand \@@startlink[1]{}%
\providecommand \@@endlink[0]{}%
\providecommand \url  [0]{\begingroup\@sanitize@url \@url }%
\providecommand \@url [1]{\endgroup\@href {#1}{\urlprefix }}%
\providecommand \urlprefix  [0]{URL }%
\providecommand \Eprint [0]{\href }%
\providecommand \doibase [0]{http://dx.doi.org/}%
\providecommand \selectlanguage [0]{\@gobble}%
\providecommand \bibinfo  [0]{\@secondoftwo}%
\providecommand \bibfield  [0]{\@secondoftwo}%
\providecommand \translation [1]{[#1]}%
\providecommand \BibitemOpen [0]{}%
\providecommand \bibitemStop [0]{}%
\providecommand \bibitemNoStop [0]{.\EOS\space}%
\providecommand \EOS [0]{\spacefactor3000\relax}%
\providecommand \BibitemShut  [1]{\csname bibitem#1\endcsname}%
\let\auto@bib@innerbib\@empty
\bibitem [{\citenamefont {Wiklund}\ \emph {et~al.}(2012)\citenamefont
  {Wiklund}, \citenamefont {Green},\ and\ \citenamefont
  {Ohlin}}]{Wiklund2012c}%
  \BibitemOpen
  \bibfield  {author} {\bibinfo {author} {\bibfnamefont {M.}~\bibnamefont
  {Wiklund}}, \bibinfo {author} {\bibfnamefont {R.}~\bibnamefont {Green}}, \
  and\ \bibinfo {author} {\bibfnamefont {M.}~\bibnamefont {Ohlin}},\
  }\href@noop {} {\bibfield  {journal} {\bibinfo  {journal} {Lab on a Chip}\
  }\textbf {\bibinfo {volume} {12}},\ \bibinfo {pages} {2438} (\bibinfo {year}
  {2012})}\BibitemShut {NoStop}%
\bibitem [{\citenamefont {King}(1934)}]{King1934}%
  \BibitemOpen
  \bibfield  {author} {\bibinfo {author} {\bibfnamefont {L.~V.}\ \bibnamefont
  {King}},\ }\href@noop {} {\bibfield  {journal} {\bibinfo  {journal}
  {Proceedings of the Royal Society of London. Series A-Mathematical and
  Physical Sciences}\ }\textbf {\bibinfo {volume} {147}},\ \bibinfo {pages}
  {212} (\bibinfo {year} {1934})}\BibitemShut {NoStop}%
\bibitem [{\citenamefont {Yosioka}\ and\ \citenamefont
  {Kawasima}(1955)}]{Yosioka1955}%
  \BibitemOpen
  \bibfield  {author} {\bibinfo {author} {\bibfnamefont {K.}~\bibnamefont
  {Yosioka}}\ and\ \bibinfo {author} {\bibfnamefont {Y.}~\bibnamefont
  {Kawasima}},\ }\href@noop {} {\bibfield  {journal} {\bibinfo  {journal} {Acta
  Acustica united with Acustica}\ }\textbf {\bibinfo {volume} {5}},\ \bibinfo
  {pages} {167} (\bibinfo {year} {1955})}\BibitemShut {NoStop}%
\bibitem [{\citenamefont {Gor'Kov}(1962)}]{GorKov1962}%
  \BibitemOpen
  \bibfield  {author} {\bibinfo {author} {\bibfnamefont {L.~P.}\ \bibnamefont
  {Gor'Kov}},\ }in\ \href@noop {} {\emph {\bibinfo {booktitle} {Sov. Phys.
  Dokl.}}},\ Vol.~\bibinfo {volume} {6}\ (\bibinfo {year} {1962})\ pp.\
  \bibinfo {pages} {773--775}\BibitemShut {NoStop}%
\bibitem [{\citenamefont {Doinikov}(1997)}]{Doinikov1997}%
  \BibitemOpen
  \bibfield  {author} {\bibinfo {author} {\bibfnamefont {A.~A.}\ \bibnamefont
  {Doinikov}},\ }\href@noop {} {\bibfield  {journal} {\bibinfo  {journal} {The
  Journal of the Acoustical Society of America}\ }\textbf {\bibinfo {volume}
  {101}},\ \bibinfo {pages} {713} (\bibinfo {year} {1997})}\BibitemShut
  {NoStop}%
\bibitem [{\citenamefont {Rayleigh}(1884)}]{Rayleigh1884a}%
  \BibitemOpen
  \bibfield  {author} {\bibinfo {author} {\bibfnamefont {L.}~\bibnamefont
  {Rayleigh}},\ }\href@noop {} {\bibfield  {journal} {\bibinfo  {journal}
  {Philosophical Transactions of the Royal Society of London}\ }\textbf
  {\bibinfo {volume} {175}},\ \bibinfo {pages} {1} (\bibinfo {year}
  {1884})}\BibitemShut {NoStop}%
\bibitem [{\citenamefont {Schlichting}\ \emph {et~al.}(1955)\citenamefont
  {Schlichting}, \citenamefont {Gersten}, \citenamefont {Krause},\ and\
  \citenamefont {Oertel}}]{Schlichting1955a}%
  \BibitemOpen
  \bibfield  {author} {\bibinfo {author} {\bibfnamefont {H.}~\bibnamefont
  {Schlichting}}, \bibinfo {author} {\bibfnamefont {K.}~\bibnamefont
  {Gersten}}, \bibinfo {author} {\bibfnamefont {E.}~\bibnamefont {Krause}}, \
  and\ \bibinfo {author} {\bibfnamefont {H.}~\bibnamefont {Oertel}},\
  }\href@noop {} {\emph {\bibinfo {title} {{Boundary-layer theory}}}},\
  Vol.~\bibinfo {volume} {7}\ (\bibinfo  {publisher} {Springer},\ \bibinfo
  {year} {1955})\BibitemShut {NoStop}%
\bibitem [{\citenamefont {Nyborg}(1953)}]{Nyborg1953}%
  \BibitemOpen
  \bibfield  {author} {\bibinfo {author} {\bibfnamefont {W.~L.}\ \bibnamefont
  {Nyborg}},\ }\href@noop {} {\bibfield  {journal} {\bibinfo  {journal} {The
  Journal of the Acoustical Society of America}\ }\textbf {\bibinfo {volume}
  {25}},\ \bibinfo {pages} {68} (\bibinfo {year} {1953})}\BibitemShut {NoStop}%
\bibitem [{\citenamefont {Nyborg}(1958)}]{Nyborg1958}%
  \BibitemOpen
  \bibfield  {author} {\bibinfo {author} {\bibfnamefont {W.~L.}\ \bibnamefont
  {Nyborg}},\ }\href@noop {} {\bibfield  {journal} {\bibinfo  {journal} {The
  Journal of the Acoustical Society of America}\ }\textbf {\bibinfo {volume}
  {30}},\ \bibinfo {pages} {329} (\bibinfo {year} {1958})}\BibitemShut
  {NoStop}%
\bibitem [{\citenamefont {Spengler}\ \emph {et~al.}(2003)\citenamefont
  {Spengler}, \citenamefont {Coakley},\ and\ \citenamefont
  {Christensen}}]{Spengler2003}%
  \BibitemOpen
  \bibfield  {author} {\bibinfo {author} {\bibfnamefont {J.~F.}\ \bibnamefont
  {Spengler}}, \bibinfo {author} {\bibfnamefont {W.~T.}\ \bibnamefont
  {Coakley}}, \ and\ \bibinfo {author} {\bibfnamefont {K.~T.}\ \bibnamefont
  {Christensen}},\ }\href@noop {} {\bibfield  {journal} {\bibinfo  {journal}
  {AIChE journal}\ }\textbf {\bibinfo {volume} {49}},\ \bibinfo {pages} {2773}
  (\bibinfo {year} {2003})}\BibitemShut {NoStop}%
\bibitem [{\citenamefont {Barnkob}\ \emph {et~al.}(2012)\citenamefont
  {Barnkob}, \citenamefont {Augustsson}, \citenamefont {Laurell},\ and\
  \citenamefont {Bruus}}]{Barnkob2012}%
  \BibitemOpen
  \bibfield  {author} {\bibinfo {author} {\bibfnamefont {R.}~\bibnamefont
  {Barnkob}}, \bibinfo {author} {\bibfnamefont {P.}~\bibnamefont {Augustsson}},
  \bibinfo {author} {\bibfnamefont {T.}~\bibnamefont {Laurell}}, \ and\
  \bibinfo {author} {\bibfnamefont {H.}~\bibnamefont {Bruus}},\ }\href@noop {}
  {\bibfield  {journal} {\bibinfo  {journal} {Physical Review E}\ }\textbf
  {\bibinfo {volume} {86}},\ \bibinfo {pages} {56307} (\bibinfo {year}
  {2012})}\BibitemShut {NoStop}%
\bibitem [{\citenamefont {Westervelt}(1953)}]{Westervelt1953a}%
  \BibitemOpen
  \bibfield  {author} {\bibinfo {author} {\bibfnamefont {P.~J.}\ \bibnamefont
  {Westervelt}},\ }\href@noop {} {\bibfield  {journal} {\bibinfo  {journal}
  {The Journal of the Acoustical Society of America}\ }\textbf {\bibinfo
  {volume} {25}},\ \bibinfo {pages} {60} (\bibinfo {year} {1953})}\BibitemShut
  {NoStop}%
\bibitem [{\citenamefont {Hamilton}\ \emph {et~al.}(2003)\citenamefont
  {Hamilton}, \citenamefont {Ilinskii},\ and\ \citenamefont
  {Zabolotskaya}}]{Hamilton2003}%
  \BibitemOpen
  \bibfield  {author} {\bibinfo {author} {\bibfnamefont {M.~F.}\ \bibnamefont
  {Hamilton}}, \bibinfo {author} {\bibfnamefont {Y.~A.}\ \bibnamefont
  {Ilinskii}}, \ and\ \bibinfo {author} {\bibfnamefont {E.~A.}\ \bibnamefont
  {Zabolotskaya}},\ }\href@noop {} {\bibfield  {journal} {\bibinfo  {journal}
  {The Journal of the Acoustical Society of America}\ }\textbf {\bibinfo
  {volume} {113}},\ \bibinfo {pages} {153} (\bibinfo {year}
  {2003})}\BibitemShut {NoStop}%
\bibitem [{\citenamefont {Rednikov}\ and\ \citenamefont
  {Sadhal}(2011)}]{Rednikov2011a}%
  \BibitemOpen
  \bibfield  {author} {\bibinfo {author} {\bibfnamefont {A.~Y.}\ \bibnamefont
  {Rednikov}}\ and\ \bibinfo {author} {\bibfnamefont {S.~S.}\ \bibnamefont
  {Sadhal}},\ }\href@noop {} {\bibfield  {journal} {\bibinfo  {journal}
  {Journal of Fluid Mechanics}\ }\textbf {\bibinfo {volume} {667}},\ \bibinfo
  {pages} {426} (\bibinfo {year} {2011})}\BibitemShut {NoStop}%
\bibitem [{\citenamefont {Muller}\ \emph {et~al.}(2013)\citenamefont {Muller},
  \citenamefont {Rossi}, \citenamefont {Mar{\'{i}}n}, \citenamefont {Barnkob},
  \citenamefont {Augustsson}, \citenamefont {Laurell}, \citenamefont
  {Kaehler},\ and\ \citenamefont {Bruus}}]{Muller2013}%
  \BibitemOpen
  \bibfield  {author} {\bibinfo {author} {\bibfnamefont {P.~B.}\ \bibnamefont
  {Muller}}, \bibinfo {author} {\bibfnamefont {M.}~\bibnamefont {Rossi}},
  \bibinfo {author} {\bibfnamefont {{\'{A}}.~G.}\ \bibnamefont {Mar{\'{i}}n}},
  \bibinfo {author} {\bibfnamefont {R.}~\bibnamefont {Barnkob}}, \bibinfo
  {author} {\bibfnamefont {P.}~\bibnamefont {Augustsson}}, \bibinfo {author}
  {\bibfnamefont {T.}~\bibnamefont {Laurell}}, \bibinfo {author} {\bibfnamefont
  {C.~J.}\ \bibnamefont {Kaehler}}, \ and\ \bibinfo {author} {\bibfnamefont
  {H.}~\bibnamefont {Bruus}},\ }\href@noop {} {\bibfield  {journal} {\bibinfo
  {journal} {Physical Review E}\ }\textbf {\bibinfo {volume} {88}},\ \bibinfo
  {pages} {23006} (\bibinfo {year} {2013})}\BibitemShut {NoStop}%
\bibitem [{\citenamefont {Evander}\ and\ \citenamefont
  {Nilsson}(2012)}]{Evander2012a}%
  \BibitemOpen
  \bibfield  {author} {\bibinfo {author} {\bibfnamefont {M.}~\bibnamefont
  {Evander}}\ and\ \bibinfo {author} {\bibfnamefont {J.}~\bibnamefont
  {Nilsson}},\ }\href@noop {} {\bibfield  {journal} {\bibinfo  {journal} {Lab
  on a Chip}\ }\textbf {\bibinfo {volume} {12}},\ \bibinfo {pages} {4667}
  (\bibinfo {year} {2012})}\BibitemShut {NoStop}%
\bibitem [{\citenamefont {Nama}\ \emph {et~al.}(2014)\citenamefont {Nama},
  \citenamefont {Huang}, \citenamefont {Huang},\ and\ \citenamefont
  {Costanzo}}]{Nama2014}%
  \BibitemOpen
  \bibfield  {author} {\bibinfo {author} {\bibfnamefont {N.}~\bibnamefont
  {Nama}}, \bibinfo {author} {\bibfnamefont {P.-H.}\ \bibnamefont {Huang}},
  \bibinfo {author} {\bibfnamefont {T.~J.}\ \bibnamefont {Huang}}, \ and\
  \bibinfo {author} {\bibfnamefont {F.}~\bibnamefont {Costanzo}},\ }\href@noop
  {} {\bibfield  {journal} {\bibinfo  {journal} {Lab on a Chip}\ }\textbf
  {\bibinfo {volume} {14}},\ \bibinfo {pages} {2824} (\bibinfo {year}
  {2014})}\BibitemShut {NoStop}%
\bibitem [{\citenamefont {Feng}\ \emph {et~al.}(2018)\citenamefont {Feng},
  \citenamefont {Song}, \citenamefont {Zhang}, \citenamefont {Jiang},\ and\
  \citenamefont {Arai}}]{Feng2018}%
  \BibitemOpen
  \bibfield  {author} {\bibinfo {author} {\bibfnamefont {L.}~\bibnamefont
  {Feng}}, \bibinfo {author} {\bibfnamefont {B.}~\bibnamefont {Song}}, \bibinfo
  {author} {\bibfnamefont {D.}~\bibnamefont {Zhang}}, \bibinfo {author}
  {\bibfnamefont {Y.}~\bibnamefont {Jiang}}, \ and\ \bibinfo {author}
  {\bibfnamefont {F.}~\bibnamefont {Arai}},\ }\href@noop {} {\bibfield
  {journal} {\bibinfo  {journal} {Micromachines}\ }\textbf {\bibinfo {volume}
  {9}},\ \bibinfo {pages} {596} (\bibinfo {year} {2018})}\BibitemShut {NoStop}%
\bibitem [{\citenamefont {Huang}\ \emph {et~al.}(2013)\citenamefont {Huang},
  \citenamefont {Xie}, \citenamefont {Ahmed}, \citenamefont {Rufo},
  \citenamefont {Nama}, \citenamefont {Chen}, \citenamefont {Chan},\ and\
  \citenamefont {Huang}}]{Huang2013a}%
  \BibitemOpen
  \bibfield  {author} {\bibinfo {author} {\bibfnamefont {P.-H.}\ \bibnamefont
  {Huang}}, \bibinfo {author} {\bibfnamefont {Y.}~\bibnamefont {Xie}}, \bibinfo
  {author} {\bibfnamefont {D.}~\bibnamefont {Ahmed}}, \bibinfo {author}
  {\bibfnamefont {J.}~\bibnamefont {Rufo}}, \bibinfo {author} {\bibfnamefont
  {N.}~\bibnamefont {Nama}}, \bibinfo {author} {\bibfnamefont {Y.}~\bibnamefont
  {Chen}}, \bibinfo {author} {\bibfnamefont {C.~Y.}\ \bibnamefont {Chan}}, \
  and\ \bibinfo {author} {\bibfnamefont {T.~J.}\ \bibnamefont {Huang}},\
  }\href@noop {} {\bibfield  {journal} {\bibinfo  {journal} {Lab on a Chip}\
  }\textbf {\bibinfo {volume} {13}},\ \bibinfo {pages} {3847} (\bibinfo {year}
  {2013})}\BibitemShut {NoStop}%
\bibitem [{\citenamefont {Huang}\ \emph {et~al.}(2014)\citenamefont {Huang},
  \citenamefont {Nama}, \citenamefont {Mao}, \citenamefont {Li}, \citenamefont
  {Rufo}, \citenamefont {Chen}, \citenamefont {Xie}, \citenamefont {Wei},
  \citenamefont {Wang},\ and\ \citenamefont {Huang}}]{Huang2014}%
  \BibitemOpen
  \bibfield  {author} {\bibinfo {author} {\bibfnamefont {P.-H.}\ \bibnamefont
  {Huang}}, \bibinfo {author} {\bibfnamefont {N.}~\bibnamefont {Nama}},
  \bibinfo {author} {\bibfnamefont {Z.}~\bibnamefont {Mao}}, \bibinfo {author}
  {\bibfnamefont {P.}~\bibnamefont {Li}}, \bibinfo {author} {\bibfnamefont
  {J.}~\bibnamefont {Rufo}}, \bibinfo {author} {\bibfnamefont {Y.}~\bibnamefont
  {Chen}}, \bibinfo {author} {\bibfnamefont {Y.}~\bibnamefont {Xie}}, \bibinfo
  {author} {\bibfnamefont {C.-H.}\ \bibnamefont {Wei}}, \bibinfo {author}
  {\bibfnamefont {L.}~\bibnamefont {Wang}}, \ and\ \bibinfo {author}
  {\bibfnamefont {T.~J.}\ \bibnamefont {Huang}},\ }\href@noop {} {\bibfield
  {journal} {\bibinfo  {journal} {Lab on a Chip}\ }\textbf {\bibinfo {volume}
  {14}},\ \bibinfo {pages} {4319} (\bibinfo {year} {2014})}\BibitemShut
  {NoStop}%
\bibitem [{\citenamefont {Ahmed}\ \emph {et~al.}(2009)\citenamefont {Ahmed},
  \citenamefont {Mao}, \citenamefont {Juluri},\ and\ \citenamefont
  {Huang}}]{Ahmed2009}%
  \BibitemOpen
  \bibfield  {author} {\bibinfo {author} {\bibfnamefont {D.}~\bibnamefont
  {Ahmed}}, \bibinfo {author} {\bibfnamefont {X.}~\bibnamefont {Mao}}, \bibinfo
  {author} {\bibfnamefont {B.~K.}\ \bibnamefont {Juluri}}, \ and\ \bibinfo
  {author} {\bibfnamefont {T.~J.}\ \bibnamefont {Huang}},\ }\href@noop {}
  {\bibfield  {journal} {\bibinfo  {journal} {Microfluidics and Nanofluidics}\
  }\textbf {\bibinfo {volume} {7}},\ \bibinfo {pages} {727} (\bibinfo {year}
  {2009})}\BibitemShut {NoStop}%
\bibitem [{\citenamefont {Yazdi}\ and\ \citenamefont
  {Ardekani}(2012)}]{Yazdi2012}%
  \BibitemOpen
  \bibfield  {author} {\bibinfo {author} {\bibfnamefont {S.}~\bibnamefont
  {Yazdi}}\ and\ \bibinfo {author} {\bibfnamefont {A.~M.}\ \bibnamefont
  {Ardekani}},\ }\href@noop {} {\bibfield  {journal} {\bibinfo  {journal}
  {Biomicrofluidics}\ }\textbf {\bibinfo {volume} {6}},\ \bibinfo {pages}
  {44114} (\bibinfo {year} {2012})}\BibitemShut {NoStop}%
\bibitem [{\citenamefont {Antfolk}\ \emph {et~al.}(2014)\citenamefont
  {Antfolk}, \citenamefont {Muller}, \citenamefont {Augustsson}, \citenamefont
  {Bruus},\ and\ \citenamefont {Laurell}}]{Antfolk2014}%
  \BibitemOpen
  \bibfield  {author} {\bibinfo {author} {\bibfnamefont {M.}~\bibnamefont
  {Antfolk}}, \bibinfo {author} {\bibfnamefont {P.~B.}\ \bibnamefont {Muller}},
  \bibinfo {author} {\bibfnamefont {P.}~\bibnamefont {Augustsson}}, \bibinfo
  {author} {\bibfnamefont {H.}~\bibnamefont {Bruus}}, \ and\ \bibinfo {author}
  {\bibfnamefont {T.}~\bibnamefont {Laurell}},\ }\href@noop {} {\bibfield
  {journal} {\bibinfo  {journal} {Lab on a Chip}\ }\textbf {\bibinfo {volume}
  {14}},\ \bibinfo {pages} {2791} (\bibinfo {year} {2014})}\BibitemShut
  {NoStop}%
\bibitem [{\citenamefont {Bernassau}\ \emph {et~al.}(2013)\citenamefont
  {Bernassau}, \citenamefont {Courtney}, \citenamefont {Beeley}, \citenamefont
  {Drinkwater},\ and\ \citenamefont {Cumming}}]{bernassau2013interactive}%
  \BibitemOpen
  \bibfield  {author} {\bibinfo {author} {\bibfnamefont {A.}~\bibnamefont
  {Bernassau}}, \bibinfo {author} {\bibfnamefont {C.}~\bibnamefont {Courtney}},
  \bibinfo {author} {\bibfnamefont {J.}~\bibnamefont {Beeley}}, \bibinfo
  {author} {\bibfnamefont {B.}~\bibnamefont {Drinkwater}}, \ and\ \bibinfo
  {author} {\bibfnamefont {D.}~\bibnamefont {Cumming}},\ }\href@noop {}
  {\bibfield  {journal} {\bibinfo  {journal} {Applied Physics Letters}\
  }\textbf {\bibinfo {volume} {102}},\ \bibinfo {pages} {164101} (\bibinfo
  {year} {2013})}\BibitemShut {NoStop}%
\bibitem [{\citenamefont {Lei}\ \emph {et~al.}(2018)\citenamefont {Lei},
  \citenamefont {Hill}, \citenamefont {{de Le{\'{o}}n Albarr{\'{a}}n}},\ and\
  \citenamefont {Glynne-Jones}}]{Lei2018}%
  \BibitemOpen
  \bibfield  {author} {\bibinfo {author} {\bibfnamefont {J.}~\bibnamefont
  {Lei}}, \bibinfo {author} {\bibfnamefont {M.}~\bibnamefont {Hill}}, \bibinfo
  {author} {\bibfnamefont {C.~P.}\ \bibnamefont {{de Le{\'{o}}n
  Albarr{\'{a}}n}}}, \ and\ \bibinfo {author} {\bibfnamefont {P.}~\bibnamefont
  {Glynne-Jones}},\ }\href@noop {} {\bibfield  {journal} {\bibinfo  {journal}
  {Microfluidics and Nanofluidics}\ }\textbf {\bibinfo {volume} {22}},\
  \bibinfo {pages} {140} (\bibinfo {year} {2018})}\BibitemShut {NoStop}%
\bibitem [{\citenamefont {Muller}\ and\ \citenamefont
  {Bruus}(2014)}]{Muller2014}%
  \BibitemOpen
  \bibfield  {author} {\bibinfo {author} {\bibfnamefont {P.~B.}\ \bibnamefont
  {Muller}}\ and\ \bibinfo {author} {\bibfnamefont {H.}~\bibnamefont {Bruus}},\
  }\href@noop {} {\bibfield  {journal} {\bibinfo  {journal} {Physical Review
  E}\ }\textbf {\bibinfo {volume} {90}},\ \bibinfo {pages} {43016} (\bibinfo
  {year} {2014})}\BibitemShut {NoStop}%
\bibitem [{\citenamefont {Landau}\ and\ \citenamefont
  {Lifshitz}(1967)}]{Landau1967}%
  \BibitemOpen
  \bibfield  {author} {\bibinfo {author} {\bibfnamefont {L.~D.}\ \bibnamefont
  {Landau}}\ and\ \bibinfo {author} {\bibfnamefont {E.~M.}\ \bibnamefont
  {Lifshitz}},\ }\href@noop {} {\bibfield  {journal} {\bibinfo  {journal}
  {Editions de Moscou}\ } (\bibinfo {year} {1967})}\BibitemShut {NoStop}%
\bibitem [{\citenamefont {Muller}\ \emph {et~al.}(2012)\citenamefont {Muller},
  \citenamefont {Barnkob}, \citenamefont {Jensen},\ and\ \citenamefont
  {Bruus}}]{Muller2012}%
  \BibitemOpen
  \bibfield  {author} {\bibinfo {author} {\bibfnamefont {P.~B.}\ \bibnamefont
  {Muller}}, \bibinfo {author} {\bibfnamefont {R.}~\bibnamefont {Barnkob}},
  \bibinfo {author} {\bibfnamefont {M.~J.~H.}\ \bibnamefont {Jensen}}, \ and\
  \bibinfo {author} {\bibfnamefont {H.}~\bibnamefont {Bruus}},\ }\href@noop {}
  {\bibfield  {journal} {\bibinfo  {journal} {Lab on a Chip}\ }\textbf
  {\bibinfo {volume} {12}},\ \bibinfo {pages} {4617} (\bibinfo {year}
  {2012})}\BibitemShut {NoStop}%
\bibitem [{\citenamefont {Muller}\ and\ \citenamefont
  {Bruus}(2015)}]{Muller2015}%
  \BibitemOpen
  \bibfield  {author} {\bibinfo {author} {\bibfnamefont {P.~B.}\ \bibnamefont
  {Muller}}\ and\ \bibinfo {author} {\bibfnamefont {H.}~\bibnamefont {Bruus}},\
  }\href@noop {} {\bibfield  {journal} {\bibinfo  {journal} {Physical Review
  E}\ }\textbf {\bibinfo {volume} {92}},\ \bibinfo {pages} {63018} (\bibinfo
  {year} {2015})}\BibitemShut {NoStop}%
\bibitem [{\citenamefont {Settnes}\ and\ \citenamefont
  {Bruus}(2012)}]{Settnes2012}%
  \BibitemOpen
  \bibfield  {author} {\bibinfo {author} {\bibfnamefont {M.}~\bibnamefont
  {Settnes}}\ and\ \bibinfo {author} {\bibfnamefont {H.}~\bibnamefont
  {Bruus}},\ }\href@noop {} {\bibfield  {journal} {\bibinfo  {journal}
  {Physical Review E}\ }\textbf {\bibinfo {volume} {85}},\ \bibinfo {pages}
  {16327} (\bibinfo {year} {2012})}\BibitemShut {NoStop}%
\end{thebibliography}%

\newpage

\end{document}